\documentclass[%
aip,jcp,amsmath,amssymb,
preprint,%
citeautoscript,
]{revtex4-2}

\usepackage[utf8]{inputenc}
\usepackage{multirow}
\usepackage[dvipsnames]{xcolor}
\usepackage{tikz}
\usetikzlibrary{shapes}

\usepackage{graphicx}

\begin{document}

\preprint{}

\title{Predicting aggregate morphology of sequence-defined macromolecules with Recurrent Neural Networks}
\author{Debjyoti Bhattacharya}
\affiliation{Materials Science and Engineering, Pennsylvania State University, University Park, PA 16802}

\author{Devon C. Kleeblatt}
\affiliation{Materials Science and Engineering, Pennsylvania State University, University Park, PA 16802}

\author{Antonia Statt}
\affiliation{Materials Science and Engineering, University of Illinois, Urbana-Champaign, IL 61801}

\author{Wesley F. Reinhart}
\email[email:]{reinhart@psu.edu}
\affiliation{Materials Science and Engineering, Pennsylvania State University, University Park, PA 16802}
\affiliation{Institute for Computational and Data Sciences, Pennsylvania State University, University Park, PA 16802}

\date{\today}

\begin{abstract}
Self-assembly of dilute sequence-defined macromolecules is a complex phenomenon in which the local arrangement of chemical moieties can lead to the formation of long-range structure.
The dependence of this structure on the sequence necessarily implies that a mapping between the two exists, yet it has been difficult to model so far.
Predicting the aggregation behavior of these macromolecules is challenging due to the lack of effective order parameters, a vast design space, inherent variability, and high computational costs associated with currently available simulation techniques.
Here, we accurately predict the morphology of aggregates self-assembled from sequence-defined macromolecules using supervised machine learning.
We find that regression models with implicit representation learning perform significantly better than those based on engineered features such as $k$-mer counting, and a Recurrent-Neural-Network-based regressor performs the best out of nine model architectures we tested.
Furthermore, we demonstrate the high-throughput screening of monomer sequences using the regression model to identify candidates for self-assembly into selected morphologies.
Our strategy is shown to successfully identify multiple suitable sequences in every test we performed, so we hope the insights gained here can be extended to other increasingly complex design scenarios in the future, such as the design of sequences under polydispersity and at varying environmental conditions.
\end{abstract}


\maketitle

\section{Introduction}
The self-assembly of block copolymers driven by microphase separation of different constituent monomers is a well-studied phenomenon that has received sustained attention for decades.\cite{meier_theory_1969,malmsten_self-assembly_1992,matsen_origins_1996,fredrickson_dynamics_1996,calleja_block_2000,spaeth_comparison_2011,Mai2012,Matsen2012,noshay2013block,feng_block_2017,sternhagen_solution_2018}
This is due to both the fundamental interest in thermodynamics of soft materials and their ability to form a rich array of aggregate morphologies such as spherical micelles\cite{willner_l_micellization_2000,won_giant_1999,forster_micellization_1996,read_recent_2007}, rods\cite{zhou_synthesis_2010,olsen_self-assembly_2008,hayashi_rod-shaped_2019}, strings\cite{guo_self-assembly_2020}, vesicles\cite{jiang_hyperbranched_2015,araste_self-assembled_2021}, spindles\cite{wang_spindle-like_2019}, tubules\cite{guo_self-assembly_2020,cisse_light-fuelled_2020,he_self-assembly_2012}, toroids\cite{kim_development_2013,xu_helical_2020,xu_polymeric_2022}, membranes\cite{monnard_membrane_2002,yurchenco_models_1986,zhang_nanoporous_2015,zhou_self-assembly_2010}, worm-like micelles\cite{willner_l_micellization_2000,won_giant_1999,forster_micellization_1996,read_recent_2007}, and other complex structures\cite{torring_dna_2011,chakraborty_amino_2018}.
As a result of the structural variations that can be obtained (and in some cases also due to their biocompatibility), block copolymers offer a broad range of possible technological applications, including biomaterials\cite{stupp_self-assembly_2010}, photovoltaic devices\cite{bdarling_block_2009,shah_correlations_2010,hadziioannou_semiconducting_2002}, pharmaceuticals\cite{kwon_soluble_1999,rosler_advanced_2002}, nanoreactors\cite{khullar_block_2013,wurbser_chemically_2021}, microelectronics\cite{5388674}, and many more.

Due to the astounding number of patterns now possible to realize in synthetic copolymers, effectively leveraging the possibilities of polymer self-assembly requires predictive tools.
Fortunately, computational approaches are available across the many lengths and time scales relevant to this problem, including Density Functional Theory \cite{terao_machine_2020}, Molecular Dynamics (MD) (from atomistic to coarse-grained) \cite{srinivas_self-assembly_2004,li_brownian_2012,huang_dissipative_2019}, Monte Carlo\cite{patti_monte_2010}, and mean field theories \cite{Matsen2002,zhang_aggregate_2007,guo_self-assembly_2020}.
Mean-field theories, in particular, have emerged as a staple approach to understanding the mechanism of self-assembly and have provided critical guidance for the experimental investigation of related phenomena.

While very successful in describing bulk systems, mean-field approaches are less effective in the dilute case, where the collective arrangement of individual chains becomes relevant.\cite{Mccarty2019}
When theories are available, they require detailed prior information about the polymer chains, such as Flory Huggins parameter and effective chemical potential fields -- which in the end forces one to resort to MD simulations to obtain estimates -- while also not providing any insight into the microscopic details of aggregate morphology.~\cite{Lyubimov2017, Gartner2019}
At the same time, even coarse-grained MD simulations are still too expensive to perform an exhaustive search over all possible monomer patterns to identify a suitable polymer design for a given application.
Recent work\cite{bale_sequence-defined} addressed this challenge by combining coarse-grained Molecular Dynamics (MD) and evolutionary computation to study structure-property relationships of a model copolymer and subsequently performed screening of single chain conformations, thereby identifying sequences spanning a wide range of radius of gyration values.
This illustrates the potential of efficient search schemes to predict the aggregate morphology resulting from copolymer self-assembly using relatively few simulations (compared to the overall design space).

Machine Learning (ML) is well suited to address this need and has emerged as an effective tool for solving various problems in soft matter over the last decade.\cite{Ferguson2011, Reinhart2017, Chen2018, Sun2020, terao_machine_2020, Bhattacharya2021, sclegg_characterising_2021, kim2021polymer} 
Furthermore, ML-based screening has shown promising results in the past wherein it has been used for tasks such as screening of complex molecules for polymer solar cells\cite{jorgensen2018machine}, exploration of nanomedicine design space\cite{yamankurt_exploration_2019}, and systematically identifying novel cancer drug targets\cite{jeon_systematic_2014}, among others\cite{wang2022machine,mehta2021enhanced}.
The premise for this strategy is that from a small number of high-quality training data points, the behavior of a much more extensive collection of possible system configurations can be predicted.
Thus, MD simulation's relatively high computational cost can be amortized over many thousands of predictions from the ML model, reducing the overall compute time.

In other contexts such as natural language processing, it has been shown that the quality of the data encoding is critical to the performance of ML models; feature leaning is regarded as an essential step for creating effective and explainable ML models.\cite{verdonck_special_2021}
Unsurprisingly, the importance of featurization also extends to materials informatics.\cite{jing2019amino, patel_featurization_2022}
Previous applications of ML to materials design have explored features based on combinations of thermodynamic, chemical, and topological information to manually create engineered features\cite{jablonka2021bias}, representing macromolecules as chemistry-informed graph based features\cite{mohapatra2022chemistry}, converting monomeric sequences to image-based features\cite{webb2020targeted}, or simple one-hot encoded features\cite{shi2021predicting}. 

In this work, we apply supervised ML to predict the aggregation behavior of a model copolymer. 
Motivated by the variety of featurization techniques described in the literature, we consider three different encoding schemes to encode the monomer sequences that require information about only the monomer arrangement.
Notably, even the formulation of this problem requires order parameters for the aggregate morphology, which have only recently been developed.\cite{Reinhart2021,statt2021unsupervised}
Using these order parameters, we evaluate several classes of regression models and identify features that lead to the most successful ones.
Since self-assembly is a stochastic process, we also quantify the intrinsic uncertainty as a benchmark for our model evaluation.

After identifying a suitable predictive model, we perform high-throughput screening of monomer sequences with a fixed composition.
This amounts to proposing a small number of target morphologies and then identifying the best candidate sequences from the collection of all possible sequences based on the model prediction.
We find that the model always finds sequences that yield qualitatively similar structures, and in most cases, the quantitative agreement is within the variance expected due to intrinsic variability.
The proposed method is also attractive because the inference is very fast after the model has been trained.
Our results show that even highly complex behaviors of soft matter systems can be predicted using ML with relatively few training data, given that sufficient attention is given to identifying suitable representations of the inputs and outputs.

\begin{figure*}
    \centering
    \includegraphics[width=0.8\textwidth]{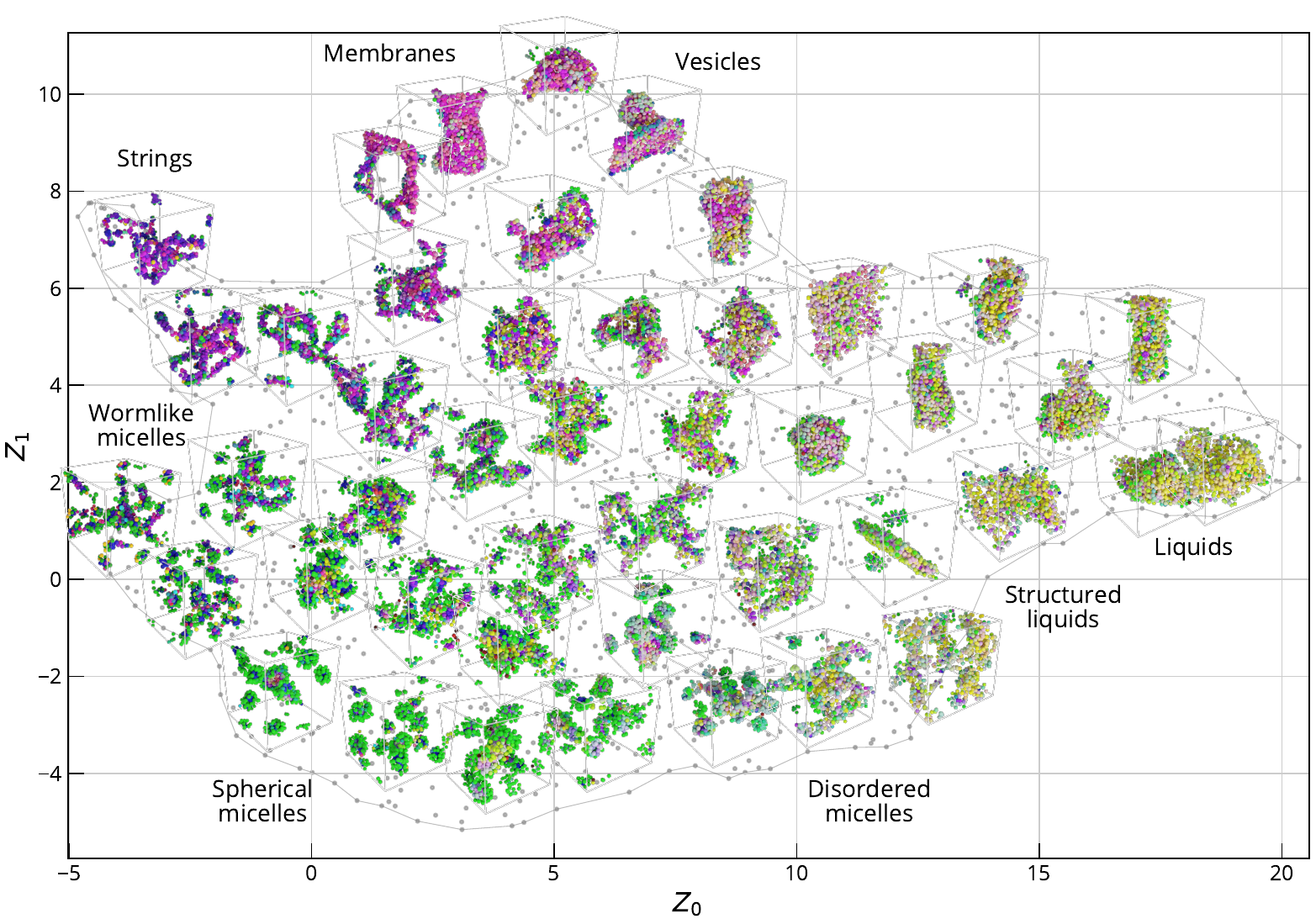}
    \caption{
    Learned manifold describing the $2\,038$ morphologies from Ref.~\citenum{statt2021unsupervised}.
    45 selected snapshots are shown to illustrate the spatial variation in morphology across the manifold.
    $Z_0$ roughly corresponds to strings versus droplets, while $Z_1$ roughly corresponds to micelles versus sheets.
    }
    \label{fig:dataset}
\end{figure*}

\section{Methods \label{sec:methods}}
\subsection{Molecular Dynamics simulations}

We use the same model as our other recent work,\cite{statt2021unsupervised} adapted from disordered protein liquid-liquid phase separation.\cite{Statt2020}
In the model, polymer chains contain a fixed number of coarse-grained beads that belong to one of two chemical groups: attractive $A$ beads and purely repulsive $B$ beads.
The attractive beads interact via the standard Lennard-Jones potential ($\sigma, \varepsilon, r_\mathrm{cut} = 3 \sigma$),\cite{Jones1924} while the repulsive beads use the Weeks-Chandler-Andersen potential.\cite{Weeks1971}
Bonds between adjacent monomers are handled by the standard finitely extensible nonlinear elastic (FENE) potential ($R_0 = 1.5 \sigma, K = 30 \varepsilon/\sigma^2$).\cite{Kremer1990}
Additional details are available in Ref.~\citenum{Statt2020}.

In these simulations, only the monomer sequence was varied while all other parameters were fixed.
We considered a fixed composition with 60\% $A$ beads and 40\% $B$ beads, with a chain length of 20 monomers for all polymers.
The reason for keeping the composition fixed is that the order parameters developed in our previous work considered only structures formed by this composition; new compositions may lead to distinct structures not present in our training set and therefore produce anomalous results.\cite{statt2021unsupervised}
The volume of the box was $V \approx (40 \sigma)^3$ with $N = 500$ chains; the monomer volume fraction of approximately 8\% is below the overlap concentration.
Equilibration was performed with HOOMD-blue\cite{Glaser2015, Anderson2008} (version 2.9.4) in the NVT ensemble using a Langevin thermostat to enforce a temperature of $T = 0.5 \varepsilon$.

Configurations were allowed to relax for $2 \times 10^5 \tau$ from an initially random state before the structure was evaluated.
We examined the kinetics of the self-assembly process in our prior work and this simulation time was the standard used in Ref.~\citenum{statt2021unsupervised}.
However, we did observe metastable intermediate morphologies from some sequences, especially vesicle formers.
In this work, we therefore consider only the aggregate morphology at time $t = 2 \times 10^5 \tau$ rather than the true equilibrium state.
The stochastic nature of the self-assembly process is explicitly addressed in Section~\ref{sec:variability}.

\subsection{Unsupervised learning}

After equilibrating the polymer chains, we apply our recently developed unsupervised learning approach to obtain order parameters for the aggregates.\cite{statt2021unsupervised, Reinhart2021}
In short, this approach uses an extensive collection of local geometric features to create a low-dimensional embedding of each bead's neighborhood via nonlinear manifold learning.
These features are then aggregated in a global pooling scheme and reduced a second time to yield two snapshot-wide order parameters, $Z_0$ and $Z_1$.
This $\mathbf{Z}$ is a rotation-, translation-, and permutation-invariant measure of the aggregate morphology.
The algorithm is described in detail in Refs.~\citenum{statt2021unsupervised} and \citenum{Reinhart2021}, and relies primarily on the Uniform Manifold Approximation and Projection (UMAP)\cite{Ziolek2021} approach.
UMAP is a non-linear, unsupervised dimensionality reduction (i.e., manifold learning) technique.

As introduced in Ref.~\citenum{statt2021unsupervised}, the $\mathbf{Z}$ manifold reveals the possible aggregate morphologies formed from the chosen 60\%-$A$-type composition.
The structure of the manifold is shown in Fig.~\ref{fig:dataset} with some representative snapshots overlaid to illustrate the shape of the aggregates in each region of $\mathbf{Z}$.
It can be seen that aggregates around the periphery of the manifold correspond to ``archetypes'' while those closer to the middle is mixtures of multiple types.
This is a result of the manifold learning scheme preserving the topology of the data during projection, such that the distance between archetypes is maximized while centrally located aggregates are equidistant to the locations of their respective pure archetypes.
The $2\,038$ points shown in Fig.~\ref{fig:dataset} comprise the training data for our supervised learning task described below.

\subsection{Supervised learning}

The main objective of this work is to approximate the function $f: X \to Z$ where $X$ corresponds to the monomer sequence of a polymer chain and $Z$ is a quantitative description of the aggregate morphology as provided by our unsupervised learning scheme.
To this end, we consider a large number of common regression algorithms, including linear regression and its regularized variants, the ensemble method Random Forest, the kernel method K-Neighbors, and three variants of Neural Networks (NNs): Multi-Layer Perceptron (MLP), Convolutional Neural Network (CNN), and Recurrent Neural Network (RNN).
The \texttt{scikit-learn}\cite{scikit-learn} package is used for all except the NNs, which are implemented with \texttt{pytorch}.\cite{pytorch}
In all cases, we use 10-fold cross-validation to train an ensemble of models on the $2\,038$ labeled samples generated in Ref.~\citenum{statt2021unsupervised} (and publicly available via the Zenodo repository\cite{stattData2021}).

In order to apply regression models to the problem at hand, we must encode the monomer sequences in a machine-readable form (i.e., as a vector $X$).
However, the sequences themselves are not suitable for regression as they are natively written as a string of characters ($A$ or $B$).
Therefore,  in this paper, we focus on three methods of featurization: sequence vectors, token counting, and implicit feature learning.

One unusual feature of this problem is the symmetry invariance of the polymer sequences; each sequence represents a physical object (i.e., a polymer chain) without a concept of directionality.
In fact, in the coarse-grained model employed here, it would be nonsensical to have a discrepancy between the model prediction when feeding the sequence ``forward'' versus ``reversed.''
As a result, we always consider the prediction of regression model to be $\tilde{f}(X) = \frac{1}{2} \left[f(X) + f(X') \right]$,
where $X$ is a sequence and $X'$ is the mirror of that sequence.
This choice ensures that all models yield exactly one result for each polymer sequence.

For all models, we have performed systematic hyperparameter tuning using a Bayesian optimization approach.
The optimization was performed using the \texttt{bayesian-optimization} package \cite{bayesopt}.
In each case, we optimized based on only one fold of a ten-fold cross-validation set.
Final hyperparameter selection was validated by training on all ten folds and evaluating the average performance.
The RMSE on the test data was taken as the objective, and the Expected Improvement acquisition function was used.
We ran at least 50 iterations for all models, but we also ran some for up to 100 iterations if they did not converge after 50.

\subsection{Sequence vector encoding}

\begin{figure}
    \centering
    \includegraphics[width=0.5\columnwidth]{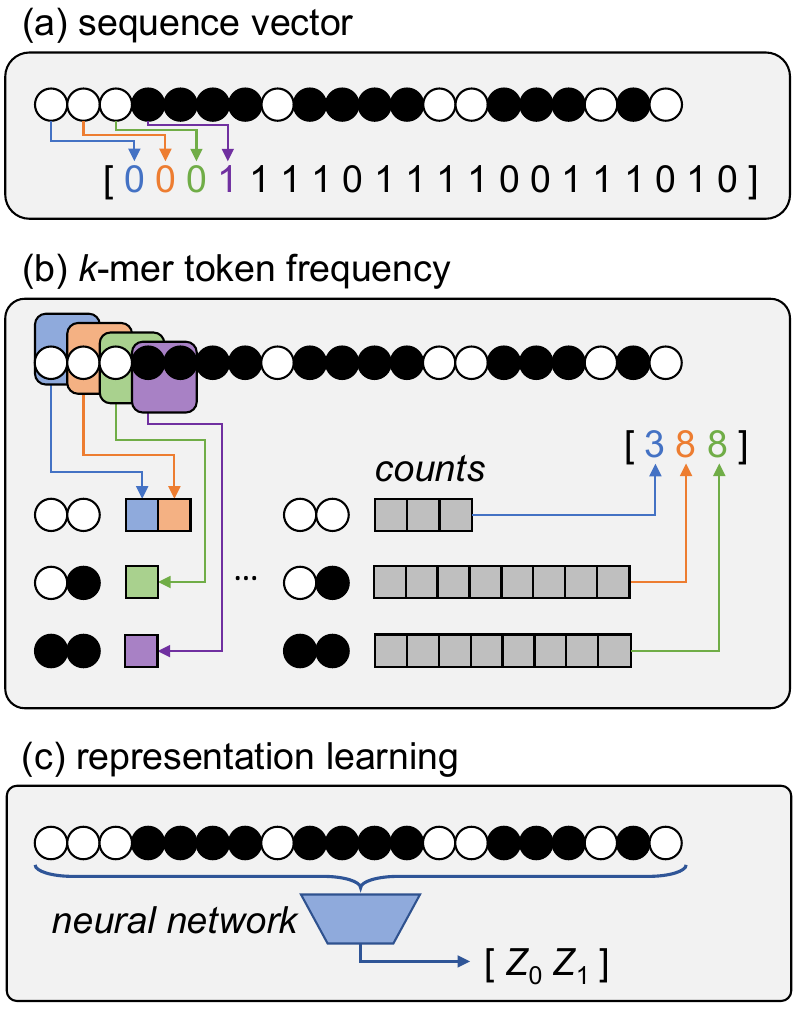}
    \caption{Schematic illustration of the three representation strategies used in the regression task.}
    \label{fig:representations}
\end{figure}

The most straightforward featurization technique we investigated was encoding the sequence directly as a feature vector.
In this scheme, the monomer chemical identity is represented as a vector of class labels (illustrated in Fig.~\ref{fig:representations}(a)).
Here, $A$ is represented as 0, and $B$ is represented as 1, yielding a vector of binary values corresponding to each monomer position in the chain.
When provided with sequences encoded in this vector form, the regression models do not receive any physically meaningful information about the topology of the chain.
That is, the performance would be identical if all the sequences were permuted in the same way (e.g., swapping beads 1 and 5).
For instance, when a Random Forest Regressor was trained on ten random permutations of the sequence vectors, the coefficient of determination (R-squared) was $0.6596 \pm 0.0011$ compared to $0.6604 \pm 0.0028$ using the correct sequence ordering (i.e., statistically indistinguishable results).
This presents a serious limitation in that the model is totally insensitive to the relative position of the monomers to each other, and can only consider the absolute position of $A$ and $B$ type monomers.

\subsection{Token counting (\textit{k}-mers)}

In order to imbue the models with a sense of the relative positions of different monomers, we consider counting the frequency of $k$-mer tokens in the sequence (illustrated in Fig.~\ref{fig:representations}(b)).
In this context, a $k$-mer is a substring of length $k$ taken from the complete monomer sequence.
Because there is no sense of direction in the physical polymer chain, we count both ``forward'' and ``reverse'' copies of the $k$-mer substring.
For instance, the substring \texttt{AB} is a $k$-mer of length $k = 2$ that appears twice in the string \texttt{AAAAAABBBBBBBBAAAAAA} -- once on each side of the $B$ block, even though the order of the $k$-mer is reversed in the two occurrences.
Note that in this case the features $X = X'$, so $\tilde{f} = f(X) = f(X')$ always, and there is no need to average the two for consistency.
After counting the $k$-mers, a frequency vector is passed as input to the regression model.

Here we have considered $k$-mers with length $k = 2$ to $k = 10$.
The lower limit of 2 ensures the model receives sufficient information to make a prediction, as $k = 1$ would reduce the count of $A$ and $B$ type beads in the chain, which is fixed in our study.
The upper limit is a practical one to limit the number of tokens that must be counted; there are $1\,085$ possible $10$-mers which makes the token frequency slow to compute.
This kind of engineered feature has been known to increase the predictive power of the ML models allowing the flexibility to use less complex models that are faster. \cite{wen_classification_2019,solis-reyes_open-source_2018}
We do not need to train on mirrored copies of the sequences because the $k$-mers are combined with their mirrored versions in the frequency vector; the token frequency of a reversed sequence is identical to the original sequence.

\subsection{Representation learning}

Representation learning is an integral part of natural language processing to understand the representations of raw tokens like words or characters in a collection of texts.
In this encoding scheme, we rely on learned embeddings of the raw monomer tokens to derive meaning from the sequence (illustrated in Fig.~\ref{fig:representations}(c)).
While this can also be thought of as using the sequence vector scheme together with a Neural Network(NN) -based regression model, we discuss this class of encodings separately due to the ability of these regressors to apply nonlinear transformations on the input data to yield more meaningful representations inside the model.
The impact of this capability will be clearly visible in the Results.

For the Representation Learning encoding scheme, we consider three types of NNs: MLP, CNN, and RNN.
The MLP is a feedforward artificial NN with fully connected layers.
While relatively simple, this architecture is able to learn relationships between the different monomer positions on the chain through connections between the input layer and hidden neurons.
On the other hand, CNN and RNN explicitly consider the structure of the sequence data.
CNNs learn filters that leverage the spatial distribution of data in a sample; the sequence is represented as a 1D ``image'' and the spatial filters identify patterns such as blocky regions.
Meanwhile, RNNs read each monomer in the chain one at a time and use a memory scheme to encode patterns.
RNNs have been widely used for modeling sequence data in ordinal or temporal problems such as image captioning~\cite{wang_evolutionary_2020}, language to language translation~\cite{auli_joint_2013}, natural language processing~\cite{8588934}, speech recognition~\cite{8588934} and so on.
This should result in a greater ability to abstract sequence-based patterns from the input.

After hyperparameter tuning, the best MLP model was a 12-layer fully connected NN with neurons in each successive hidden layer decreasing evenly from 128 in the first layer to 57 in the last (i.e., each layer had 7 fewer weights).
We used the ReLU activation function (it was not optimized as a hyperparameter).
This model is referred to as MLP-12 throughout this article to distinguish it from the shallow MLP-1 that had only one layer (to demonstrate the importance of multiple nonlinear transforms).

Our CNN was comprised of a variable number of convolutional layers, each with variable kernel width and variable number of channels.
After hyperparameter optimization, the optimal architecture was found to be 8 convolutional layers with 13 channels coming from kernels 12 monomers wide.
We again used the ReLU activation function.

\begin{figure}
    \centering
    \includegraphics[width=0.5\columnwidth]{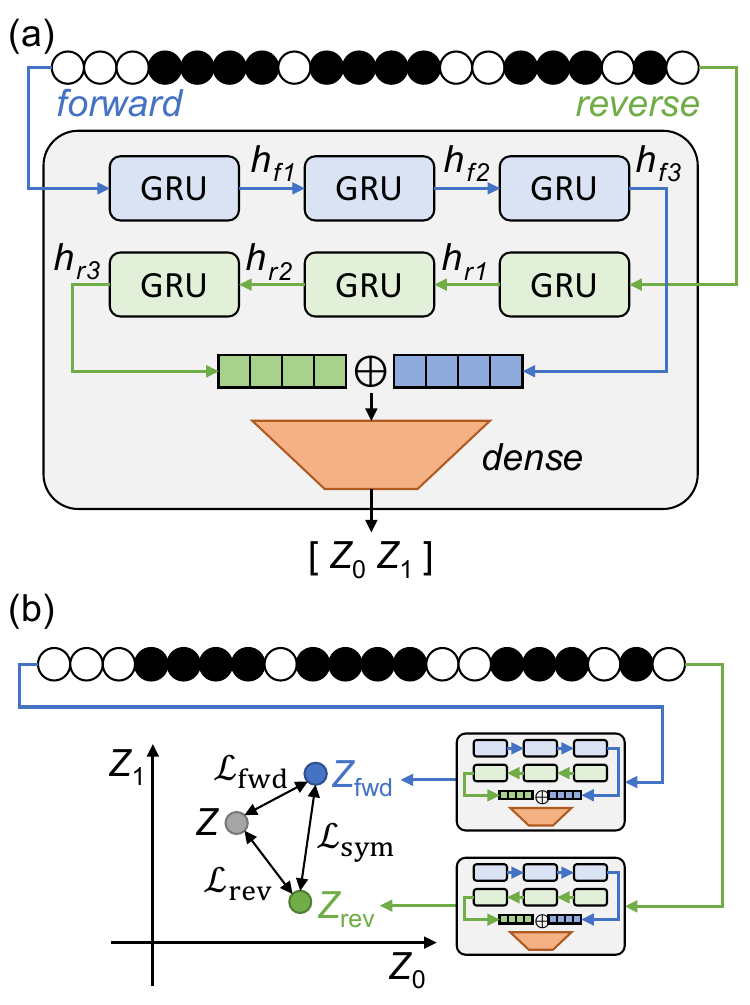}
    \caption{Schematic of the RNN regressor architecture and training procedure.}
    \label{fig:rnn-arch}
\end{figure}

For the RNN, we use Gated Recurrent Units (GRUs).\cite{cho2014learning}
GRUs are a relatively recent class of RNNs introduced in 2014 and is an advancement over a standard RNN because it overcomes the problems of vanishing gradients by using gates and memory cells. Additionally, they are computationally more efficient than Long Short Term Memory (LSTM)\cite{hochreiter_long_1997} networks because GRUs have simpler architecture and fewer gates.
Bidirectional GRUs further train two networks, where one traverses the input sequence from left to right (``forward'') while another traverses from right to left (``reverse'').
The bidirectional model thereby yields two low-dimensional embeddings of the sequence, which are merged via concatenation.
This enables the network to have the sequence information in both directions at each token in the sequence and is known to perform better than unidirectional GRU models in other sequence modeling tasks.\cite{ju2019study,du2018biomedical}
The architecture is illustrated in Fig.~\ref{fig:rnn-arch}(a).

The final RNN model employed for sequence prediction had three bidirectional GRU layers with 7-dimensional hidden states followed by a fully connected linear layer.
We again used the ReLU activation function.
The model was again trained on the sequence vectors.
We trained with a batch size of 128 for $1\,000$ epochs using the Adam optimizer with a learning rate of $0.01$.
Furthermore, a custom loss function was used to ensure the model learned a symmetric representation of the polymer sequences:
\begin{equation}
\mathcal{L} = \mathcal{L}_\mathrm{fwd} + \mathcal{L}_\mathrm{rev} + \mathcal{L}_\mathrm{sym},
\label{eq:symmetry}
\end{equation}
where $\mathcal{L}_\mathrm{fwd}$ indicates the ``forward'' pass, $\mathcal{L}_\mathrm{rev}$ indicates the ``reverse'' pass, and $\mathcal{L}_\mathrm{sym}$ is a penalty for violating symmetry.
Each of these terms is simply a Mean Squared Error loss between respective observations and targets:
\begin{align}
    \mathcal{L}_\mathrm{fwd} &= \mathbb{E}_X [ (f(X) - Z)^2 ] \\
    \mathcal{L}_\mathrm{rev} &= \mathbb{E}_X [ (f(X') - Z)^2 ] \\
    \mathcal{L}_\mathrm{sym} &= \mathbb{E}_X [ (f(X) - f(X'))^2 ]
\end{align}
This loss function is illustrated schematically in Fig.~\ref{fig:rnn-arch}(b).

\section{Results\label{sec:results}}

\subsection{Stochasticity in observed aggregate morphology}\label{sec:variability}

\begin{figure}
    \centering
    \includegraphics[width=0.6\columnwidth]{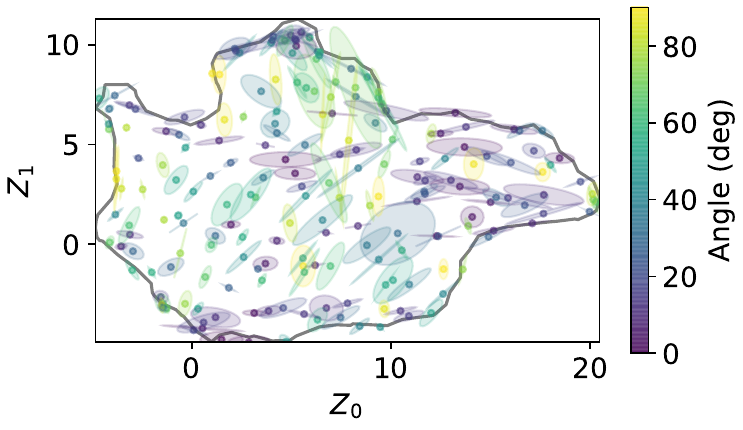}
    \includegraphics[width=0.6\columnwidth]{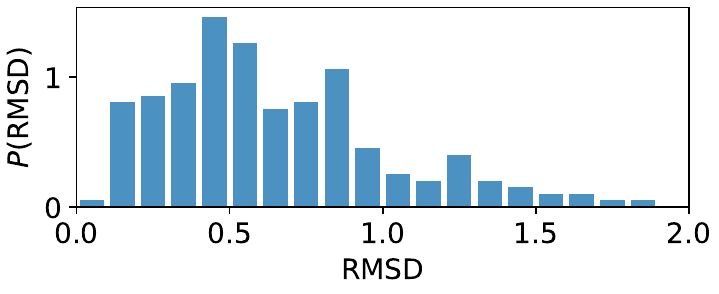}
    \caption{Intrinsic variability for the same sequence over the latent space.
    (top) Covariance ellipses from 3 replicas.
    The ellipses provide a visual representation of the anisotropic uncertainty; they are calculated from the eigenvectors of the covariance matrix from replica MD simulations with the same monomer sequence.
    (bottom) Histogram of Root Mean Squared Distance from an average of replicas.}
    \label{fig:uncertainty}
\end{figure}

Our prior work\cite{statt2021unsupervised} observed that repeated self-assembly simulations of the identical sequences at the same thermodynamic conditions resulted in considerable variability.
However, we did not perform an exhaustive analysis of the intrinsic variability of the self-assembly process that results from these stochasticity in the initial state and random thermal forces.
Here, we simulate three replicas each of 200 different sequences to systematically evaluate the expected amount of variability from self-assembly of the same sequence.
In essence, the objective is to bound the best possible performance we should expect from a deterministic regression model of a stochastic process.

The sequences represent a subset of the $2\,038$ labeled sequences shown in Fig.~\ref{fig:dataset}, chosen to sample the learned manifold approximately uniformly using K-Means clustering.
Each chosen sequence was repeated three times, and the covariance between samples was analyzed to provide context for the regression models.
The eigenvectors of the covariance matrix for each set of simulations were used to render representative ellipses in Fig.~\ref{fig:uncertainty}.
These can be thought of as the smallest ellipse encompassing all the points observed from a single sequence.

As shown in the figure, there is a large range of variances observed in the simulations.
The distribution is reflected in the histogram of root-mean-square deviation (RMSD) in the bottom panel of Fig.~\ref{fig:uncertainty}.
The median of observed deviations is $\mathrm{RMSD} = 0.57$, while the mean is $\mathrm{RMSD} = 0.67$, confirming the skew visible in the histogram. 
This should provide a lower bound on the expected performance of any regression model since even the generating function itself (i.e., the MD simulation) will have a typical uncertainty of around $0.5$ to $0.7$.

There may also be some spatial dependence on the uncertainty, such as reduced variance for $Z_0 < 5$ compared to $Z_0 > 5$.
The results are colored by angle to show the spatial dependence on the direction of variance, such as those in the liquid-like $Z_0 > 10$ region being biased towards higher variance in the horizontal direction compared to those in the vesicle-like $Z_1 > 5$ region being biased towards higher variance in the vertical direction.

\subsection{Regression with sequence vectors}

We evaluated the performance of the following regression models using the sequence vector input scheme: Linear\cite{maulud_review_2020}, Lasso\cite{1556215}, Ridge\cite{rajan_efficient_2022}, K-Neighbors\cite{ATM10170}, Random Forest\cite{7919549}, and MLP\cite{7919549}.
Lasso and Ridge regressions are linear regression schemes that use $L_1$ and $L_2$ regularization, respectively, which results in Lasso preferring to shrink some weights to zero.
Evaluating these three models gives insight into how effectively each input feature produces accurate output.
The other models all have some nonlinear capability.
K-Neighbors use a distance metric to identify the most similar observations and make the inference by aggregating those similar samples; the nonlinearity comes from switching which neighbors are included in the aggregation.
This model allows us to evaluate whether the input encoding exhibits a similar topology to the $Z$ labels.
Random Forest uses an ensemble of decision trees to predict the output, with the trees being optimized according to information-theoretic measures.
Its nonlinearity comes from branching in the trees.
Finally, the MLP uses repeated linear transforms modified by a nonlinear activation function.
It has many trainable weights and cannot be readily interpreted.

\begin{figure}
    \centering
    \includegraphics[width=0.5\columnwidth]{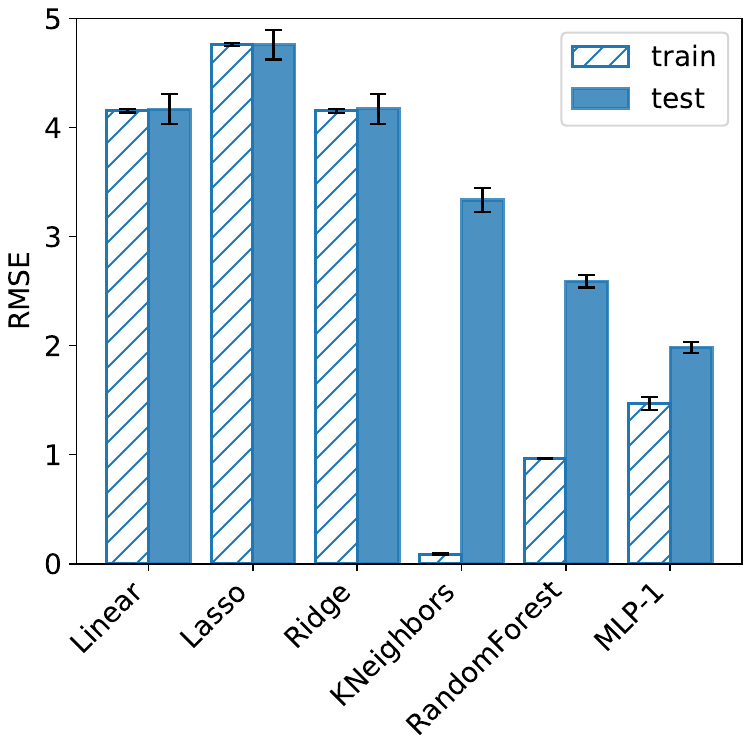}
    \includegraphics[width=0.5\columnwidth]{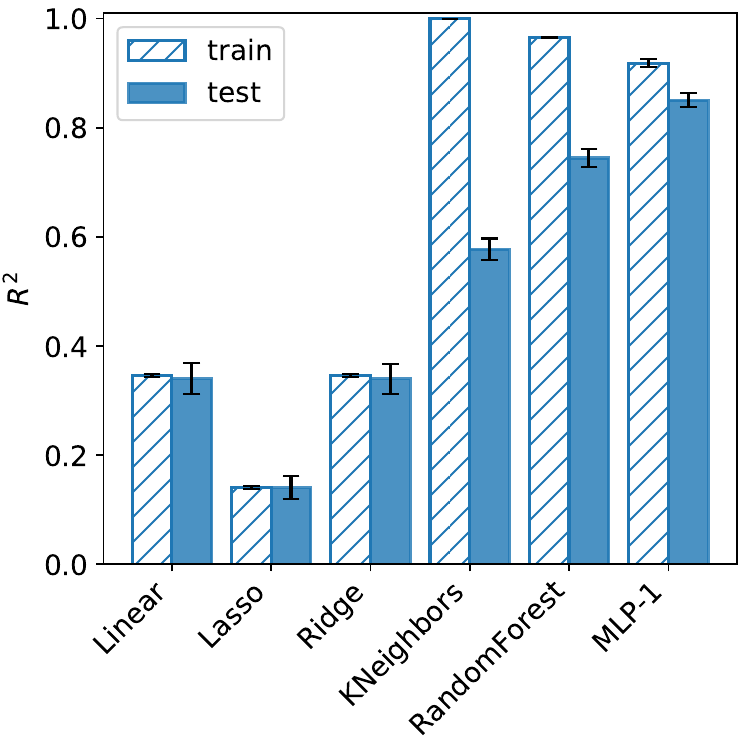}
    \caption{
    Comparison of model performance on the sequence vectors, both root-mean-square error (RMSE, lower is better) and  model coefficient of determination ($R^2$, higher is better) are shown.
    Here, the error bars show averaged standard deviation representing the test performance with respect to the 10 trained models of each class of models  obtained by cross-fold validation.
    }
    \label{fig:sequence-barchart}
\end{figure}

The root-mean-square error (RMSE) and coefficient of determination ($R^2$) for different models on the label-encoded features are shown in Fig.~\ref{fig:sequence-barchart}.
For ML models, the training set represents data seen by the model during fitting, while the testing set represents unseen data that should indicate the performance in general.
Significant discrepancies in train and test performance, therefore, indicate overfitting.

Starting with the linear models, we observe that Linear and Ridge regression perform about the same, while Lasso performs worse.
On the other hand, the nonlinear models appear to be much more susceptible to overfitting, with K-Neighbors being the worst offender.
The poor performance of K-Neighbors in testing suggests that Euclidean distance in the ``sequence vector space'' does not approximate the structure of the aggregate morphology space well.
We tried other metrics, such as Hamming distance and cosine similarity, but the results did not vary significantly.
Random Forest performs better in testing and also has less egregious overfitting.
Finally, we see that MLP-1 performs the best in testing while also shrinking the gap between test and train performance.

It is noteworthy that the nonlinear methods strictly outperform the linear methods in both training and testing.
All evidence supports the notion that the sequence vector poorly represents the physics at play in the aggregation process.
For instance, the overfitting of the K-Neighbors shows how unreliable it is to use chemical identity of individual monomers as features.
Likewise, regularization of the linear models reduces performance without reducing overfitting (because no overfitting is observed).

\subsection{Regression with token frequency}

\begin{figure}
    \centering
    \includegraphics[width=0.7\columnwidth]{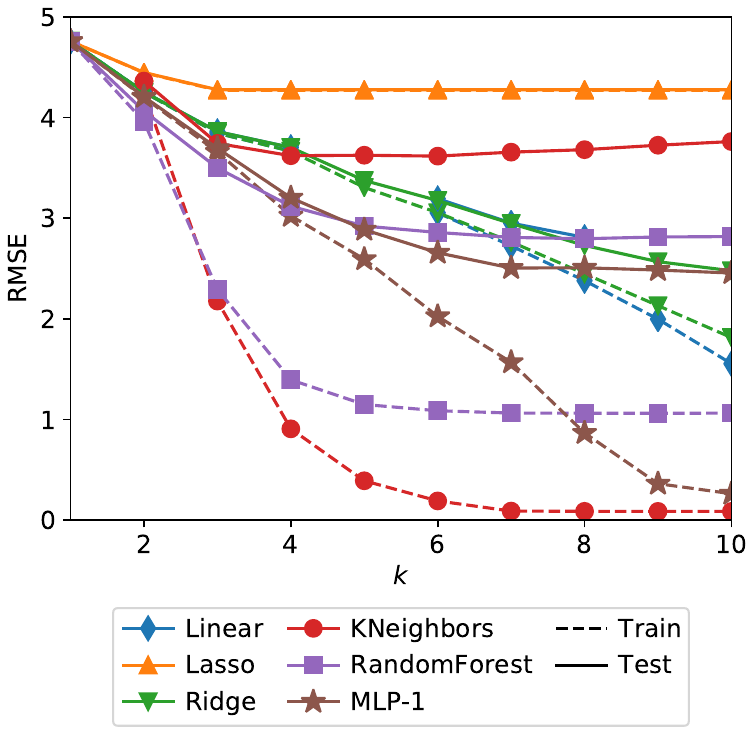}
    \caption{
    Model RMSE as a function of $k$-mer length in train (dashed lines) and test (solid lines).
    Error bars from cross-validation are omitted as they are small relative to the plotted values, with mean and standard deviation $0.074 \pm 0.057$ across all points.
    }
    \label{fig:regression-sklearn}
\end{figure}

Next, we evaluated the performance of the same regression methods using the token frequency of $k$-mers.
Critically, this provides a nonlinear featurization that captures interactions between monomers up to $k-1$ beads away in the chain, which should improve the performance relative to using sequence vectors.
The results for varying $k$ are shown in Fig.~\ref{fig:regression-sklearn}.
As before, lower RMSE is better, and the discrepancy between train and test performance indicates overfitting.

In general, RMSE decreases as $k$ increases, except for K-Neighbors, which shows a local minimum at $k=4$ before increasing again with high $k$.
This is likely because the distance metric becomes more meaningful as additional tokens are added to the vector up to a point, whereafter the space becomes too high-dimensional, and the neighborhood is less meaningful.
For the other models, having access to more features and features that convey more non-local information improves the predictive performance.
At the same time, this also results in increased overfitting for higher $k$, as indicated by the growing gap between test and train curves for each model.

Again evaluating the linear model variants, we see that Lasso performs significantly worse than Linear or Ridge,  which are nearly identical.
It is interesting to observe this in the $k$-mer featurization strategy in addition to the sequence vector one since the $k$-mer tokens are supposed to include nonlinearity.
We assumed at the outset that certain patterns would indicate different aggregate morphology, such as alternating monomers corresponding to more liquid-like structures, while repeated monomers of the same type would probably correspond to micelle-like structures.
Based on Fig.~\ref{fig:regression-sklearn}, this is not the case; the feature selection induced by Lasso's $L_1$ regularization leads to clearly worse performance than the weight sharing induced by Ridge's $L_2$ regularization.
In essence, the regressors actually suffer from regularization, indicating that utilizing more of the available features is necessary to generalize to unseen data.
The only time either of the regularized models outperform Linear Regression  is for $k > 8$, but even here Lasso regression performs quite poorly.

To corroborate this observation, we also performed hyperparameter tuning for the regularized linear models.
These have a single parameter $\alpha$ that controls the relative strength of the regularization term.
For both Lasso and Ridge, the optimal solution was to drive $\alpha$ towards zero; the regularized models never performed better than vanilla Linear Regression for $k \le 8$.
However, when $k > 8$ the Linear Regression failed to generalize to the unseen test data.
On these cases Lasso performed exactly the same as it did for $k = 3$, which was generally very poor but insensitive to the proliferation of input features.
As a result of these observations, the data reported here are for the default value of $\alpha = 1$ rather than the optimized $\alpha \approx 0$ which approximate Linear Regression.

MLP-1 and Random Forest perform about the same as each other across the $k$ range, with MLP-1 performing slightly better overall.
Random Forest can be seen significantly overfitting even at low $k$, indicating again that $k$-mer token counts are not very informative.
This is also reinforced by the improved performance of the nonlinear models on the label encoded vectors compared to any $k$.
Hyperparameter tuning did not yield statistically significant improvement in the Random Forest model.
MLP-1 was deliberately left unoptimized to compare to the optimized DNN version described in the following section.

\subsection{Representation learning\label{sec:regression}}

Here we investigated several variants of Deep Neural Networks to attempt to learn more meaningful representations from the sequences.
These included a deep MLP with 12 layers (MLP-12), a Convolutional Neural Network (CNN), and a Recurrent Neural Network (RNN).
Our most successful model was a bidirectional-GRU-based RNN, with an RMSE of $1.407 \pm 0.091$ on training data.
This was substantially better than the other DNNs, where the MLP-12 gave a RMSE of $1.72 \pm 0.16$ and the CNN gave a RMSE of $1.75 \pm 0.13$.
While the MLP-12 does have the capacity for representation learning, it cannot utilize the structural information in the sequences.

It was surprising that the CNN performs only slightly better than the MLP-12, while the RNN shows a strong advantage over either of those models.
To the former point, we suspect that the 1D nature of the problem allows the MLP-12 to match the performance of the CNN, whereas in higher dimensions the MLP would lack the capacity to effectively learn convolutions.
The latter point may be related to the number of trainable weights in the different models: $101 \, 617$ for the MLP-12, $15 \, 186$ for the CNN, and $3 \, 058$ for the RNN.
Thus the RNN model may be able to better learn generalizable representations using (80\% of) $2 \, 038$ labeled data during training.

Note that the hyperparameter optimization was based on performance on the test set, so the strong overfitting shown in Fig.~\ref{fig:regression-comparison} is not a result of optimizing based on training performance.
Instead, these models with many trainable weights still generalized (i.e., gave lower RMSE on unseen data) better than smaller models with similar architectures.
Thus, it is possible that the performance gap between RNN and CNN or MLP-12 may shrink with more training data.

On a related note, our hyperparameter optimization was tightly constrained for the CNN to control the dimensionality of the problem -- we kept the kernel width and number of channels constant across all the layers.
It is possible that an unconstrained (higher dimensional) hyperparameter optimization could find a CNN model that matches or exceeds the performance of RNN.
However, our result is still significant in the sense that the RNN performs better than the CNN given a similar hyperparameter tuning program (i.e., we also kept the hyperparameters of the GRU layers the same throughout the network).

We also performed an ablation study on the custom symmetry loss term in Eq.~\ref{eq:symmetry}.
In the case of the RNN, the performance with a simple MSE loss term (while still respecting the equivalence of the reversed sequences) deteriorated to RMSE = $2.116 \pm 0.078$, a 50\% increase in error.
On the other hand, the RMSE of the MLP-12 and CNN models did not result in a statistically significant change in the RMSE.
This is likely a result of the sequential read-in of the RNN, which cannot be made symmetric by construction, whereas the CNN and MLP-12 can learn to produce symmetric mappings.

As previously noted, the MLP-1 has a limited ability to learn representations with its single layer, so its performance is reduced at RMSE of $1.983 \pm 0.053$.
The next-best model that has no representation learning capability is Ridge regression for $k=10$ with RMSE $2.48 \pm 0.15$, which is only slightly better than Random Forest and Linear regression.
The dominance of the DNN models demonstrates the unsuitability of both the sequence vector and token frequency features for the regression task.

\subsection{Comparison of regression models}

\begin{figure}
    \centering
    \includegraphics[width=0.6\columnwidth]{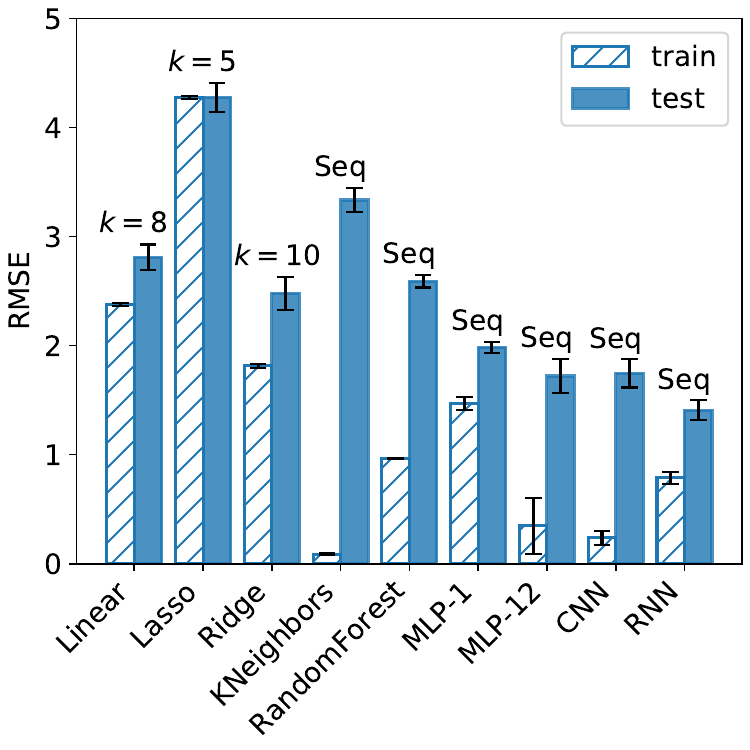}
    \caption{Comparison of the best result for each class of overall model representations considered, where lower values are better.
    The label \texttt{Seq} indicates the full sequence vector, while \texttt{k=L} indicates $k$-mers of length $L$. Similar to  Fig.~\ref{fig:sequence-barchart}, the error bars for the respective models in this figure show the averaged standard deviation representing the test performance with respect to the 10 trained models of each class of models  obtained by cross-fold validation.
    }
    \label{fig:regression-comparison}
\end{figure}

To summarize, we considered three types of features as input to the regression models in this study: sequence vectors, $k$-mer token counts, and implicit representation learning by NN.
Six different regressors comprising three linear and three nonlinear models were trained on the sequence vector features and on the token counts with varying lengths of pattern from $k=2$ to $k=10$.
Three additional NN-based architectures with implicit representation learning were also tested on the sequence vectors.
Of the encoding schemes, representation learning by NN consistently performed the best.
Supporting this, the RNN model performed the best of all the models, with MLP-12 and CNN tied for a statistically indistinguishable second place.

Beyond accuracy, it may be worth considering the training and inference time for the different models.
The only models that took more than 1 s to train were Random Forest, MLP-1, and the DNN models (MLP-12, CNN, and RNN).
Random Forest took 1.01 s to train on sequence vector and 17.5 s to train on tokens with $k=10$ but had inference times of 10-20 ms in both cases (to predict 203 samples).
MLP-1 took 27.5 s to train on sequence vector and 199 s to train on tokens with $k=10$ but had inference times of 0.5 ms and 2.42 ms, respectively.
Training times for the DNNs varied, but a representative figure is 10 epochs/s when training on an NVIDIA Tesla P100 GPU.
For 1,000 epochs this corresponds to a 100 s training time for a single model.
Inference times for the DNNs depended on complexity, with MLP-12 taking about 0.9 ms but the CNN taking 10.5 ms and RNN taking 13.1 ms (again, for a single model to predict 203 samples).
Note that inference was performed on CPU; GPU hardware was only used for model training.
For reference, MD simulations typically took 6-13 minutes on the same GPU hardware.
Therefore avoiding even a single MD simulation by use of the DNN predictions would save time during screening.

We summarize the results in Fig.~\ref{fig:regression-comparison} wherein the best version of each model over all input representations is considered.
The plot shows that token counts perform better than sequence vectors for linear models, while for nonlinear models, the performance is best with the sequence vector as input.
The RNN reading full sequence vectors outperforms all other regression schemes across all representation learning techniques, achieving RMSE of $1.407 \pm 0.091$ in testing.
Note that this compares favorably to the RMSD of $0.67$ reported in Fig.~\ref{fig:uncertainty}, which would be the minimum error we could theoretically expect.
As the best-performing model, the RNN is used in the high-throughput screening procedure that follows in the remainder of this article.

\subsection{High-throughput screening}

The obvious practical application of a well-trained regression model is to identify sequences that assemble into a target aggregate morphology from among the many thousands of possible permutations.
Accordingly, we deploy the ensemble of ten trained RNN models (one for each fold in the cross-validation of  Fig.~\ref{fig:regression-comparison}) to predict the aggregate morphology (i.e., $Z$ embedding) for all $63 \, 090$ unique sequences with 60\% $A$ monomers (the origin of this number is explained in the ESI\dag).
From these predictions, we perform high-throughput screening to select the most promising candidates for exhibiting the desired aggregate structure.

We present the results from two demonstrations of this high-throughput screening approach.
First, we seek sequences with morphology similar to the six sequences reported in Table 2 of Ref.~\citenum{statt2021unsupervised}, corresponding to the ``archetypal'' structures in the learned manifold.
We refer to these structures as strings, membranes, vesicles, liquid droplets, spherical micelles, and wormlike micelles, for labels (a)-(f) respectively.
Second, we select eight morphologies from the $2 \, 038$ labeled data points using K-Means clustering, such that they are  well dispersed across the latent space.
In each case, we choose the five sequences with the smallest Euclidean distance in $Z$ between the prediction of the RNN model $\tilde{Z}$ and the target location $Z_t$, not permitting sequences that are already present in the $2 \, 038$ labeled set.
For each sequence, five replicas were simulated to quantify stochasticity in the self-assembly process.

\begin{figure}
    \centering
    \includegraphics[width=0.7\textwidth]{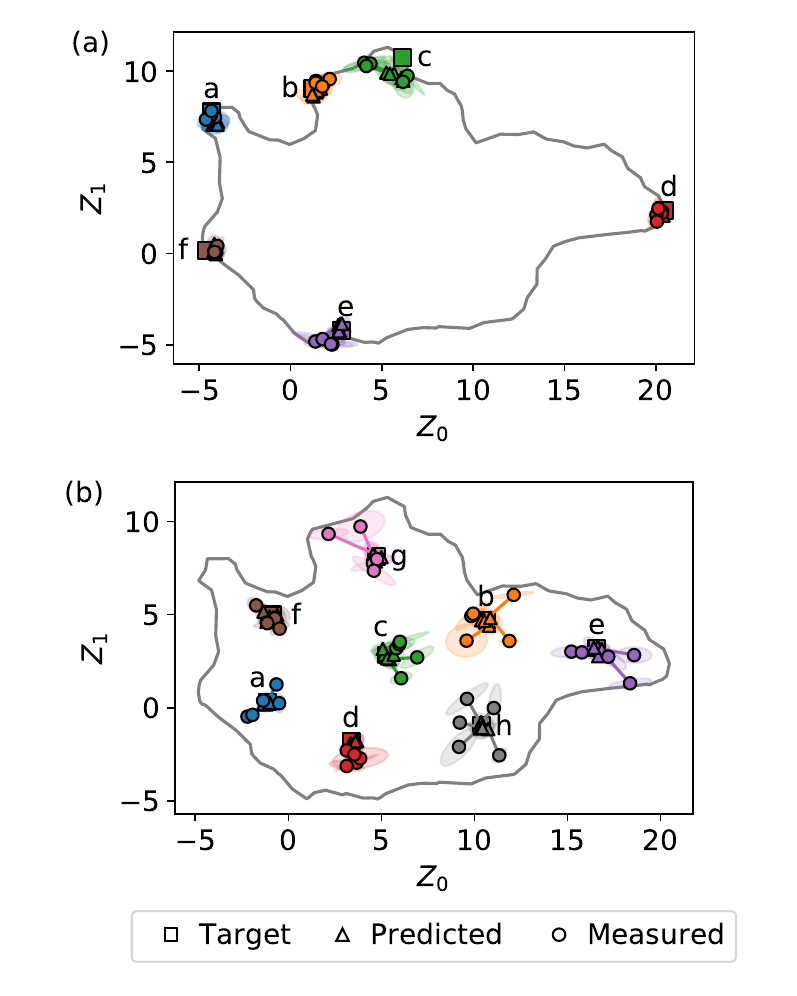}
    \caption{Results of the high-throughput screening using the RNN regressor trained on 2038 labeled data from Ref~\citenum{statt2021unsupervised}.
    (top) Targets based on structures/sequences from Table 1 of Ref~\citenum{statt2021unsupervised}.
    (bottom) Targets based on k-means clustering of the labeled sequences.
    Similar to Fig.~\ref{fig:uncertainty}(top), in both the cases in this figure, the ellipses give a visual representation of the anisotropic uncertainty; they are calculated from the eigenvectors of the covariance matrix from replica MD simulations with the same monomer sequence.}
    \label{fig:full-design}
\end{figure}

The results of the high-throughput screening are shown in Fig.~\ref{fig:full-design}.
At least one of the five candidate sequences includes the target structure within one standard deviation of its mean for all but a few of the targets.
As indicated by the ellipses, the covariance among replicas is significantly smaller in the archetypes (top) compared to the K-Means clusters (bottom).
This is consistent with Fig.~\ref{fig:uncertainty}, which shows reduced variance on the periphery of the latent space compared to the center;
Since UMAP assumes a uniform density, morphologies with the higher disorder (and thus more possible states) are distorted to occupy a larger area than more ordered ones (with fewer possible states).

Note that variance is shown for both the simulation results (circles) and RNN predictions (triangles) in Fig.~\ref{fig:full-design}; for the regression model, the variance comes from cross-fold validation.
Large uncertainty around the targets -- especially (h) -- reveals low confidence in the ensemble to predict successful candidates.
The ensemble prediction uncertainty generally seems to be smaller than the actual error observed in MD simulations.

\begin{figure}
    \centering
    \includegraphics[width=0.7\columnwidth]{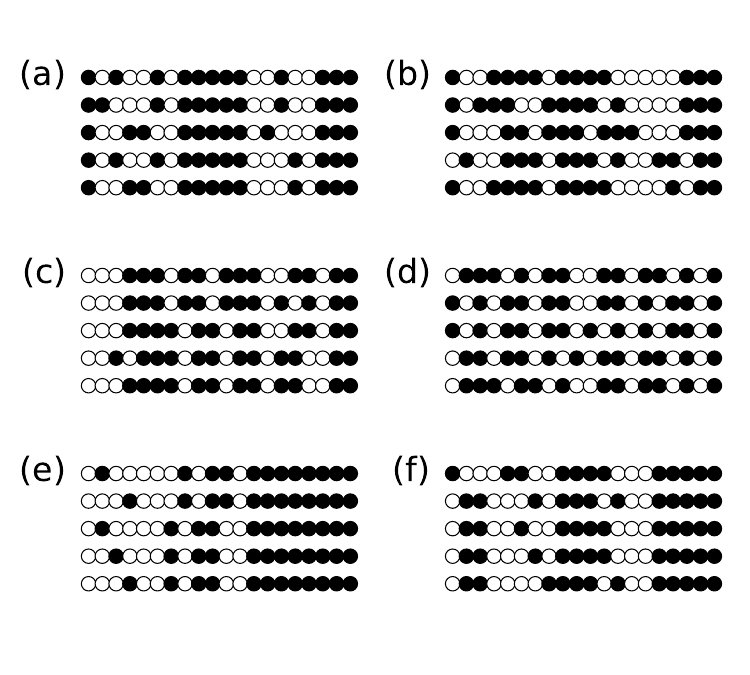}
    \caption{Ranked list of sequences used to generate Fig.~\ref{fig:full-design}a.
    }
    \label{fig:sequences-archetypes}
\end{figure}

The sequences selected by the screening procedure the archetypal structures in Fig.~\ref{fig:full-design}(a) are shown in Fig.~\ref{fig:sequences-archetypes}.
To reiterate, these five sequences are the ones predicted by the RNN to lie closest to a target $Z_t$ from among all possible candidates.
In the figure, they are ordered by how close they appeared to the target in validation MD simulations, but in many cases the results are well within the radius expected due to stochastic variation.
Therefore, these figures are best read as five nearly-equally-good candidates to produce the target morphology, and the general patterns spanning all five should be noted.

In Fig.~\ref{fig:sequences-archetypes}, these patterns are quite clear.
For panel (a), corresponding to strings, we see A-blocks of length 5 in the center of the chain together with another A-block of length 3 at one end.
For panels (b)-(c), corresponding to membranes and vesicles, we see regularly spaced A-blocks of length 3-4 throughout the chain broken by single or paired B monomers.
For panel (d), corresponding to liquids, the sequences are nearly random, with only a few blocks of length 3 and none longer than that.
For panel (e), corresponding to spherical micelles, the RNN favors long A-blocks with only a few A monomers dispersed throughout the rest of the chain.
Finally, for panel (f), corresponding to wormlike micelles, sequences with two long A-blocks and one short A-block are chosen.
Similar patterns can be observed in Fig.~S1 in the ESI\dag, but we refrain from listing them in the main text since they largely follow the archetypes described above but are less regular.

\begin{figure*}
    \centering
    \includegraphics[width=0.8\textwidth]{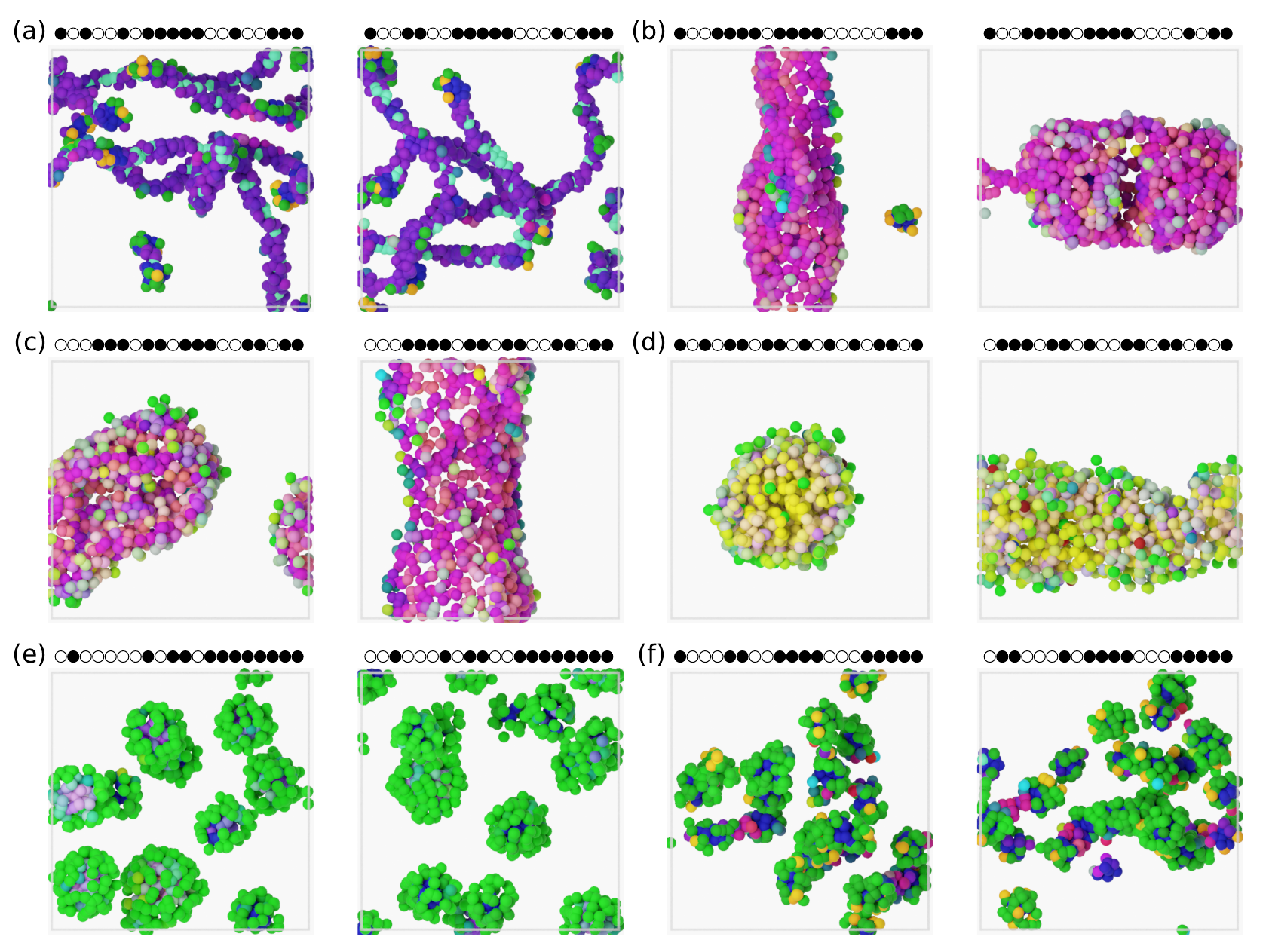}
    \caption{Snapshots from the simulations in the bottom row of Fig.~\ref{fig:full-design}, corresponding to the targets examined in our prior work.
    Coloring is determined by the local environment around each particle, as described in Ref.~\citenum{statt2021unsupervised}.
    The left panels are closest to the target in the batch of candidates (the best result of 25 samples), right panels are farthest away (the worst result of 25 samples).
    Labels correspond to those in Fig.~\ref{fig:full-design}.}
    \label{fig:design-snapshots}
\end{figure*}

To further illustrate the successful outcome of this procedure, we show the best and worst snapshots for each archetypal structure target in Fig.~\ref{fig:design-snapshots}.
Visual inspection reveals that tightly clustered points in latent space such as (a) appear very similar, while those with a higher variance like (c) exhibit slightly different morphology.
To be clear, we emphasize that each target was matched with five sequences with five replicas each, so the configuration shown to the right of each panel (a)-(f) is the worst of 25 possible snapshots targeting that morphology.
An equivalent figure is available in the ESI\dag for the structures depicted in Fig.~\ref{fig:full-design}(b).
Note that those tend to exhibit greater variance as indicated by the larger ellipses compared to Fig.~\ref{fig:full-design}(a).

\subsection{Interpretation of RNN inference}

Fig.~\ref{fig:sequences-archetypes} shows that while general patterns hold for the five selected sequences, it may be difficult to develop simple rules to achieve high accuracy on the classification task.
For instance, the strings in Fig.~\ref{fig:sequences-archetypes}(a) appear to be characterized by A-blocks of length 5 in the center and length 3 at one end, but some sequences in Fig.~\ref{fig:sequences-archetypes}(b) are only one or two edits away from meeting this definition, yet we know they form membranes instead.
While a human expert could reasonably learn to classify the final observed morphologies based on characteristics of the sequence such as the example given here, the task performed by the RNN model is far more difficult than the one just described.
The RNN is performing a quantitative regression on the sequences and predicting precisely where it will appear in a continuous structural space.
Even though the prediction is not always accurate, the model is always precise, providing numerical values for the two order parameters that follow some internal logic.
For instance, the model is converting the raw sequence into something like ``A-blocks of length 5 in the center and length 3 at one end, except with some random bit flips throughout'' and then into a value for $\tilde{Z}$ as the data passes through the layers of the network.

While the details of the RNN are mostly hidden behind its complexity, we can gain some qualitative understanding by evaluating its predictions in some case studies.
To develop an intuition for these predictions we consider a paradigm of contrast: placing sequences that the RNN model considers good matches for a target structure side by side with sequences predicted to be bad matches.
Following this approach, we can describe the model predictions using both the presence and absence of recognizable features.

To maximize the information gained through this contrastive approach, we search for the most similar sequences that are predicted to have the greatest discrepancy in morphology.
In effect, we ask the following question: how could a target sequence be modified with to achieve (i) the least change in predicted morphology and (ii) the greatest change in predicted morphology.
We measure the using the Levenshtein edit distance\cite{levenshtein1966binary} (as implemented in this\cite{levenshtein} Python package), which is slightly more sophisticated than a Euclidean distance as it considers operations like shifting a substring by one unit to be the same as a bit flip.
Due to the symmetry of the sequences we also define our distance metric to be the minimum of the edit distance on the forward and reversed target.

\begin{figure}[h]
    \centering
    \includegraphics[width=0.48\textwidth]{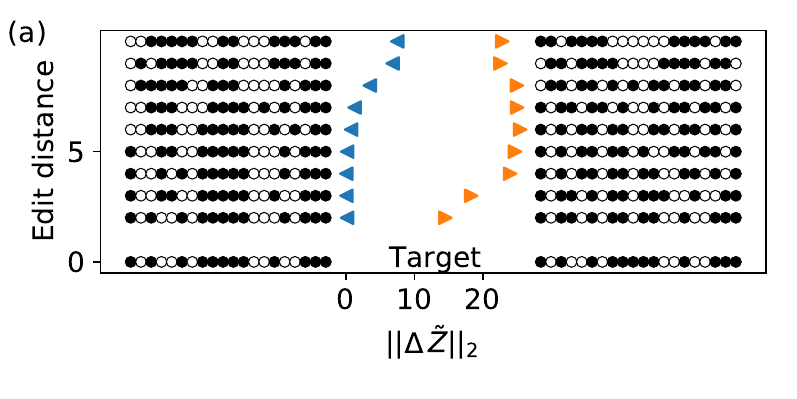}
    \includegraphics[width=0.48\textwidth]{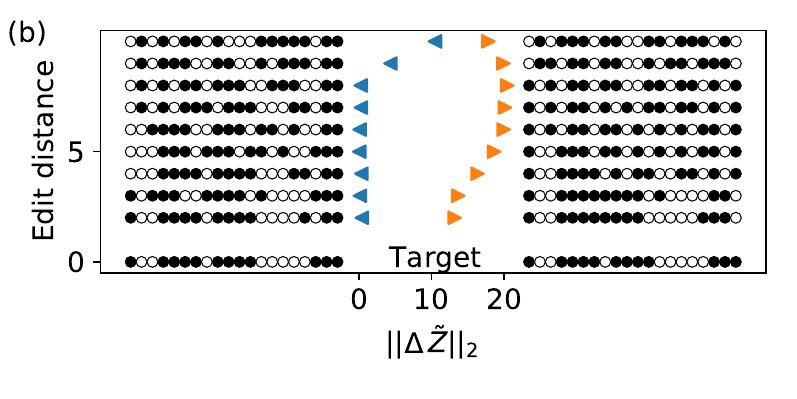}
    \includegraphics[width=0.48\textwidth]{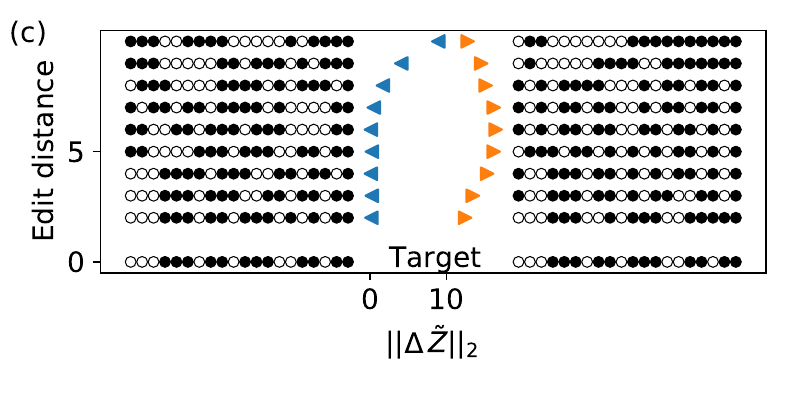}
    \includegraphics[width=0.48\textwidth]{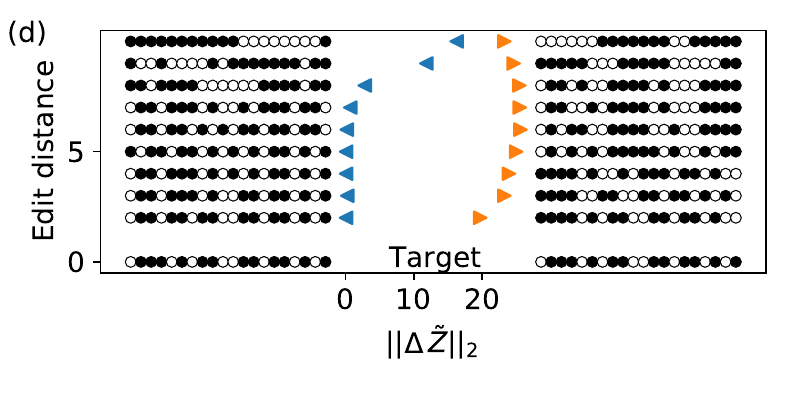}
    \includegraphics[width=0.48\textwidth]{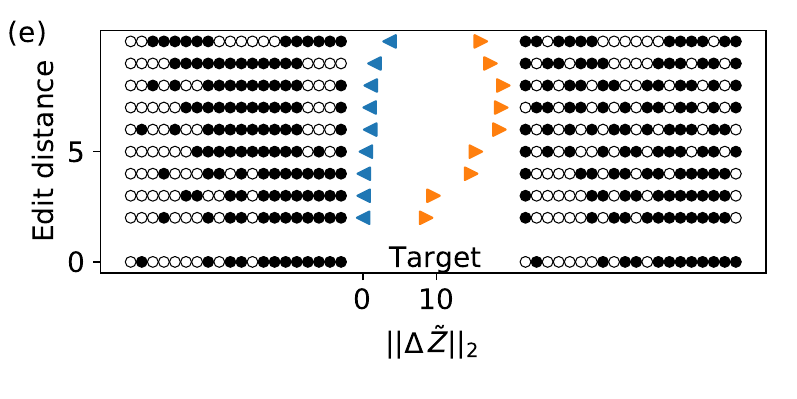}
    \includegraphics[width=0.48\textwidth]{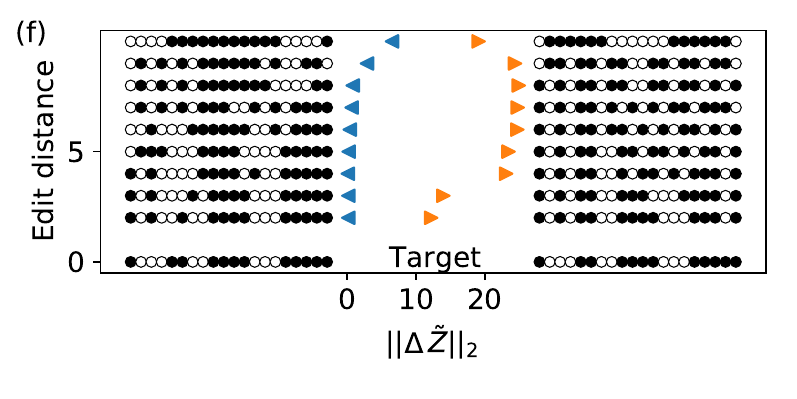}
    \caption{ Contrastive analysis of RNN inference on selected sequences from Fig.~\ref{fig:full-design}(a).
    Symbols show Levenshtein edit distance versus predicted distance from target sequence in $Z$ space.
    Blue left triangles show minimum $||\Delta \tilde{Z}||_2$ at fixed edit distance from among all possible sequences (at fixed composition), while orange right triangles show maximum.}
\end{figure}

The results of this analysis are shown in Fig.~\ref{fig:edits-archetypes}.
In each panel, a single sequence is analyzed by contrasting the sequences predicted to have the most similar and least similar morphology at a fixed edit distance.
Thus, moving up the y axis results shows sequences that resemble the target sequence less, while the x axis separates sequences that the RNN predicts will form a similar morphology (left) from those predicted to form a very different morphology (right).
Note there are no sequences at edit distance of one due to fixed composition (i.e., any edit of A-to-B necessitates a corresponding B-to-A edit).

Taking Fig.~\ref{fig:edits-archetypes}(a) as an example, we again see the ``A-blocks of length 5 in the center and length 3 at one end'' motif characteristic of the string-forming sequences at the bottom (at zero edit distance and labeled ``Target'').
As we move up the left column of sequences, various changes are made to the sequence without disrupting this motif, resulting in little change to the predicted distance from the Target -- until the number of edits exceeds seven, where a corresponding drift is observed in the predicted distance.
Meanwhile, moving up the right column shows the fewest number of edits that can maximally disrupt the motif: first breaking the central block, then the end block.
With only four edits the motif is completely erased, resulting in a plateau in maximum $Z$ distance.

Nearly the same trend can be observed in Fig.~\ref{fig:edits-archetypes}(e), with a diblock motif corresponding to spherical micelles.
In this case, the minimally disruptive edits correspond to shifting the A-block from the edge to the center, while maximally disruptive edits shorten the A-block and lead to an essentially random sequence.
Precisely the reverse effect can be observed in Fig.~\ref{fig:edits-archetypes}(d), wherein an initially non-blocky sequence is given several short blocks that coalesce into longer blocks with additional edits.

While it is impossible to fully explain the decisions of the RNN, this analysis should offer some insight into the trends that are especially favorable or unfavorable for particular morphologies.
An equivalent analysis is presented for the sequences from Fig.~\ref{fig:full-design}(b) in the ESI\dag.

\section{Conclusions\label{sec:conclusions}}

In this work, we demonstrate the use of supervised ML to identify monomer sequences for a model copolymer that self-assembles into aggregates with prescribed morphology.
The objective of the supervised regression task is to learn a mapping between the monomer sequence and the morphology of the self-assembled aggregates under fixed environmental conditions.
Since this target morphology is used to fit a regression model, we require a quantitative order parameter, which was only recently developed in Ref.~\citenum{statt2021unsupervised}.
We emphasize that this provides a continuous description of morphology throughout the entire structural space, rather than consolidating aggregates into discrete classes.

To obtain an accurate predictive model, we evaluated three different representation schemes for the monomer sequences: raw sequence vectors, $k$-mer token counts, and feature learning by NNs.
The effectiveness of each scheme was determined by considering the accuracy (via R-squared and RMSE metrics) of regression models using that representation and trained to predict the morphology of $2\,038$ sequences simulated in our prior work.\cite{statt2021unsupervised}
The results clearly show that feature learning is the most effective strategy, whereas raw sequence vectors perform poorly for linear models (low accuracy) and nonlinear kernel-based models (overfitting).
Surprisingly, $k$-mer token counting was only more effective than raw sequence vectors for linear models, presumably because these models have no other way of accounting for the nonlinearity of the problem.
The raw sequence vectors were more effective in all other cases, and it was most effective when feature learning was available via NN.
We also found that RNNs with GRU layers performed better than NNs with only fully connected layers (i.e., MLPs) or convolutional layers (CNN).

The best regression model we trained was a bidirectional-GRU-based RNN, which gave an RMSE of $1.407 \pm 0.091$ when deployed on unseen test data.
This compares favorably with the intrinsic uncertainty observed in the self-assembly process, which had $\mathrm{RMSD} = 0.67$.
Using this model, we performed high throughput screening to identify suitable sequences from the over $63\,000$ possible sequences at the chosen composition.
For each of 14 target aggregate morphologies, represented as a point in the continuous morphology manifold, we selected the five sequences predicted to be closest to that point by the RNN model.
MD simulations of these sequences showed that the model successfully identified sequences that matched the qualitative morphology of the target in the vast majority of cases, and at least one of the five sequences matched quantitatively in every case.
Furthermore, the accuracy was much better than the RMSE when targeting archetypes than the disordered structures, indicating that the high variability in the center of the manifold may dominate estimates of model accuracy.
Both MD simulation data and the trained RNN models are available on Zenodo\cite{this_data}, and our source code (mostly in the form of Jupyter notebooks) is available on Github\cite{this_github}.

A major limitation of the present study is the restriction to fixed composition, density, temperature, and so on, with only the monomer sequence varying between simulations.
This was primarily driven by the availability of existing data from prior work, but we also anticipate additional challenges when training the model on a much larger domain.
However, the practical usefulness of the proposed approach will depend on its ability to scale up to increasingly complex design scenarios.
We also limited our study to monodisperse chains, which does not reflect the reality of most experimental copolymer systems. 
Thus, it will be crucial to include polydispersity in the future to capture the physical reality of these materials.
Our work here clearly demonstrates that these open challenges are worth pursuing with an RNN-based regression model.

Finally, we speculate that our results will hold for targets other than our learned order parameters if the property of interest are functions of the aggregate morphology.
Essentially, the results will be most applicable to implicit functions of Z.
In principle it is possible to predict any property that is a function of the polymer sequence, but in these cases it is not clear whether the same conclusions will be found regarding effectiveness of different model architectures.
We also do not know a priori how different environments will affect the model performance.
For instance, for copolymer melts we speculate that the local variation in monomer sequence of individual chains will be less influential to the chain packing, perhaps decreasing or even reversing the superiority of the RNN over CNN and MLP models.
We leave this investigation for future work.

\section*{Author contributions}
Debjyoti Bhattacharya: conceptualization, software, data curation, writing – original draft, writing – review and editing, visualization.
Devon Kleeblatt: software, investigation.
Antonia Statt: conceptualization, writing – original draft, writing – review and editing.
Wesley Reinhart: conceptualization, software, data curation, writing – original draft, writing – review and editing, visualization, supervision.

\section*{Conflicts of interest}
There are no conflicts to declare.

\begin{acknowledgments}
This work was supported by the Department of Materials Science and Engineering, the Institute for Computational and Data Sciences, and the Materials Research Institute at the Pennsylvania State University.
\end{acknowledgments}

\bibliography{references}

\begin{thebibliography}{103}%
\makeatletter
\providecommand \@ifxundefined [1]{%
 \@ifx{#1\undefined}
}%
\providecommand \@ifnum [1]{%
 \ifnum #1\expandafter \@firstoftwo
 \else \expandafter \@secondoftwo
 \fi
}%
\providecommand \@ifx [1]{%
 \ifx #1\expandafter \@firstoftwo
 \else \expandafter \@secondoftwo
 \fi
}%
\providecommand \natexlab [1]{#1}%
\providecommand \enquote  [1]{``#1''}%
\providecommand \bibnamefont  [1]{#1}%
\providecommand \bibfnamefont [1]{#1}%
\providecommand \citenamefont [1]{#1}%
\providecommand \href@noop [0]{\@secondoftwo}%
\providecommand \href [0]{\begingroup \@sanitize@url \@href}%
\providecommand \@href[1]{\@@startlink{#1}\@@href}%
\providecommand \@@href[1]{\endgroup#1\@@endlink}%
\providecommand \@sanitize@url [0]{\catcode `\\12\catcode `\$12\catcode
  `\&12\catcode `\#12\catcode `\^12\catcode `\_12\catcode `\%12\relax}%
\providecommand \@@startlink[1]{}%
\providecommand \@@endlink[0]{}%
\providecommand \url  [0]{\begingroup\@sanitize@url \@url }%
\providecommand \@url [1]{\endgroup\@href {#1}{\urlprefix }}%
\providecommand \urlprefix  [0]{URL }%
\providecommand \Eprint [0]{\href }%
\providecommand \doibase [0]{https://doi.org/}%
\providecommand \selectlanguage [0]{\@gobble}%
\providecommand \bibinfo  [0]{\@secondoftwo}%
\providecommand \bibfield  [0]{\@secondoftwo}%
\providecommand \translation [1]{[#1]}%
\providecommand \BibitemOpen [0]{}%
\providecommand \bibitemStop [0]{}%
\providecommand \bibitemNoStop [0]{.\EOS\space}%
\providecommand \EOS [0]{\spacefactor3000\relax}%
\providecommand \BibitemShut  [1]{\csname bibitem#1\endcsname}%
\let\auto@bib@innerbib\@empty
\bibitem [{\citenamefont {Meier}(1969)}]{meier_theory_1969}%
  \BibitemOpen
  \bibfield  {author} {\bibinfo {author} {\bibfnamefont {D.~J.}\ \bibnamefont
  {Meier}},\ }\bibfield  {title} {\enquote {\bibinfo {title} {Theory of block
  copolymers. {I}. {Domain} formation in {A}-{B} block copolymers},}\ }\href
  {https://doi.org/https://doi.org/10.1002/polc.5070260106} {\bibfield
  {journal} {\bibinfo  {journal} {Journal of Polymer Science Part C: Polymer
  Symposia}\ }\textbf {\bibinfo {volume} {26}},\ \bibinfo {pages} {81--98}
  (\bibinfo {year} {1969})},\ \bibinfo {note} {\_eprint:
  https://onlinelibrary.wiley.com/doi/pdf/10.1002/polc.5070260106}\BibitemShut
  {NoStop}%
\bibitem [{\citenamefont {Malmsten}\ and\ \citenamefont
  {Lindman}(1992)}]{malmsten_self-assembly_1992}%
  \BibitemOpen
  \bibfield  {author} {\bibinfo {author} {\bibfnamefont {M.}~\bibnamefont
  {Malmsten}}\ and\ \bibinfo {author} {\bibfnamefont {B.}~\bibnamefont
  {Lindman}},\ }\bibfield  {title} {\enquote {\bibinfo {title} {Self-assembly
  in aqueous block copolymer solutions},}\ }\href
  {https://doi.org/10.1021/ma00046a049} {\bibfield  {journal} {\bibinfo
  {journal} {Macromolecules}\ }\textbf {\bibinfo {volume} {25}},\ \bibinfo
  {pages} {5440--5445} (\bibinfo {year} {1992})}\BibitemShut {NoStop}%
\bibitem [{\citenamefont {Matsen}\ and\ \citenamefont
  {Bates}(1996)}]{matsen_origins_1996}%
  \BibitemOpen
  \bibfield  {author} {\bibinfo {author} {\bibfnamefont {M.~W.}\ \bibnamefont
  {Matsen}}\ and\ \bibinfo {author} {\bibfnamefont {F.~S.}\ \bibnamefont
  {Bates}},\ }\bibfield  {title} {\enquote {\bibinfo {title} {Origins of
  {Complex} {Self}-{Assembly} in {Block} {Copolymers}},}\ }\href
  {https://doi.org/10.1021/ma960744q} {\bibfield  {journal} {\bibinfo
  {journal} {Macromolecules}\ }\textbf {\bibinfo {volume} {29}},\ \bibinfo
  {pages} {7641--7644} (\bibinfo {year} {1996})},\ \bibinfo {note} {publisher:
  American Chemical Society}\BibitemShut {NoStop}%
\bibitem [{\citenamefont {Fredrickson}\ and\ \citenamefont
  {Bates}(1996)}]{fredrickson_dynamics_1996}%
  \BibitemOpen
  \bibfield  {author} {\bibinfo {author} {\bibfnamefont {G.~H.}\ \bibnamefont
  {Fredrickson}}\ and\ \bibinfo {author} {\bibfnamefont {F.~S.}\ \bibnamefont
  {Bates}},\ }\bibfield  {title} {\enquote {\bibinfo {title} {Dynamics of
  {Block} {Copolymers}: {Theory} and {Experiment}},}\ }\href
  {https://doi.org/10.1146/annurev.ms.26.080196.002441} {\bibfield  {journal}
  {\bibinfo  {journal} {Annual Review of Materials Science}\ }\textbf {\bibinfo
  {volume} {26}},\ \bibinfo {pages} {501--550} (\bibinfo {year} {1996})},\
  \bibinfo {note} {\_eprint:
  https://doi.org/10.1146/annurev.ms.26.080196.002441}\BibitemShut {NoStop}%
\bibitem [{\citenamefont {Calleja}\ and\ \citenamefont
  {Roslaniec}(2000)}]{calleja_block_2000}%
  \BibitemOpen
  \bibfield  {author} {\bibinfo {author} {\bibfnamefont {F.}~\bibnamefont
  {Calleja}}\ and\ \bibinfo {author} {\bibfnamefont {Z.}~\bibnamefont
  {Roslaniec}},\ }\href {https://books.google.com/books?id=2-4vnJ7kqXAC} {\emph
  {\bibinfo {title} {Block {Copolymers}}}}\ (\bibinfo  {publisher} {Taylor \&
  Francis},\ \bibinfo {year} {2000})\BibitemShut {NoStop}%
\bibitem [{\citenamefont {Spaeth}, \citenamefont {Kevrekidis},\ and\
  \citenamefont {Panagiotopoulos}(2011)}]{spaeth_comparison_2011}%
  \BibitemOpen
  \bibfield  {author} {\bibinfo {author} {\bibfnamefont {J.~R.}\ \bibnamefont
  {Spaeth}}, \bibinfo {author} {\bibfnamefont {I.~G.}\ \bibnamefont
  {Kevrekidis}},\ and\ \bibinfo {author} {\bibfnamefont {A.~Z.}\ \bibnamefont
  {Panagiotopoulos}},\ }\bibfield  {title} {\enquote {\bibinfo {title} {A
  comparison of implicit- and explicit-solvent simulations of self-assembly in
  block copolymer and solute systems},}\ }\href
  {https://doi.org/10.1063/1.3580293} {\bibfield  {journal} {\bibinfo
  {journal} {The Journal of Chemical Physics}\ }\textbf {\bibinfo {volume}
  {134}},\ \bibinfo {pages} {164902} (\bibinfo {year} {2011})},\ \bibinfo
  {note} {publisher: American Institute of PhysicsAIP}\BibitemShut {NoStop}%
\bibitem [{\citenamefont {Mai}\ and\ \citenamefont
  {Eisenberg}(2012)}]{Mai2012}%
  \BibitemOpen
  \bibfield  {author} {\bibinfo {author} {\bibfnamefont {Y.}~\bibnamefont
  {Mai}}\ and\ \bibinfo {author} {\bibfnamefont {A.}~\bibnamefont
  {Eisenberg}},\ }\bibfield  {title} {\enquote {\bibinfo {title} {Self-assembly
  of block copolymers},}\ }\href {https://doi.org/10.1039/C2CS35115C}
  {\bibfield  {journal} {\bibinfo  {journal} {Chem. Soc. Rev.}\ }\textbf
  {\bibinfo {volume} {41}},\ \bibinfo {pages} {5969--5985} (\bibinfo {year}
  {2012})}\BibitemShut {NoStop}%
\bibitem [{\citenamefont {Matsen}(2012)}]{Matsen2012}%
  \BibitemOpen
  \bibfield  {author} {\bibinfo {author} {\bibfnamefont {M.~W.}\ \bibnamefont
  {Matsen}},\ }\bibfield  {title} {\enquote {\bibinfo {title} {Effect of
  architecture on the phase behavior of ab-type block copolymer melts},}\
  }\href {https://doi.org/10.1021/ma202782s} {\bibfield  {journal} {\bibinfo
  {journal} {Macromolecules}\ }\textbf {\bibinfo {volume} {45}},\ \bibinfo
  {pages} {2161--2165} (\bibinfo {year} {2012})}\BibitemShut {NoStop}%
\bibitem [{\citenamefont {Noshay}\ and\ \citenamefont
  {McGrath}(2013)}]{noshay2013block}%
  \BibitemOpen
  \bibfield  {author} {\bibinfo {author} {\bibfnamefont {A.}~\bibnamefont
  {Noshay}}\ and\ \bibinfo {author} {\bibfnamefont {J.}~\bibnamefont
  {McGrath}},\ }\href {https://books.google.com/books?id=AEcvBQAAQBAJ} {\emph
  {\bibinfo {title} {Block Copolymers: Overview and Critical Survey}}}\
  (\bibinfo  {publisher} {Elsevier Science},\ \bibinfo {year}
  {2013})\BibitemShut {NoStop}%
\bibitem [{\citenamefont {Feng}\ \emph {et~al.}(2017)\citenamefont {Feng},
  \citenamefont {Lu}, \citenamefont {Wang}, \citenamefont {Kang},\ and\
  \citenamefont {Mays}}]{feng_block_2017}%
  \BibitemOpen
  \bibfield  {author} {\bibinfo {author} {\bibfnamefont {H.}~\bibnamefont
  {Feng}}, \bibinfo {author} {\bibfnamefont {X.}~\bibnamefont {Lu}}, \bibinfo
  {author} {\bibfnamefont {W.}~\bibnamefont {Wang}}, \bibinfo {author}
  {\bibfnamefont {N.-G.}\ \bibnamefont {Kang}},\ and\ \bibinfo {author}
  {\bibfnamefont {J.~W.}\ \bibnamefont {Mays}},\ }\bibfield  {title} {\enquote
  {\bibinfo {title} {Block {Copolymers}: {Synthesis}, {Self}-{Assembly}, and
  {Applications}},}\ }\href {https://doi.org/10.3390/polym9100494} {\bibfield
  {journal} {\bibinfo  {journal} {Polymers}\ }\textbf {\bibinfo {volume} {9}},\
  \bibinfo {pages} {494} (\bibinfo {year} {2017})}\BibitemShut {NoStop}%
\bibitem [{\citenamefont {Sternhagen}\ \emph {et~al.}(2018)\citenamefont
  {Sternhagen}, \citenamefont {Gupta}, \citenamefont {Zhang}, \citenamefont
  {John}, \citenamefont {Schneider},\ and\ \citenamefont
  {Zhang}}]{sternhagen_solution_2018}%
  \BibitemOpen
  \bibfield  {author} {\bibinfo {author} {\bibfnamefont {G.~L.}\ \bibnamefont
  {Sternhagen}}, \bibinfo {author} {\bibfnamefont {S.}~\bibnamefont {Gupta}},
  \bibinfo {author} {\bibfnamefont {Y.}~\bibnamefont {Zhang}}, \bibinfo
  {author} {\bibfnamefont {V.}~\bibnamefont {John}}, \bibinfo {author}
  {\bibfnamefont {G.~J.}\ \bibnamefont {Schneider}},\ and\ \bibinfo {author}
  {\bibfnamefont {D.}~\bibnamefont {Zhang}},\ }\bibfield  {title} {\enquote
  {\bibinfo {title} {Solution {Self}-{Assemblies} of {Sequence}-{Defined}
  {Ionic} {Peptoid} {Block} {Copolymers}},}\ }\href
  {https://doi.org/10.1021/jacs.8b00461} {\bibfield  {journal} {\bibinfo
  {journal} {Journal of the American Chemical Society}\ }\textbf {\bibinfo
  {volume} {140}},\ \bibinfo {pages} {4100--4109} (\bibinfo {year} {2018})},\
  \bibinfo {note} {publisher: American Chemical Society}\BibitemShut {NoStop}%
\bibitem [{\citenamefont {{Willner, L.}}\ \emph {et~al.}(2000)\citenamefont
  {{Willner, L.}}, \citenamefont {{Poppe, A.}}, \citenamefont {{Allgaier, J.}},
  \citenamefont {{Monkenbusch, M.}}, \citenamefont {{Lindner, P.}},\ and\
  \citenamefont {{Richter, D.}}}]{willner_l_micellization_2000}%
  \BibitemOpen
  \bibfield  {author} {\bibinfo {author} {\bibnamefont {{Willner, L.}}},
  \bibinfo {author} {\bibnamefont {{Poppe, A.}}}, \bibinfo {author}
  {\bibnamefont {{Allgaier, J.}}}, \bibinfo {author} {\bibnamefont
  {{Monkenbusch, M.}}}, \bibinfo {author} {\bibnamefont {{Lindner, P.}}},\ and\
  \bibinfo {author} {\bibnamefont {{Richter, D.}}},\ }\bibfield  {title}
  {\enquote {\bibinfo {title} {Micellization of amphiphilic diblock copolymers:
  {Corona} shape and mean-field to scaling crossover},}\ }\href
  {https://doi.org/10.1209/epl/i2000-00384-1} {\bibfield  {journal} {\bibinfo
  {journal} {Europhys. Lett.}\ }\textbf {\bibinfo {volume} {51}},\ \bibinfo
  {pages} {628--634} (\bibinfo {year} {2000})}\BibitemShut {NoStop}%
\bibitem [{\citenamefont {Won}, \citenamefont {Davis},\ and\ \citenamefont
  {Bates}(1999)}]{won_giant_1999}%
  \BibitemOpen
  \bibfield  {author} {\bibinfo {author} {\bibfnamefont {n.}~\bibnamefont
  {Won}}, \bibinfo {author} {\bibfnamefont {n.}~\bibnamefont {Davis}},\ and\
  \bibinfo {author} {\bibfnamefont {n.}~\bibnamefont {Bates}},\ }\bibfield
  {title} {\enquote {\bibinfo {title} {Giant wormlike rubber micelles},}\
  }\href {https://doi.org/10.1126/science.283.5404.960} {\bibfield  {journal}
  {\bibinfo  {journal} {Science (New York, N.Y.)}\ }\textbf {\bibinfo {volume}
  {283}},\ \bibinfo {pages} {960--963} (\bibinfo {year} {1999})}\BibitemShut
  {NoStop}%
\bibitem [{\citenamefont {Förster}\ \emph {et~al.}(1996)\citenamefont
  {Förster}, \citenamefont {Zisenis}, \citenamefont {Wenz},\ and\
  \citenamefont {Antonietti}}]{forster_micellization_1996}%
  \BibitemOpen
  \bibfield  {author} {\bibinfo {author} {\bibfnamefont {S.}~\bibnamefont
  {Förster}}, \bibinfo {author} {\bibfnamefont {M.}~\bibnamefont {Zisenis}},
  \bibinfo {author} {\bibfnamefont {E.}~\bibnamefont {Wenz}},\ and\ \bibinfo
  {author} {\bibfnamefont {M.}~\bibnamefont {Antonietti}},\ }\bibfield  {title}
  {\enquote {\bibinfo {title} {Micellization of strongly segregated block
  copolymers},}\ }\href {https://doi.org/10.1063/1.471723} {\bibfield
  {journal} {\bibinfo  {journal} {The Journal of Chemical Physics}\ }\textbf
  {\bibinfo {volume} {104}},\ \bibinfo {pages} {9956--9970} (\bibinfo {year}
  {1996})},\ \bibinfo {note} {publisher: American Institute of
  Physics}\BibitemShut {NoStop}%
\bibitem [{\citenamefont {Read}\ and\ \citenamefont
  {Armes}(2007)}]{read_recent_2007}%
  \BibitemOpen
  \bibfield  {author} {\bibinfo {author} {\bibfnamefont {E.~S.}\ \bibnamefont
  {Read}}\ and\ \bibinfo {author} {\bibfnamefont {S.~P.}\ \bibnamefont
  {Armes}},\ }\bibfield  {title} {\enquote {\bibinfo {title} {Recent advances
  in shell cross-linked micelles},}\ }\href {https://doi.org/10.1039/B701217A}
  {\bibfield  {journal} {\bibinfo  {journal} {Chemical Communications}\ ,\
  \bibinfo {pages} {3021--3035}} (\bibinfo {year} {2007})},\ \bibinfo {note}
  {publisher: The Royal Society of Chemistry}\BibitemShut {NoStop}%
\bibitem [{\citenamefont {Zhou}\ \emph
  {et~al.}(2010{\natexlab{a}})\citenamefont {Zhou}, \citenamefont {Zheng},
  \citenamefont {Shen}, \citenamefont {Fan}, \citenamefont {Chen},\ and\
  \citenamefont {Zhou}}]{zhou_synthesis_2010}%
  \BibitemOpen
  \bibfield  {author} {\bibinfo {author} {\bibfnamefont {Q.-H.}\ \bibnamefont
  {Zhou}}, \bibinfo {author} {\bibfnamefont {J.-K.}\ \bibnamefont {Zheng}},
  \bibinfo {author} {\bibfnamefont {Z.}~\bibnamefont {Shen}}, \bibinfo {author}
  {\bibfnamefont {X.-H.}\ \bibnamefont {Fan}}, \bibinfo {author} {\bibfnamefont
  {X.-F.}\ \bibnamefont {Chen}},\ and\ \bibinfo {author} {\bibfnamefont
  {Q.-F.}\ \bibnamefont {Zhou}},\ }\bibfield  {title} {\enquote {\bibinfo
  {title} {Synthesis and hierarchical self-assembly of rod- rod block
  copolymers via click chemistry between mesogen-jacketed liquid crystalline
  polymers and helical polypeptides},}\ }\href@noop {} {\bibfield  {journal}
  {\bibinfo  {journal} {Macromolecules}\ }\textbf {\bibinfo {volume} {43}},\
  \bibinfo {pages} {5637--5646} (\bibinfo {year}
  {2010}{\natexlab{a}})}\BibitemShut {NoStop}%
\bibitem [{\citenamefont {Olsen}\ and\ \citenamefont
  {Segalman}(2008)}]{olsen_self-assembly_2008}%
  \BibitemOpen
  \bibfield  {author} {\bibinfo {author} {\bibfnamefont {B.~D.}\ \bibnamefont
  {Olsen}}\ and\ \bibinfo {author} {\bibfnamefont {R.~A.}\ \bibnamefont
  {Segalman}},\ }\bibfield  {title} {\enquote {\bibinfo {title} {Self-assembly
  of rod–coil block copolymers},}\ }\href
  {https://doi.org/https://doi.org/10.1016/j.mser.2008.04.001} {\bibfield
  {journal} {\bibinfo  {journal} {Materials Science and Engineering: R:
  Reports}\ }\textbf {\bibinfo {volume} {62}},\ \bibinfo {pages} {37--66}
  (\bibinfo {year} {2008})}\BibitemShut {NoStop}%
\bibitem [{\citenamefont {Hayashi}\ \emph {et~al.}(2019)\citenamefont
  {Hayashi}, \citenamefont {Kikuchi}, \citenamefont {Kumai}, \citenamefont
  {Takeguchi},\ and\ \citenamefont {Goto}}]{hayashi_rod-shaped_2019}%
  \BibitemOpen
  \bibfield  {author} {\bibinfo {author} {\bibfnamefont {H.}~\bibnamefont
  {Hayashi}}, \bibinfo {author} {\bibfnamefont {R.}~\bibnamefont {Kikuchi}},
  \bibinfo {author} {\bibfnamefont {R.}~\bibnamefont {Kumai}}, \bibinfo
  {author} {\bibfnamefont {M.}~\bibnamefont {Takeguchi}},\ and\ \bibinfo
  {author} {\bibfnamefont {H.}~\bibnamefont {Goto}},\ }\bibfield  {title}
  {\enquote {\bibinfo {title} {Rod-shaped {1D} polymer-assisted anisotropic
  self-assembly of {0D} nanoparticles by a solution-drying method},}\ }\href
  {https://doi.org/10.1039/C9TC00702D} {\bibfield  {journal} {\bibinfo
  {journal} {Journal of Materials Chemistry C}\ }\textbf {\bibinfo {volume}
  {7}},\ \bibinfo {pages} {7442--7453} (\bibinfo {year} {2019})},\ \bibinfo
  {note} {publisher: The Royal Society of Chemistry}\BibitemShut {NoStop}%
\bibitem [{\citenamefont {Guo}, \citenamefont {Ravensteijn},\ and\
  \citenamefont {Kegel}(2020)}]{guo_self-assembly_2020}%
  \BibitemOpen
  \bibfield  {author} {\bibinfo {author} {\bibfnamefont {Y.}~\bibnamefont
  {Guo}}, \bibinfo {author} {\bibfnamefont {B.~G. P.~v.}\ \bibnamefont
  {Ravensteijn}},\ and\ \bibinfo {author} {\bibfnamefont {W.~K.}\ \bibnamefont
  {Kegel}},\ }\bibfield  {title} {\enquote {\bibinfo {title} {Self-assembly of
  isotropic colloids into colloidal strings, {Bernal} spiral-like, and tubular
  clusters},}\ }\href {https://doi.org/10.1039/D0CC00948B} {\bibfield
  {journal} {\bibinfo  {journal} {Chemical Communications}\ }\textbf {\bibinfo
  {volume} {56}},\ \bibinfo {pages} {6309--6312} (\bibinfo {year} {2020})},\
  \bibinfo {note} {publisher: The Royal Society of Chemistry}\BibitemShut
  {NoStop}%
\bibitem [{\citenamefont {Jiang}, \citenamefont {Zhou},\ and\ \citenamefont
  {Yan}(2015)}]{jiang_hyperbranched_2015}%
  \BibitemOpen
  \bibfield  {author} {\bibinfo {author} {\bibfnamefont {W.}~\bibnamefont
  {Jiang}}, \bibinfo {author} {\bibfnamefont {Y.}~\bibnamefont {Zhou}},\ and\
  \bibinfo {author} {\bibfnamefont {D.}~\bibnamefont {Yan}},\ }\bibfield
  {title} {\enquote {\bibinfo {title} {Hyperbranched polymer vesicles: from
  self-assembly, characterization, mechanisms, and properties to
  applications},}\ }\href {https://doi.org/10.1039/C4CS00274A} {\bibfield
  {journal} {\bibinfo  {journal} {Chemical Society Reviews}\ }\textbf {\bibinfo
  {volume} {44}},\ \bibinfo {pages} {3874--3889} (\bibinfo {year} {2015})},\
  \bibinfo {note} {publisher: Royal Society of Chemistry}\BibitemShut {NoStop}%
\bibitem [{\citenamefont {Araste}\ \emph {et~al.}(2021)\citenamefont {Araste},
  \citenamefont {Aliabadi}, \citenamefont {Abnous}, \citenamefont {Taghdisi},
  \citenamefont {Ramezani},\ and\ \citenamefont
  {Alibolandi}}]{araste_self-assembled_2021}%
  \BibitemOpen
  \bibfield  {author} {\bibinfo {author} {\bibfnamefont {F.}~\bibnamefont
  {Araste}}, \bibinfo {author} {\bibfnamefont {A.}~\bibnamefont {Aliabadi}},
  \bibinfo {author} {\bibfnamefont {K.}~\bibnamefont {Abnous}}, \bibinfo
  {author} {\bibfnamefont {S.~M.}\ \bibnamefont {Taghdisi}}, \bibinfo {author}
  {\bibfnamefont {M.}~\bibnamefont {Ramezani}},\ and\ \bibinfo {author}
  {\bibfnamefont {M.}~\bibnamefont {Alibolandi}},\ }\bibfield  {title}
  {\enquote {\bibinfo {title} {Self-assembled polymeric vesicles: {Focus} on
  polymersomes in cancer treatment},}\ }\href
  {https://doi.org/10.1016/j.jconrel.2020.12.027} {\bibfield  {journal}
  {\bibinfo  {journal} {Journal of Controlled Release}\ }\textbf {\bibinfo
  {volume} {330}},\ \bibinfo {pages} {502--528} (\bibinfo {year}
  {2021})}\BibitemShut {NoStop}%
\bibitem [{\citenamefont {Wang}\ \emph {et~al.}(2019)\citenamefont {Wang},
  \citenamefont {Choi}, \citenamefont {Sun}, \citenamefont {Wei}, \citenamefont
  {Feng},\ and\ \citenamefont {Thang}}]{wang_spindle-like_2019}%
  \BibitemOpen
  \bibfield  {author} {\bibinfo {author} {\bibfnamefont {M.}~\bibnamefont
  {Wang}}, \bibinfo {author} {\bibfnamefont {B.}~\bibnamefont {Choi}}, \bibinfo
  {author} {\bibfnamefont {Z.}~\bibnamefont {Sun}}, \bibinfo {author}
  {\bibfnamefont {X.}~\bibnamefont {Wei}}, \bibinfo {author} {\bibfnamefont
  {A.}~\bibnamefont {Feng}},\ and\ \bibinfo {author} {\bibfnamefont {S.~H.}\
  \bibnamefont {Thang}},\ }\bibfield  {title} {\enquote {\bibinfo {title}
  {Spindle-like and telophase-like self-assemblies mediated by complementary
  nucleobase molecular recognition},}\ }\href
  {https://doi.org/10.1039/C8CC09923E} {\bibfield  {journal} {\bibinfo
  {journal} {Chemical Communications}\ }\textbf {\bibinfo {volume} {55}},\
  \bibinfo {pages} {1462--1465} (\bibinfo {year} {2019})},\ \bibinfo {note}
  {publisher: The Royal Society of Chemistry}\BibitemShut {NoStop}%
\bibitem [{\citenamefont {Cissé}\ and\ \citenamefont
  {Kudernac}(2020)}]{cisse_light-fuelled_2020}%
  \BibitemOpen
  \bibfield  {author} {\bibinfo {author} {\bibfnamefont {N.}~\bibnamefont
  {Cissé}}\ and\ \bibinfo {author} {\bibfnamefont {T.}~\bibnamefont
  {Kudernac}},\ }\bibfield  {title} {\enquote {\bibinfo {title}
  {Light-{Fuelled} {Self}-{Assembly} of {Cyclic} {Peptides} into
  {Supramolecular} {Tubules}},}\ }\href
  {https://doi.org/10.1002/syst.202000012} {\bibfield  {journal} {\bibinfo
  {journal} {ChemSystemsChem}\ }\textbf {\bibinfo {volume} {2}},\ \bibinfo
  {pages} {e2000012} (\bibinfo {year} {2020})},\ \bibinfo {note} {\_eprint:
  https://onlinelibrary.wiley.com/doi/pdf/10.1002/syst.202000012}\BibitemShut
  {NoStop}%
\bibitem [{\citenamefont {He}\ \emph {et~al.}(2012)\citenamefont {He},
  \citenamefont {Liu}, \citenamefont {Babu}, \citenamefont {Wei},\ and\
  \citenamefont {Nie}}]{he_self-assembly_2012}%
  \BibitemOpen
  \bibfield  {author} {\bibinfo {author} {\bibfnamefont {J.}~\bibnamefont
  {He}}, \bibinfo {author} {\bibfnamefont {Y.}~\bibnamefont {Liu}}, \bibinfo
  {author} {\bibfnamefont {T.}~\bibnamefont {Babu}}, \bibinfo {author}
  {\bibfnamefont {Z.}~\bibnamefont {Wei}},\ and\ \bibinfo {author}
  {\bibfnamefont {Z.}~\bibnamefont {Nie}},\ }\bibfield  {title} {\enquote
  {\bibinfo {title} {Self-{Assembly} of {Inorganic} {Nanoparticle} {Vesicles}
  and {Tubules} {Driven} by {Tethered} {Linear} {Block} {Copolymers}},}\ }\href
  {https://doi.org/10.1021/ja3032295} {\bibfield  {journal} {\bibinfo
  {journal} {Journal of the American Chemical Society}\ }\textbf {\bibinfo
  {volume} {134}},\ \bibinfo {pages} {11342--11345} (\bibinfo {year} {2012})},\
  \bibinfo {note} {publisher: American Chemical Society}\BibitemShut {NoStop}%
\bibitem [{\citenamefont {Kim}\ \emph {et~al.}(2013)\citenamefont {Kim},
  \citenamefont {Li}, \citenamefont {Shin},\ and\ \citenamefont
  {Lee}}]{kim_development_2013}%
  \BibitemOpen
  \bibfield  {author} {\bibinfo {author} {\bibfnamefont {Y.}~\bibnamefont
  {Kim}}, \bibinfo {author} {\bibfnamefont {W.}~\bibnamefont {Li}}, \bibinfo
  {author} {\bibfnamefont {S.}~\bibnamefont {Shin}},\ and\ \bibinfo {author}
  {\bibfnamefont {M.}~\bibnamefont {Lee}},\ }\bibfield  {title} {\enquote
  {\bibinfo {title} {Development of {Toroidal} {Nanostructures} by
  {Self}-{Assembly}: {Rational} {Designs} and {Applications}},}\ }\href
  {https://doi.org/10.1021/ar400027c} {\bibfield  {journal} {\bibinfo
  {journal} {Accounts of Chemical Research}\ }\textbf {\bibinfo {volume}
  {46}},\ \bibinfo {pages} {2888--2897} (\bibinfo {year} {2013})},\ \bibinfo
  {note} {publisher: American Chemical Society}\BibitemShut {NoStop}%
\bibitem [{\citenamefont {Xu}\ \emph {et~al.}(2020)\citenamefont {Xu},
  \citenamefont {Gao}, \citenamefont {Cai}, \citenamefont {Lin}, \citenamefont
  {Wang},\ and\ \citenamefont {Tian}}]{xu_helical_2020}%
  \BibitemOpen
  \bibfield  {author} {\bibinfo {author} {\bibfnamefont {P.}~\bibnamefont
  {Xu}}, \bibinfo {author} {\bibfnamefont {L.}~\bibnamefont {Gao}}, \bibinfo
  {author} {\bibfnamefont {C.}~\bibnamefont {Cai}}, \bibinfo {author}
  {\bibfnamefont {J.}~\bibnamefont {Lin}}, \bibinfo {author} {\bibfnamefont
  {L.}~\bibnamefont {Wang}},\ and\ \bibinfo {author} {\bibfnamefont
  {X.}~\bibnamefont {Tian}},\ }\bibfield  {title} {\enquote {\bibinfo {title}
  {Helical {Toroids} {Self}-{Assembled} from a {Binary} {System} of
  {Polypeptide} {Homopolymer} and its {Block} {Copolymer}},}\ }\href
  {https://doi.org/https://doi.org/10.1002/anie.202004102} {\bibfield
  {journal} {\bibinfo  {journal} {Angewandte Chemie International Edition}\
  }\textbf {\bibinfo {volume} {59}},\ \bibinfo {pages} {14281--14285} (\bibinfo
  {year} {2020})},\ \bibinfo {note} {\_eprint:
  https://onlinelibrary.wiley.com/doi/pdf/10.1002/anie.202004102}\BibitemShut
  {NoStop}%
\bibitem [{\citenamefont {Xu}\ \emph {et~al.}(2022)\citenamefont {Xu},
  \citenamefont {Gao}, \citenamefont {Cai}, \citenamefont {Lin}, \citenamefont
  {Wang},\ and\ \citenamefont {Tian}}]{xu_polymeric_2022}%
  \BibitemOpen
  \bibfield  {author} {\bibinfo {author} {\bibfnamefont {P.}~\bibnamefont
  {Xu}}, \bibinfo {author} {\bibfnamefont {L.}~\bibnamefont {Gao}}, \bibinfo
  {author} {\bibfnamefont {C.}~\bibnamefont {Cai}}, \bibinfo {author}
  {\bibfnamefont {J.}~\bibnamefont {Lin}}, \bibinfo {author} {\bibfnamefont
  {L.}~\bibnamefont {Wang}},\ and\ \bibinfo {author} {\bibfnamefont
  {X.}~\bibnamefont {Tian}},\ }\bibfield  {title} {\enquote {\bibinfo {title}
  {Polymeric {Toroidal} {Self}-{Assemblies}: {Diverse} {Formation} {Mechanisms}
  and {Functions}},}\ }\href
  {https://doi.org/https://doi.org/10.1002/adfm.202106036} {\bibfield
  {journal} {\bibinfo  {journal} {Advanced Functional Materials}\ }\textbf
  {\bibinfo {volume} {32}},\ \bibinfo {pages} {2106036} (\bibinfo {year}
  {2022})},\ \bibinfo {note} {\_eprint:
  https://onlinelibrary.wiley.com/doi/pdf/10.1002/adfm.202106036}\BibitemShut
  {NoStop}%
\bibitem [{\citenamefont {Monnard}\ and\ \citenamefont
  {Deamer}(2002)}]{monnard_membrane_2002}%
  \BibitemOpen
  \bibfield  {author} {\bibinfo {author} {\bibfnamefont {P.-A.}\ \bibnamefont
  {Monnard}}\ and\ \bibinfo {author} {\bibfnamefont {D.~W.}\ \bibnamefont
  {Deamer}},\ }\bibfield  {title} {\enquote {\bibinfo {title} {Membrane
  self-assembly processes: {Steps} toward the first cellular life},}\ }\href
  {https://doi.org/10.1002/ar.10154} {\bibfield  {journal} {\bibinfo  {journal}
  {The Anatomical Record}\ }\textbf {\bibinfo {volume} {268}},\ \bibinfo
  {pages} {196--207} (\bibinfo {year} {2002})}\BibitemShut {NoStop}%
\bibitem [{\citenamefont {Yurchenco}\ \emph {et~al.}(1986)\citenamefont
  {Yurchenco}, \citenamefont {Tsilibary}, \citenamefont {Charonis},\ and\
  \citenamefont {Furthmayr}}]{yurchenco_models_1986}%
  \BibitemOpen
  \bibfield  {author} {\bibinfo {author} {\bibfnamefont {P.~D.}\ \bibnamefont
  {Yurchenco}}, \bibinfo {author} {\bibfnamefont {E.~C.}\ \bibnamefont
  {Tsilibary}}, \bibinfo {author} {\bibfnamefont {A.~S.}\ \bibnamefont
  {Charonis}},\ and\ \bibinfo {author} {\bibfnamefont {H.}~\bibnamefont
  {Furthmayr}},\ }\bibfield  {title} {\enquote {\bibinfo {title} {Models for
  the self-assembly of basement membrane.}}\ }\href
  {https://doi.org/10.1177/34.1.3510247} {\bibfield  {journal} {\bibinfo
  {journal} {Journal of Histochemistry \& Cytochemistry}\ }\textbf {\bibinfo
  {volume} {34}},\ \bibinfo {pages} {93--102} (\bibinfo {year} {1986})},\
  \bibinfo {note} {publisher: Journal of Histochemistry \&
  Cytochemistry}\BibitemShut {NoStop}%
\bibitem [{\citenamefont {Zhang}\ \emph {et~al.}(2015)\citenamefont {Zhang},
  \citenamefont {Sargent}, \citenamefont {Boudouris},\ and\ \citenamefont
  {Phillip}}]{zhang_nanoporous_2015}%
  \BibitemOpen
  \bibfield  {author} {\bibinfo {author} {\bibfnamefont {Y.}~\bibnamefont
  {Zhang}}, \bibinfo {author} {\bibfnamefont {J.~L.}\ \bibnamefont {Sargent}},
  \bibinfo {author} {\bibfnamefont {B.~W.}\ \bibnamefont {Boudouris}},\ and\
  \bibinfo {author} {\bibfnamefont {W.~A.}\ \bibnamefont {Phillip}},\
  }\bibfield  {title} {\enquote {\bibinfo {title} {Nanoporous membranes
  generated from self-assembled block polymer precursors: {Quo} {Vadis}?}}\
  }\href {https://doi.org/https://doi.org/10.1002/app.41683} {\bibfield
  {journal} {\bibinfo  {journal} {Journal of Applied Polymer Science}\ }\textbf
  {\bibinfo {volume} {132}} (\bibinfo {year} {2015}),\
  https://doi.org/10.1002/app.41683},\ \bibinfo {note} {\_eprint:
  https://onlinelibrary.wiley.com/doi/pdf/10.1002/app.41683}\BibitemShut
  {NoStop}%
\bibitem [{\citenamefont {Zhou}\ \emph
  {et~al.}(2010{\natexlab{b}})\citenamefont {Zhou}, \citenamefont {Huang},
  \citenamefont {Liu}, \citenamefont {Zhu},\ and\ \citenamefont
  {Yan}}]{zhou_self-assembly_2010}%
  \BibitemOpen
  \bibfield  {author} {\bibinfo {author} {\bibfnamefont {Y.}~\bibnamefont
  {Zhou}}, \bibinfo {author} {\bibfnamefont {W.}~\bibnamefont {Huang}},
  \bibinfo {author} {\bibfnamefont {J.}~\bibnamefont {Liu}}, \bibinfo {author}
  {\bibfnamefont {X.}~\bibnamefont {Zhu}},\ and\ \bibinfo {author}
  {\bibfnamefont {D.}~\bibnamefont {Yan}},\ }\bibfield  {title} {\enquote
  {\bibinfo {title} {Self-{Assembly} of {Hyperbranched} {Polymers} and {Its}
  {Biomedical} {Applications}},}\ }\href
  {https://doi.org/10.1002/adma.201000369} {\bibfield  {journal} {\bibinfo
  {journal} {Advanced Materials}\ }\textbf {\bibinfo {volume} {22}},\ \bibinfo
  {pages} {4567--4590} (\bibinfo {year} {2010}{\natexlab{b}})},\ \bibinfo
  {note} {\_eprint:
  https://onlinelibrary.wiley.com/doi/pdf/10.1002/adma.201000369}\BibitemShut
  {NoStop}%
\bibitem [{\citenamefont {Tørring}\ \emph {et~al.}(2011)\citenamefont
  {Tørring}, \citenamefont {V. Voigt}, \citenamefont {Nangreave},
  \citenamefont {Yan},\ and\ \citenamefont {V. Gothelf}}]{torring_dna_2011}%
  \BibitemOpen
  \bibfield  {author} {\bibinfo {author} {\bibfnamefont {T.}~\bibnamefont
  {Tørring}}, \bibinfo {author} {\bibfnamefont {N.}~\bibnamefont {V. Voigt}},
  \bibinfo {author} {\bibfnamefont {J.}~\bibnamefont {Nangreave}}, \bibinfo
  {author} {\bibfnamefont {H.}~\bibnamefont {Yan}},\ and\ \bibinfo {author}
  {\bibfnamefont {K.}~\bibnamefont {V. Gothelf}},\ }\bibfield  {title}
  {\enquote {\bibinfo {title} {{DNA} origami: a quantum leap for self-assembly
  of complex structures},}\ }\href {https://doi.org/10.1039/C1CS15057J}
  {\bibfield  {journal} {\bibinfo  {journal} {Chemical Society Reviews}\
  }\textbf {\bibinfo {volume} {40}},\ \bibinfo {pages} {5636--5646} (\bibinfo
  {year} {2011})},\ \bibinfo {note} {publisher: Royal Society of
  Chemistry}\BibitemShut {NoStop}%
\bibitem [{\citenamefont {Chakraborty}\ and\ \citenamefont
  {Gazit}(2018)}]{chakraborty_amino_2018}%
  \BibitemOpen
  \bibfield  {author} {\bibinfo {author} {\bibfnamefont {P.}~\bibnamefont
  {Chakraborty}}\ and\ \bibinfo {author} {\bibfnamefont {E.}~\bibnamefont
  {Gazit}},\ }\bibfield  {title} {\enquote {\bibinfo {title} {Amino {Acid}
  {Based} {Self}-assembled {Nanostructures}: {Complex} {Structures} from
  {Remarkably} {Simple} {Building} {Blocks}},}\ }\href
  {https://doi.org/10.1002/cnma.201800147} {\bibfield  {journal} {\bibinfo
  {journal} {ChemNanoMat}\ }\textbf {\bibinfo {volume} {4}},\ \bibinfo {pages}
  {730--740} (\bibinfo {year} {2018})},\ \bibinfo {note} {\_eprint:
  https://onlinelibrary.wiley.com/doi/pdf/10.1002/cnma.201800147}\BibitemShut
  {NoStop}%
\bibitem [{\citenamefont {Stupp}(2010)}]{stupp_self-assembly_2010}%
  \BibitemOpen
  \bibfield  {author} {\bibinfo {author} {\bibfnamefont {S.~I.}\ \bibnamefont
  {Stupp}},\ }\bibfield  {title} {\enquote {\bibinfo {title} {Self-{Assembly}
  and {Biomaterials}},}\ }\href {https://doi.org/10.1021/nl103567y} {\bibfield
  {journal} {\bibinfo  {journal} {Nano Letters}\ }\textbf {\bibinfo {volume}
  {10}},\ \bibinfo {pages} {4783--4786} (\bibinfo {year} {2010})},\ \bibinfo
  {note} {publisher: American Chemical Society}\BibitemShut {NoStop}%
\bibitem [{\citenamefont {B. Darling}(2009)}]{bdarling_block_2009}%
  \BibitemOpen
  \bibfield  {author} {\bibinfo {author} {\bibfnamefont {S.}~\bibnamefont
  {B. Darling}},\ }\bibfield  {title} {\enquote {\bibinfo {title} {Block
  copolymers for photovoltaics},}\ }\href {https://doi.org/10.1039/B912086F}
  {\bibfield  {journal} {\bibinfo  {journal} {Energy \& Environmental Science}\
  }\textbf {\bibinfo {volume} {2}},\ \bibinfo {pages} {1266--1273} (\bibinfo
  {year} {2009})},\ \bibinfo {note} {publisher: Royal Society of
  Chemistry}\BibitemShut {NoStop}%
\bibitem [{\citenamefont {Shah}\ and\ \citenamefont
  {Ganesan}(2010)}]{shah_correlations_2010}%
  \BibitemOpen
  \bibfield  {author} {\bibinfo {author} {\bibfnamefont {M.}~\bibnamefont
  {Shah}}\ and\ \bibinfo {author} {\bibfnamefont {V.}~\bibnamefont {Ganesan}},\
  }\bibfield  {title} {\enquote {\bibinfo {title} {Correlations between
  morphologies and photovoltaic properties of rod- coil block copolymers},}\
  }\href@noop {} {\bibfield  {journal} {\bibinfo  {journal} {Macromolecules}\
  }\textbf {\bibinfo {volume} {43}},\ \bibinfo {pages} {543--552} (\bibinfo
  {year} {2010})}\BibitemShut {NoStop}%
\bibitem [{\citenamefont
  {Hadziioannou}(2002)}]{hadziioannou_semiconducting_2002}%
  \BibitemOpen
  \bibfield  {author} {\bibinfo {author} {\bibfnamefont {G.}~\bibnamefont
  {Hadziioannou}},\ }\bibfield  {title} {\enquote {\bibinfo {title}
  {Semiconducting {Block} {Copolymers} for {Self}-{Assembled} {Photovoltaic}
  {Devices}},}\ }\href {https://doi.org/10.1557/mrs2002.145} {\bibfield
  {journal} {\bibinfo  {journal} {MRS Bulletin}\ }\textbf {\bibinfo {volume}
  {27}},\ \bibinfo {pages} {456--460} (\bibinfo {year} {2002})},\ \bibinfo
  {note} {publisher: Cambridge University Press}\BibitemShut {NoStop}%
\bibitem [{\citenamefont {Kwon}\ and\ \citenamefont
  {Okano}(1999)}]{kwon_soluble_1999}%
  \BibitemOpen
  \bibfield  {author} {\bibinfo {author} {\bibfnamefont {G.~S.}\ \bibnamefont
  {Kwon}}\ and\ \bibinfo {author} {\bibfnamefont {T.}~\bibnamefont {Okano}},\
  }\bibfield  {title} {\enquote {\bibinfo {title} {Soluble {Self}-{Assembled}
  {Block} {Copolymers} for {Drug} {Delivery}},}\ }\href
  {https://doi.org/10.1023/A:1011991617857} {\bibfield  {journal} {\bibinfo
  {journal} {Pharmaceutical Research}\ }\textbf {\bibinfo {volume} {16}},\
  \bibinfo {pages} {597--600} (\bibinfo {year} {1999})}\BibitemShut {NoStop}%
\bibitem [{\citenamefont {Rösler}, \citenamefont {Vandermeulen},\ and\
  \citenamefont {Klok}(2002)}]{rosler_advanced_2002}%
  \BibitemOpen
  \bibfield  {author} {\bibinfo {author} {\bibfnamefont {A.}~\bibnamefont
  {Rösler}}, \bibinfo {author} {\bibfnamefont {G.}~\bibnamefont
  {Vandermeulen}},\ and\ \bibinfo {author} {\bibfnamefont {H.-A.}\ \bibnamefont
  {Klok}},\ }\bibfield  {title} {\enquote {\bibinfo {title} {Advanced {Drug}
  {Delivery} {Devices} via {Self}-{Assembly} of {Amphiphilic} {Block}
  {Copolymers}},}\ }\href {https://doi.org/10.1016/S0169-409X(01)00222-8}
  {\bibfield  {journal} {\bibinfo  {journal} {Advanced drug delivery reviews}\
  }\textbf {\bibinfo {volume} {53}},\ \bibinfo {pages} {95--108} (\bibinfo
  {year} {2002})}\BibitemShut {NoStop}%
\bibitem [{\citenamefont {Khullar}\ \emph {et~al.}(2013)\citenamefont
  {Khullar}, \citenamefont {Singh}, \citenamefont {Mahal}, \citenamefont
  {Kumar}, \citenamefont {Kaur},\ and\ \citenamefont
  {Bakshi}}]{khullar_block_2013}%
  \BibitemOpen
  \bibfield  {author} {\bibinfo {author} {\bibfnamefont {P.}~\bibnamefont
  {Khullar}}, \bibinfo {author} {\bibfnamefont {V.}~\bibnamefont {Singh}},
  \bibinfo {author} {\bibfnamefont {A.}~\bibnamefont {Mahal}}, \bibinfo
  {author} {\bibfnamefont {H.}~\bibnamefont {Kumar}}, \bibinfo {author}
  {\bibfnamefont {G.}~\bibnamefont {Kaur}},\ and\ \bibinfo {author}
  {\bibfnamefont {M.~S.}\ \bibnamefont {Bakshi}},\ }\bibfield  {title}
  {\enquote {\bibinfo {title} {Block {Copolymer} {Micelles} as {Nanoreactors}
  for {Self}-{Assembled} {Morphologies} of {Gold} {Nanoparticles}},}\ }\href
  {https://doi.org/10.1021/jp310507m} {\bibfield  {journal} {\bibinfo
  {journal} {The Journal of Physical Chemistry B}\ }\textbf {\bibinfo {volume}
  {117}},\ \bibinfo {pages} {3028--3039} (\bibinfo {year} {2013})},\ \bibinfo
  {note} {publisher: American Chemical Society}\BibitemShut {NoStop}%
\bibitem [{\citenamefont {Würbser}\ \emph {et~al.}(2021)\citenamefont
  {Würbser}, \citenamefont {Schwarz}, \citenamefont {Heckel}, \citenamefont
  {Bergmann}, \citenamefont {Walther},\ and\ \citenamefont
  {Boekhoven}}]{wurbser_chemically_2021}%
  \BibitemOpen
  \bibfield  {author} {\bibinfo {author} {\bibfnamefont {M.~A.}\ \bibnamefont
  {Würbser}}, \bibinfo {author} {\bibfnamefont {P.~S.}\ \bibnamefont
  {Schwarz}}, \bibinfo {author} {\bibfnamefont {J.}~\bibnamefont {Heckel}},
  \bibinfo {author} {\bibfnamefont {A.~M.}\ \bibnamefont {Bergmann}}, \bibinfo
  {author} {\bibfnamefont {A.}~\bibnamefont {Walther}},\ and\ \bibinfo {author}
  {\bibfnamefont {J.}~\bibnamefont {Boekhoven}},\ }\bibfield  {title} {\enquote
  {\bibinfo {title} {Chemically {Fueled} {Block} {Copolymer} {Self}-{Assembly}
  into {Transient} {Nanoreactors}**},}\ }\href
  {https://doi.org/https://doi.org/10.1002/syst.202100015} {\bibfield
  {journal} {\bibinfo  {journal} {ChemSystemsChem}\ }\textbf {\bibinfo {volume}
  {3}},\ \bibinfo {pages} {e2100015} (\bibinfo {year} {2021})},\ \bibinfo
  {note} {\_eprint:
  https://chemistry-europe.onlinelibrary.wiley.com/doi/pdf/10.1002/syst.202100015}\BibitemShut
  {NoStop}%
\bibitem [{\citenamefont {Black}\ \emph {et~al.}(2007)\citenamefont {Black},
  \citenamefont {Ruiz}, \citenamefont {Breyta}, \citenamefont {Cheng},
  \citenamefont {Colburn}, \citenamefont {Guarini}, \citenamefont {Kim},\ and\
  \citenamefont {Zhang}}]{5388674}%
  \BibitemOpen
  \bibfield  {author} {\bibinfo {author} {\bibfnamefont {C.~T.}\ \bibnamefont
  {Black}}, \bibinfo {author} {\bibfnamefont {R.}~\bibnamefont {Ruiz}},
  \bibinfo {author} {\bibfnamefont {G.}~\bibnamefont {Breyta}}, \bibinfo
  {author} {\bibfnamefont {J.~Y.}\ \bibnamefont {Cheng}}, \bibinfo {author}
  {\bibfnamefont {M.~E.}\ \bibnamefont {Colburn}}, \bibinfo {author}
  {\bibfnamefont {K.~W.}\ \bibnamefont {Guarini}}, \bibinfo {author}
  {\bibfnamefont {H.-C.}\ \bibnamefont {Kim}},\ and\ \bibinfo {author}
  {\bibfnamefont {Y.}~\bibnamefont {Zhang}},\ }\bibfield  {title} {\enquote
  {\bibinfo {title} {Polymer self assembly in semiconductor
  microelectronics},}\ }\href {https://doi.org/10.1147/rd.515.0605} {\bibfield
  {journal} {\bibinfo  {journal} {IBM Journal of Research and Development}\
  }\textbf {\bibinfo {volume} {51}},\ \bibinfo {pages} {605--633} (\bibinfo
  {year} {2007})}\BibitemShut {NoStop}%
\bibitem [{\citenamefont {Terao}(2020)}]{terao_machine_2020}%
  \BibitemOpen
  \bibfield  {author} {\bibinfo {author} {\bibfnamefont {T.}~\bibnamefont
  {Terao}},\ }\bibfield  {title} {\enquote {\bibinfo {title} {A machine
  learning approach to analyze the structural formation of soft matter via
  image recognition},}\ }\href {https://doi.org/10.1080/1539445X.2020.1715433}
  {\bibfield  {journal} {\bibinfo  {journal} {Soft Materials}\ }\textbf
  {\bibinfo {volume} {18}},\ \bibinfo {pages} {215--227} (\bibinfo {year}
  {2020})},\ \bibinfo {note} {publisher: Taylor \& Francis \_eprint:
  https://doi.org/10.1080/1539445X.2020.1715433}\BibitemShut {NoStop}%
\bibitem [{\citenamefont {Srinivas}, \citenamefont {Discher},\ and\
  \citenamefont {Klein}(2004)}]{srinivas_self-assembly_2004}%
  \BibitemOpen
  \bibfield  {author} {\bibinfo {author} {\bibfnamefont {G.}~\bibnamefont
  {Srinivas}}, \bibinfo {author} {\bibfnamefont {D.~E.}\ \bibnamefont
  {Discher}},\ and\ \bibinfo {author} {\bibfnamefont {M.~L.}\ \bibnamefont
  {Klein}},\ }\bibfield  {title} {\enquote {\bibinfo {title} {Self-assembly and
  properties of diblock copolymers by coarse-grain molecular dynamics},}\
  }\href {https://doi.org/10.1038/nmat1185} {\bibfield  {journal} {\bibinfo
  {journal} {Nature Materials}\ }\textbf {\bibinfo {volume} {3}},\ \bibinfo
  {pages} {638--644} (\bibinfo {year} {2004})}\BibitemShut {NoStop}%
\bibitem [{\citenamefont {Li}\ \emph {et~al.}(2012)\citenamefont {Li},
  \citenamefont {Zhu}, \citenamefont {Liu},\ and\ \citenamefont
  {Lu}}]{li_brownian_2012}%
  \BibitemOpen
  \bibfield  {author} {\bibinfo {author} {\bibfnamefont {B.}~\bibnamefont
  {Li}}, \bibinfo {author} {\bibfnamefont {Y.-L.}\ \bibnamefont {Zhu}},
  \bibinfo {author} {\bibfnamefont {H.}~\bibnamefont {Liu}},\ and\ \bibinfo
  {author} {\bibfnamefont {Z.-Y.}\ \bibnamefont {Lu}},\ }\bibfield  {title}
  {\enquote {\bibinfo {title} {Brownian dynamics simulation study on the
  self-assembly of incompatible star-like block copolymers in dilute
  solution},}\ }\href {https://doi.org/10.1039/C2CP23932A} {\bibfield
  {journal} {\bibinfo  {journal} {Phys. Chem. Chem. Phys.}\ }\textbf {\bibinfo
  {volume} {14}},\ \bibinfo {pages} {4964--4970} (\bibinfo {year} {2012})},\
  \bibinfo {note} {publisher: The Royal Society of Chemistry}\BibitemShut
  {NoStop}%
\bibitem [{\citenamefont {Huang}\ and\ \citenamefont
  {Alexander-Katz}(2019)}]{huang_dissipative_2019}%
  \BibitemOpen
  \bibfield  {author} {\bibinfo {author} {\bibfnamefont {H.}~\bibnamefont
  {Huang}}\ and\ \bibinfo {author} {\bibfnamefont {A.}~\bibnamefont
  {Alexander-Katz}},\ }\bibfield  {title} {\enquote {\bibinfo {title}
  {Dissipative particle dynamics for directed self-assembly of block
  copolymers},}\ }\href {https://doi.org/10.1063/1.5117839} {\bibfield
  {journal} {\bibinfo  {journal} {The Journal of Chemical Physics}\ }\textbf
  {\bibinfo {volume} {151}},\ \bibinfo {pages} {154905} (\bibinfo {year}
  {2019})},\ \bibinfo {note} {\_eprint:
  https://doi.org/10.1063/1.5117839}\BibitemShut {NoStop}%
\bibitem [{\citenamefont {Patti}(2010)}]{patti_monte_2010}%
  \BibitemOpen
  \bibfield  {author} {\bibinfo {author} {\bibfnamefont {A.}~\bibnamefont
  {Patti}},\ }\bibfield  {title} {\enquote {\bibinfo {title} {Monte {Carlo}
  simulations of self-assembling star-block copolymers in dilute solutions},}\
  }\href {https://doi.org/10.1016/j.colsurfa.2010.03.022} {\bibfield  {journal}
  {\bibinfo  {journal} {Colloids and Surfaces A: Physicochemical and
  Engineering Aspects}\ }\textbf {\bibinfo {volume} {361}},\ \bibinfo {pages}
  {81--89} (\bibinfo {year} {2010})}\BibitemShut {NoStop}%
\bibitem [{\citenamefont {Matsen}(2002)}]{Matsen2002}%
  \BibitemOpen
  \bibfield  {author} {\bibinfo {author} {\bibfnamefont {M.~W.}\ \bibnamefont
  {Matsen}},\ }\bibfield  {title} {\enquote {\bibinfo {title} {{The standard
  Gaussian model for block copolymer melts}},}\ }\href
  {https://doi.org/10.1088/0953-8984/14/2/201} {\bibfield  {journal} {\bibinfo
  {journal} {Journal of Physics Condensed Matter}\ }\textbf {\bibinfo {volume}
  {14}},\ \bibinfo {pages} {21--47} (\bibinfo {year} {2002})}\BibitemShut
  {NoStop}%
\bibitem [{\citenamefont {Zhang}, \citenamefont {Lin},\ and\ \citenamefont
  {Lin}(2007)}]{zhang_aggregate_2007}%
  \BibitemOpen
  \bibfield  {author} {\bibinfo {author} {\bibfnamefont {L.}~\bibnamefont
  {Zhang}}, \bibinfo {author} {\bibfnamefont {J.}~\bibnamefont {Lin}},\ and\
  \bibinfo {author} {\bibfnamefont {S.}~\bibnamefont {Lin}},\ }\bibfield
  {title} {\enquote {\bibinfo {title} {Aggregate {Morphologies} of
  {Amphiphilic} {Graft} {Copolymers} in {Dilute} {Solution} {Studied} by
  {Self}-{Consistent} {Field} {Theory}},}\ }\href
  {https://doi.org/10.1021/jp068429l} {\bibfield  {journal} {\bibinfo
  {journal} {The Journal of Physical Chemistry B}\ }\textbf {\bibinfo {volume}
  {111}},\ \bibinfo {pages} {9209--9217} (\bibinfo {year} {2007})},\ \bibinfo
  {note} {publisher: American Chemical Society}\BibitemShut {NoStop}%
\bibitem [{\citenamefont {McCarty}\ \emph {et~al.}(2019)\citenamefont
  {McCarty}, \citenamefont {Delaney}, \citenamefont {Danielsen}, \citenamefont
  {Fredrickson},\ and\ \citenamefont {Shea}}]{Mccarty2019}%
  \BibitemOpen
  \bibfield  {author} {\bibinfo {author} {\bibfnamefont {J.}~\bibnamefont
  {McCarty}}, \bibinfo {author} {\bibfnamefont {K.~T.}\ \bibnamefont
  {Delaney}}, \bibinfo {author} {\bibfnamefont {S.~P.}\ \bibnamefont
  {Danielsen}}, \bibinfo {author} {\bibfnamefont {G.~H.}\ \bibnamefont
  {Fredrickson}},\ and\ \bibinfo {author} {\bibfnamefont {J.-E.}\ \bibnamefont
  {Shea}},\ }\bibfield  {title} {\enquote {\bibinfo {title} {Complete phase
  diagram for liquid--liquid phase separation of intrinsically disordered
  proteins},}\ }\href@noop {} {\bibfield  {journal} {\bibinfo  {journal} {The
  journal of physical chemistry letters}\ }\textbf {\bibinfo {volume} {10}},\
  \bibinfo {pages} {1644--1652} (\bibinfo {year} {2019})}\BibitemShut {NoStop}%
\bibitem [{\citenamefont {Lyubimov}, \citenamefont {Beltran-Villegas},\ and\
  \citenamefont {Jayaraman}(2017)}]{Lyubimov2017}%
  \BibitemOpen
  \bibfield  {author} {\bibinfo {author} {\bibfnamefont {I.}~\bibnamefont
  {Lyubimov}}, \bibinfo {author} {\bibfnamefont {D.~J.}\ \bibnamefont
  {Beltran-Villegas}},\ and\ \bibinfo {author} {\bibfnamefont {A.}~\bibnamefont
  {Jayaraman}},\ }\bibfield  {title} {\enquote {\bibinfo {title} {Prism theory
  study of amphiphilic block copolymer solutions with varying copolymer
  sequence and composition},}\ }\href
  {https://doi.org/10.1021/acs.macromol.7b01419} {\bibfield  {journal}
  {\bibinfo  {journal} {Macromolecules}\ }\textbf {\bibinfo {volume} {50}},\
  \bibinfo {pages} {7419--7431} (\bibinfo {year} {2017})},\ \Eprint
  {https://arxiv.org/abs/https://doi.org/10.1021/acs.macromol.7b01419}
  {https://doi.org/10.1021/acs.macromol.7b01419} \BibitemShut {NoStop}%
\bibitem [{\citenamefont {Gartner}\ and\ \citenamefont
  {Jayaraman}(2019)}]{Gartner2019}%
  \BibitemOpen
  \bibfield  {author} {\bibinfo {author} {\bibfnamefont {T.~E.}\ \bibnamefont
  {Gartner}}\ and\ \bibinfo {author} {\bibfnamefont {A.}~\bibnamefont
  {Jayaraman}},\ }\bibfield  {title} {\enquote {\bibinfo {title} {Modeling and
  simulations of polymers: A roadmap},}\ }\href
  {https://doi.org/10.1021/acs.macromol.8b01836} {\bibfield  {journal}
  {\bibinfo  {journal} {Macromolecules}\ }\textbf {\bibinfo {volume} {52}},\
  \bibinfo {pages} {755--786} (\bibinfo {year} {2019})},\ \Eprint
  {https://arxiv.org/abs/https://doi.org/10.1021/acs.macromol.8b01836}
  {https://doi.org/10.1021/acs.macromol.8b01836} \BibitemShut {NoStop}%
\bibitem [{\citenamefont {Bale}, \citenamefont {Gautham},\ and\ \citenamefont
  {Patra}(2022)}]{bale_sequence-defined}%
  \BibitemOpen
  \bibfield  {author} {\bibinfo {author} {\bibfnamefont {A.~A.}\ \bibnamefont
  {Bale}}, \bibinfo {author} {\bibfnamefont {S.~M.~B.}\ \bibnamefont
  {Gautham}},\ and\ \bibinfo {author} {\bibfnamefont {T.~K.}\ \bibnamefont
  {Patra}},\ }\bibfield  {title} {\enquote {\bibinfo {title} {Sequence-defined
  {Pareto} frontier of a copolymer structure},}\ }\href
  {https://doi.org/https://doi.org/10.1002/pol.20220088} {\bibfield  {journal}
  {\bibinfo  {journal} {Journal of Polymer Science}\ } (\bibinfo {year}
  {2022}),\ https://doi.org/10.1002/pol.20220088},\ \bibinfo {note} {\_eprint:
  https://onlinelibrary.wiley.com/doi/pdf/10.1002/pol.20220088}\BibitemShut
  {NoStop}%
\bibitem [{\citenamefont {Ferguson}\ \emph {et~al.}(2011)\citenamefont
  {Ferguson}, \citenamefont {Panagiotopoulos}, \citenamefont {Kevrekidis},\
  and\ \citenamefont {Debenedetti}}]{Ferguson2011}%
  \BibitemOpen
  \bibfield  {author} {\bibinfo {author} {\bibfnamefont {A.~L.}\ \bibnamefont
  {Ferguson}}, \bibinfo {author} {\bibfnamefont {A.~Z.}\ \bibnamefont
  {Panagiotopoulos}}, \bibinfo {author} {\bibfnamefont {I.~G.}\ \bibnamefont
  {Kevrekidis}},\ and\ \bibinfo {author} {\bibfnamefont {P.~G.}\ \bibnamefont
  {Debenedetti}},\ }\bibfield  {title} {\enquote {\bibinfo {title} {Nonlinear
  dimensionality reduction in molecular simulation: The diffusion map
  approach},}\ }\href@noop {} {\bibfield  {journal} {\bibinfo  {journal}
  {Chemical Physics Letters}\ }\textbf {\bibinfo {volume} {509}},\ \bibinfo
  {pages} {1--11} (\bibinfo {year} {2011})}\BibitemShut {NoStop}%
\bibitem [{\citenamefont {Reinhart}\ \emph {et~al.}(2017)\citenamefont
  {Reinhart}, \citenamefont {Long}, \citenamefont {Howard}, \citenamefont
  {Ferguson},\ and\ \citenamefont {Panagiotopoulos}}]{Reinhart2017}%
  \BibitemOpen
  \bibfield  {author} {\bibinfo {author} {\bibfnamefont {W.~F.}\ \bibnamefont
  {Reinhart}}, \bibinfo {author} {\bibfnamefont {A.~W.}\ \bibnamefont {Long}},
  \bibinfo {author} {\bibfnamefont {M.~P.}\ \bibnamefont {Howard}}, \bibinfo
  {author} {\bibfnamefont {A.~L.}\ \bibnamefont {Ferguson}},\ and\ \bibinfo
  {author} {\bibfnamefont {A.~Z.}\ \bibnamefont {Panagiotopoulos}},\ }\bibfield
   {title} {\enquote {\bibinfo {title} {Machine learning for autonomous crystal
  structure identification},}\ }\href {https://doi.org/10.1039/C7SM00957G}
  {\bibfield  {journal} {\bibinfo  {journal} {Soft Matter}\ }\textbf {\bibinfo
  {volume} {13}},\ \bibinfo {pages} {4733--4745} (\bibinfo {year}
  {2017})}\BibitemShut {NoStop}%
\bibitem [{\citenamefont {Chen}, \citenamefont {Tan},\ and\ \citenamefont
  {Ferguson}(2018)}]{Chen2018}%
  \BibitemOpen
  \bibfield  {author} {\bibinfo {author} {\bibfnamefont {W.}~\bibnamefont
  {Chen}}, \bibinfo {author} {\bibfnamefont {A.~R.}\ \bibnamefont {Tan}},\ and\
  \bibinfo {author} {\bibfnamefont {A.~L.}\ \bibnamefont {Ferguson}},\
  }\bibfield  {title} {\enquote {\bibinfo {title} {Collective variable
  discovery and enhanced sampling using autoencoders: Innovations in network
  architecture and error function design},}\ }\href@noop {} {\bibfield
  {journal} {\bibinfo  {journal} {The Journal of chemical physics}\ }\textbf
  {\bibinfo {volume} {149}},\ \bibinfo {pages} {072312} (\bibinfo {year}
  {2018})}\BibitemShut {NoStop}%
\bibitem [{\citenamefont {Sun}\ \emph {et~al.}(2020)\citenamefont {Sun},
  \citenamefont {Li}, \citenamefont {Zhang}, \citenamefont {Gao},\ and\
  \citenamefont {Luo}}]{Sun2020}%
  \BibitemOpen
  \bibfield  {author} {\bibinfo {author} {\bibfnamefont {L.-W.}\ \bibnamefont
  {Sun}}, \bibinfo {author} {\bibfnamefont {H.}~\bibnamefont {Li}}, \bibinfo
  {author} {\bibfnamefont {X.-Q.}\ \bibnamefont {Zhang}}, \bibinfo {author}
  {\bibfnamefont {H.-B.}\ \bibnamefont {Gao}},\ and\ \bibinfo {author}
  {\bibfnamefont {M.-B.}\ \bibnamefont {Luo}},\ }\bibfield  {title} {\enquote
  {\bibinfo {title} {Identifying conformation states of polymer through
  unsupervised machine learning},}\ }\href@noop {} {\bibfield  {journal}
  {\bibinfo  {journal} {Chinese Journal of Polymer Science}\ }\textbf {\bibinfo
  {volume} {38}},\ \bibinfo {pages} {1403--1408} (\bibinfo {year}
  {2020})}\BibitemShut {NoStop}%
\bibitem [{\citenamefont {Bhattacharya}\ and\ \citenamefont
  {Patra}(2021)}]{Bhattacharya2021}%
  \BibitemOpen
  \bibfield  {author} {\bibinfo {author} {\bibfnamefont {D.}~\bibnamefont
  {Bhattacharya}}\ and\ \bibinfo {author} {\bibfnamefont {T.~K.}\ \bibnamefont
  {Patra}},\ }\bibfield  {title} {\enquote {\bibinfo {title} {dpoly: Deep
  learning of polymer phases and phase transition},}\ }\href@noop {} {\bibfield
   {journal} {\bibinfo  {journal} {Macromolecules}\ }\textbf {\bibinfo {volume}
  {54}},\ \bibinfo {pages} {3065--3074} (\bibinfo {year} {2021})}\BibitemShut
  {NoStop}%
\bibitem [{\citenamefont {S. Clegg}(2021)}]{sclegg_characterising_2021}%
  \BibitemOpen
  \bibfield  {author} {\bibinfo {author} {\bibfnamefont {P.}~\bibnamefont
  {S. Clegg}},\ }\bibfield  {title} {\enquote {\bibinfo {title}
  {Characterising soft matter using machine learning},}\ }\href
  {https://doi.org/10.1039/D0SM01686A} {\bibfield  {journal} {\bibinfo
  {journal} {Soft Matter}\ }\textbf {\bibinfo {volume} {17}},\ \bibinfo {pages}
  {3991--4005} (\bibinfo {year} {2021})},\ \bibinfo {note} {publisher: Royal
  Society of Chemistry}\BibitemShut {NoStop}%
\bibitem [{\citenamefont {Kim}\ \emph {et~al.}(2021)\citenamefont {Kim},
  \citenamefont {Batra}, \citenamefont {Chen}, \citenamefont {Tran},\ and\
  \citenamefont {Ramprasad}}]{kim2021polymer}%
  \BibitemOpen
  \bibfield  {author} {\bibinfo {author} {\bibfnamefont {C.}~\bibnamefont
  {Kim}}, \bibinfo {author} {\bibfnamefont {R.}~\bibnamefont {Batra}}, \bibinfo
  {author} {\bibfnamefont {L.}~\bibnamefont {Chen}}, \bibinfo {author}
  {\bibfnamefont {H.}~\bibnamefont {Tran}},\ and\ \bibinfo {author}
  {\bibfnamefont {R.}~\bibnamefont {Ramprasad}},\ }\bibfield  {title} {\enquote
  {\bibinfo {title} {Polymer design using genetic algorithm and machine
  learning},}\ }\href@noop {} {\bibfield  {journal} {\bibinfo  {journal}
  {Computational Materials Science}\ }\textbf {\bibinfo {volume} {186}},\
  \bibinfo {pages} {110067} (\bibinfo {year} {2021})}\BibitemShut {NoStop}%
\bibitem [{\citenamefont {J{\o}rgensen}\ \emph {et~al.}(2018)\citenamefont
  {J{\o}rgensen}, \citenamefont {Mesta}, \citenamefont {Shil}, \citenamefont
  {Garc{\'\i}a~Lastra}, \citenamefont {Jacobsen}, \citenamefont {Thygesen},\
  and\ \citenamefont {Schmidt}}]{jorgensen2018machine}%
  \BibitemOpen
  \bibfield  {author} {\bibinfo {author} {\bibfnamefont {P.~B.}\ \bibnamefont
  {J{\o}rgensen}}, \bibinfo {author} {\bibfnamefont {M.}~\bibnamefont {Mesta}},
  \bibinfo {author} {\bibfnamefont {S.}~\bibnamefont {Shil}}, \bibinfo {author}
  {\bibfnamefont {J.~M.}\ \bibnamefont {Garc{\'\i}a~Lastra}}, \bibinfo {author}
  {\bibfnamefont {K.~W.}\ \bibnamefont {Jacobsen}}, \bibinfo {author}
  {\bibfnamefont {K.~S.}\ \bibnamefont {Thygesen}},\ and\ \bibinfo {author}
  {\bibfnamefont {M.~N.}\ \bibnamefont {Schmidt}},\ }\bibfield  {title}
  {\enquote {\bibinfo {title} {Machine learning-based screening of complex
  molecules for polymer solar cells},}\ }\href@noop {} {\bibfield  {journal}
  {\bibinfo  {journal} {The Journal of chemical physics}\ }\textbf {\bibinfo
  {volume} {148}},\ \bibinfo {pages} {241735} (\bibinfo {year}
  {2018})}\BibitemShut {NoStop}%
\bibitem [{\citenamefont {Yamankurt}\ \emph {et~al.}(2019)\citenamefont
  {Yamankurt}, \citenamefont {Berns}, \citenamefont {Xue}, \citenamefont {Lee},
  \citenamefont {Bagheri}, \citenamefont {Mrksich},\ and\ \citenamefont
  {Mirkin}}]{yamankurt_exploration_2019}%
  \BibitemOpen
  \bibfield  {author} {\bibinfo {author} {\bibfnamefont {G.}~\bibnamefont
  {Yamankurt}}, \bibinfo {author} {\bibfnamefont {E.~J.}\ \bibnamefont
  {Berns}}, \bibinfo {author} {\bibfnamefont {A.}~\bibnamefont {Xue}}, \bibinfo
  {author} {\bibfnamefont {A.}~\bibnamefont {Lee}}, \bibinfo {author}
  {\bibfnamefont {N.}~\bibnamefont {Bagheri}}, \bibinfo {author} {\bibfnamefont
  {M.}~\bibnamefont {Mrksich}},\ and\ \bibinfo {author} {\bibfnamefont {C.~A.}\
  \bibnamefont {Mirkin}},\ }\bibfield  {title} {\enquote {\bibinfo {title}
  {Exploration of the nanomedicine-design space with high-throughput screening
  and machine learning},}\ }\href {https://doi.org/10.1038/s41551-019-0351-1}
  {\bibfield  {journal} {\bibinfo  {journal} {Nature Biomedical Engineering}\
  }\textbf {\bibinfo {volume} {3}},\ \bibinfo {pages} {318--327} (\bibinfo
  {year} {2019})}\BibitemShut {NoStop}%
\bibitem [{\citenamefont {Jeon}\ \emph {et~al.}(2014)\citenamefont {Jeon},
  \citenamefont {Nim}, \citenamefont {Teyra}, \citenamefont {Datti},
  \citenamefont {Wrana}, \citenamefont {Sidhu}, \citenamefont {Moffat},\ and\
  \citenamefont {Kim}}]{jeon_systematic_2014}%
  \BibitemOpen
  \bibfield  {author} {\bibinfo {author} {\bibfnamefont {J.}~\bibnamefont
  {Jeon}}, \bibinfo {author} {\bibfnamefont {S.}~\bibnamefont {Nim}}, \bibinfo
  {author} {\bibfnamefont {J.}~\bibnamefont {Teyra}}, \bibinfo {author}
  {\bibfnamefont {A.}~\bibnamefont {Datti}}, \bibinfo {author} {\bibfnamefont
  {J.~L.}\ \bibnamefont {Wrana}}, \bibinfo {author} {\bibfnamefont {S.~S.}\
  \bibnamefont {Sidhu}}, \bibinfo {author} {\bibfnamefont {J.}~\bibnamefont
  {Moffat}},\ and\ \bibinfo {author} {\bibfnamefont {P.~M.}\ \bibnamefont
  {Kim}},\ }\bibfield  {title} {\enquote {\bibinfo {title} {A systematic
  approach to identify novel cancer drug targets using machine learning,
  inhibitor design and high-throughput screening},}\ }\href
  {https://doi.org/10.1186/s13073-014-0057-7} {\bibfield  {journal} {\bibinfo
  {journal} {Genome Medicine}\ }\textbf {\bibinfo {volume} {6}},\ \bibinfo
  {pages} {57} (\bibinfo {year} {2014})}\BibitemShut {NoStop}%
\bibitem [{\citenamefont {Wang}\ \emph {et~al.}(2022)\citenamefont {Wang},
  \citenamefont {Xu}, \citenamefont {Tang},\ and\ \citenamefont
  {Jiang}}]{wang2022machine}%
  \BibitemOpen
  \bibfield  {author} {\bibinfo {author} {\bibfnamefont {M.}~\bibnamefont
  {Wang}}, \bibinfo {author} {\bibfnamefont {Q.}~\bibnamefont {Xu}}, \bibinfo
  {author} {\bibfnamefont {H.}~\bibnamefont {Tang}},\ and\ \bibinfo {author}
  {\bibfnamefont {J.}~\bibnamefont {Jiang}},\ }\bibfield  {title} {\enquote
  {\bibinfo {title} {Machine learning-enabled prediction and high-throughput
  screening of polymer membranes for pervaporation separation},}\ }\href@noop
  {} {\bibfield  {journal} {\bibinfo  {journal} {ACS Applied Materials \&
  Interfaces}\ }\textbf {\bibinfo {volume} {14}},\ \bibinfo {pages}
  {8427--8436} (\bibinfo {year} {2022})}\BibitemShut {NoStop}%
\bibitem [{\citenamefont {Mehta}\ \emph {et~al.}(2021)\citenamefont {Mehta},
  \citenamefont {Laghuvarapu}, \citenamefont {Pathak}, \citenamefont {Sethi},
  \citenamefont {Alvala},\ and\ \citenamefont
  {Priyakumar}}]{mehta2021enhanced}%
  \BibitemOpen
  \bibfield  {author} {\bibinfo {author} {\bibfnamefont {S.}~\bibnamefont
  {Mehta}}, \bibinfo {author} {\bibfnamefont {S.}~\bibnamefont {Laghuvarapu}},
  \bibinfo {author} {\bibfnamefont {Y.}~\bibnamefont {Pathak}}, \bibinfo
  {author} {\bibfnamefont {A.}~\bibnamefont {Sethi}}, \bibinfo {author}
  {\bibfnamefont {M.}~\bibnamefont {Alvala}},\ and\ \bibinfo {author}
  {\bibfnamefont {U.~D.}\ \bibnamefont {Priyakumar}},\ }\bibfield  {title}
  {\enquote {\bibinfo {title} {Enhanced sampling of chemical space for high
  throughput screening applications using machine learning},}\ }\href
  {https://doi.org/10.26434/chemrxiv.14139275.v1} {\bibfield  {journal}
  {\bibinfo  {journal} {ChemRxiv}\ } (\bibinfo {year} {2021}),\
  10.26434/chemrxiv.14139275.v1}\BibitemShut {NoStop}%
\bibitem [{\citenamefont {Verdonck}\ \emph {et~al.}(2021)\citenamefont
  {Verdonck}, \citenamefont {Baesens}, \citenamefont {{\'O}skarsd{\'o}ttir}
  \emph {et~al.}}]{verdonck_special_2021}%
  \BibitemOpen
  \bibfield  {author} {\bibinfo {author} {\bibfnamefont {T.}~\bibnamefont
  {Verdonck}}, \bibinfo {author} {\bibfnamefont {B.}~\bibnamefont {Baesens}},
  \bibinfo {author} {\bibfnamefont {M.}~\bibnamefont {{\'O}skarsd{\'o}ttir}},
  \emph {et~al.},\ }\bibfield  {title} {\enquote {\bibinfo {title} {Special
  issue on feature engineering editorial},}\ }\href@noop {} {\bibfield
  {journal} {\bibinfo  {journal} {Machine Learning}\ ,\ \bibinfo {pages}
  {1--12}} (\bibinfo {year} {2021})}\BibitemShut {NoStop}%
\bibitem [{\citenamefont {Jing}\ \emph {et~al.}(2019)\citenamefont {Jing},
  \citenamefont {Dong}, \citenamefont {Hong},\ and\ \citenamefont
  {Lu}}]{jing2019amino}%
  \BibitemOpen
  \bibfield  {author} {\bibinfo {author} {\bibfnamefont {X.}~\bibnamefont
  {Jing}}, \bibinfo {author} {\bibfnamefont {Q.}~\bibnamefont {Dong}}, \bibinfo
  {author} {\bibfnamefont {D.}~\bibnamefont {Hong}},\ and\ \bibinfo {author}
  {\bibfnamefont {R.}~\bibnamefont {Lu}},\ }\bibfield  {title} {\enquote
  {\bibinfo {title} {Amino acid encoding methods for protein sequences: a
  comprehensive review and assessment},}\ }\href@noop {} {\bibfield  {journal}
  {\bibinfo  {journal} {IEEE/ACM transactions on computational biology and
  bioinformatics}\ }\textbf {\bibinfo {volume} {17}},\ \bibinfo {pages}
  {1918--1931} (\bibinfo {year} {2019})}\BibitemShut {NoStop}%
\bibitem [{\citenamefont {Patel}, \citenamefont {Borca},\ and\ \citenamefont
  {Webb}(2022)}]{patel_featurization_2022}%
  \BibitemOpen
  \bibfield  {author} {\bibinfo {author} {\bibfnamefont {R.~A.}\ \bibnamefont
  {Patel}}, \bibinfo {author} {\bibfnamefont {C.~H.}\ \bibnamefont {Borca}},\
  and\ \bibinfo {author} {\bibfnamefont {M.~A.}\ \bibnamefont {Webb}},\
  }\bibfield  {title} {\enquote {\bibinfo {title} {Featurization strategies for
  polymer sequence or composition design by machine learning},}\ }\href
  {https://doi.org/10.1039/D1ME00160D} {\bibfield  {journal} {\bibinfo
  {journal} {Mol. Syst. Des. Eng.}\ ,\ \bibinfo {pages} {--}} (\bibinfo {year}
  {2022})},\ \bibinfo {note} {publisher: The Royal Society of
  Chemistry}\BibitemShut {NoStop}%
\bibitem [{\citenamefont {Jablonka}\ \emph {et~al.}(2021)\citenamefont
  {Jablonka}, \citenamefont {Jothiappan}, \citenamefont {Wang}, \citenamefont
  {Smit},\ and\ \citenamefont {Yoo}}]{jablonka2021bias}%
  \BibitemOpen
  \bibfield  {author} {\bibinfo {author} {\bibfnamefont {K.~M.}\ \bibnamefont
  {Jablonka}}, \bibinfo {author} {\bibfnamefont {G.~M.}\ \bibnamefont
  {Jothiappan}}, \bibinfo {author} {\bibfnamefont {S.}~\bibnamefont {Wang}},
  \bibinfo {author} {\bibfnamefont {B.}~\bibnamefont {Smit}},\ and\ \bibinfo
  {author} {\bibfnamefont {B.}~\bibnamefont {Yoo}},\ }\bibfield  {title}
  {\enquote {\bibinfo {title} {Bias free multiobjective active learning for
  materials design and discovery},}\ }\href@noop {} {\bibfield  {journal}
  {\bibinfo  {journal} {Nature communications}\ }\textbf {\bibinfo {volume}
  {12}},\ \bibinfo {pages} {1--10} (\bibinfo {year} {2021})}\BibitemShut
  {NoStop}%
\bibitem [{\citenamefont {Mohapatra}, \citenamefont {An},\ and\ \citenamefont
  {G{\'o}mez-Bombarelli}(2022)}]{mohapatra2022chemistry}%
  \BibitemOpen
  \bibfield  {author} {\bibinfo {author} {\bibfnamefont {S.}~\bibnamefont
  {Mohapatra}}, \bibinfo {author} {\bibfnamefont {J.}~\bibnamefont {An}},\ and\
  \bibinfo {author} {\bibfnamefont {R.}~\bibnamefont {G{\'o}mez-Bombarelli}},\
  }\bibfield  {title} {\enquote {\bibinfo {title} {Chemistry-informed
  macromolecule graph representation for similarity computation, unsupervised
  and supervised learning},}\ }\href@noop {} {\bibfield  {journal} {\bibinfo
  {journal} {Machine Learning: Science and Technology}\ }\textbf {\bibinfo
  {volume} {3}},\ \bibinfo {pages} {015028} (\bibinfo {year}
  {2022})}\BibitemShut {NoStop}%
\bibitem [{\citenamefont {Webb}\ \emph {et~al.}(2020)\citenamefont {Webb},
  \citenamefont {Jackson}, \citenamefont {Gil},\ and\ \citenamefont
  {de~Pablo}}]{webb2020targeted}%
  \BibitemOpen
  \bibfield  {author} {\bibinfo {author} {\bibfnamefont {M.~A.}\ \bibnamefont
  {Webb}}, \bibinfo {author} {\bibfnamefont {N.~E.}\ \bibnamefont {Jackson}},
  \bibinfo {author} {\bibfnamefont {P.~S.}\ \bibnamefont {Gil}},\ and\ \bibinfo
  {author} {\bibfnamefont {J.~J.}\ \bibnamefont {de~Pablo}},\ }\bibfield
  {title} {\enquote {\bibinfo {title} {Targeted sequence design within the
  coarse-grained polymer genome},}\ }\href@noop {} {\bibfield  {journal}
  {\bibinfo  {journal} {Science advances}\ }\textbf {\bibinfo {volume} {6}},\
  \bibinfo {pages} {eabc6216} (\bibinfo {year} {2020})}\BibitemShut {NoStop}%
\bibitem [{\citenamefont {Shi}\ \emph {et~al.}(2021)\citenamefont {Shi},
  \citenamefont {Quevillon}, \citenamefont {Valen{\c{c}}a},\ and\ \citenamefont
  {Whitmer}}]{shi2021predicting}%
  \BibitemOpen
  \bibfield  {author} {\bibinfo {author} {\bibfnamefont {J.}~\bibnamefont
  {Shi}}, \bibinfo {author} {\bibfnamefont {M.~J.}\ \bibnamefont {Quevillon}},
  \bibinfo {author} {\bibfnamefont {P.~H.~A.}\ \bibnamefont {Valen{\c{c}}a}},\
  and\ \bibinfo {author} {\bibfnamefont {J.~K.}\ \bibnamefont {Whitmer}},\
  }\bibfield  {title} {\enquote {\bibinfo {title} {Predicting adhesive free
  energies of polymer--surface interactions with machine learning},}\
  }\href@noop {} {\bibfield  {journal} {\bibinfo  {journal} {arXiv preprint
  arXiv:2110.03041}\ } (\bibinfo {year} {2021})}\BibitemShut {NoStop}%
\bibitem [{\citenamefont {Reinhart}(2021)}]{Reinhart2021}%
  \BibitemOpen
  \bibfield  {author} {\bibinfo {author} {\bibfnamefont {W.~F.}\ \bibnamefont
  {Reinhart}},\ }\bibfield  {title} {\enquote {\bibinfo {title} {Unsupervised
  learning of atomic environments from simple features},}\ }\href@noop {}
  {\bibfield  {journal} {\bibinfo  {journal} {Computational Materials Science}\
  }\textbf {\bibinfo {volume} {196}},\ \bibinfo {pages} {110511} (\bibinfo
  {year} {2021})}\BibitemShut {NoStop}%
\bibitem [{\citenamefont {Statt}, \citenamefont {Kleeblatt},\ and\
  \citenamefont {Reinhart}(2021{\natexlab{a}})}]{statt2021unsupervised}%
  \BibitemOpen
  \bibfield  {author} {\bibinfo {author} {\bibfnamefont {A.}~\bibnamefont
  {Statt}}, \bibinfo {author} {\bibfnamefont {D.~C.}\ \bibnamefont
  {Kleeblatt}},\ and\ \bibinfo {author} {\bibfnamefont {W.~F.}\ \bibnamefont
  {Reinhart}},\ }\bibfield  {title} {\enquote {\bibinfo {title} {Unsupervised
  learning of sequence-specific aggregation behavior for a model copolymer},}\
  }\href@noop {} {\bibfield  {journal} {\bibinfo  {journal} {Soft matter}\
  }\textbf {\bibinfo {volume} {17}},\ \bibinfo {pages} {7697--7707} (\bibinfo
  {year} {2021}{\natexlab{a}})}\BibitemShut {NoStop}%
\bibitem [{\citenamefont {Statt}\ \emph {et~al.}(2020)\citenamefont {Statt},
  \citenamefont {Casademunt}, \citenamefont {Brangwynne},\ and\ \citenamefont
  {Panagiotopoulos}}]{Statt2020}%
  \BibitemOpen
  \bibfield  {author} {\bibinfo {author} {\bibfnamefont {A.}~\bibnamefont
  {Statt}}, \bibinfo {author} {\bibfnamefont {H.}~\bibnamefont {Casademunt}},
  \bibinfo {author} {\bibfnamefont {C.~P.}\ \bibnamefont {Brangwynne}},\ and\
  \bibinfo {author} {\bibfnamefont {A.~Z.}\ \bibnamefont {Panagiotopoulos}},\
  }\bibfield  {title} {\enquote {\bibinfo {title} {Model for disordered
  proteins with strongly sequence-dependent liquid phase behavior},}\ }\href
  {https://doi.org/10.1063/1.5141095} {\bibfield  {journal} {\bibinfo
  {journal} {The Journal of Chemical Physics}\ }\textbf {\bibinfo {volume}
  {152}},\ \bibinfo {pages} {075101} (\bibinfo {year} {2020})},\ \Eprint
  {https://arxiv.org/abs/https://doi.org/10.1063/1.5141095}
  {https://doi.org/10.1063/1.5141095} \BibitemShut {NoStop}%
\bibitem [{\citenamefont {Jones}\ and\ \citenamefont
  {Chapman}(1924)}]{Jones1924}%
  \BibitemOpen
  \bibfield  {author} {\bibinfo {author} {\bibfnamefont {J.~E.}\ \bibnamefont
  {Jones}}\ and\ \bibinfo {author} {\bibfnamefont {S.}~\bibnamefont
  {Chapman}},\ }\bibfield  {title} {\enquote {\bibinfo {title} {On the
  determination of molecular fields. \&\#x2014;ii. from the equation of state
  of a gas},}\ }\href {https://doi.org/10.1098/rspa.1924.0082} {\bibfield
  {journal} {\bibinfo  {journal} {Proceedings of the Royal Society of London.
  Series A, Containing Papers of a Mathematical and Physical Character}\
  }\textbf {\bibinfo {volume} {106}},\ \bibinfo {pages} {463--477} (\bibinfo
  {year} {1924})},\ \Eprint
  {https://arxiv.org/abs/https://royalsocietypublishing.org/doi/pdf/10.1098/rspa.1924.0082}
  {https://royalsocietypublishing.org/doi/pdf/10.1098/rspa.1924.0082}
  \BibitemShut {NoStop}%
\bibitem [{\citenamefont {Weeks}, \citenamefont {Chandler},\ and\ \citenamefont
  {Andersen}(1971)}]{Weeks1971}%
  \BibitemOpen
  \bibfield  {author} {\bibinfo {author} {\bibfnamefont {J.~D.}\ \bibnamefont
  {Weeks}}, \bibinfo {author} {\bibfnamefont {D.}~\bibnamefont {Chandler}},\
  and\ \bibinfo {author} {\bibfnamefont {H.~C.}\ \bibnamefont {Andersen}},\
  }\bibfield  {title} {\enquote {\bibinfo {title} {Role of repulsive forces in
  determining the equilibrium structure of simple liquids},}\ }\href
  {https://doi.org/10.1063/1.1674820} {\bibfield  {journal} {\bibinfo
  {journal} {The Journal of Chemical Physics}\ }\textbf {\bibinfo {volume}
  {54}},\ \bibinfo {pages} {5237--5247} (\bibinfo {year} {1971})}\BibitemShut
  {NoStop}%
\bibitem [{\citenamefont {Kremer}\ and\ \citenamefont
  {Grest}(1990)}]{Kremer1990}%
  \BibitemOpen
  \bibfield  {author} {\bibinfo {author} {\bibfnamefont {K.}~\bibnamefont
  {Kremer}}\ and\ \bibinfo {author} {\bibfnamefont {G.~S.}\ \bibnamefont
  {Grest}},\ }\bibfield  {title} {\enquote {\bibinfo {title} {Dynamics of
  entangled linear polymer melts: A molecular‐dynamics simulation},}\ }\href
  {https://doi.org/10.1063/1.458541} {\bibfield  {journal} {\bibinfo  {journal}
  {The Journal of Chemical Physics}\ }\textbf {\bibinfo {volume} {92}},\
  \bibinfo {pages} {5057--5086} (\bibinfo {year} {1990})}\BibitemShut {NoStop}%
\bibitem [{\citenamefont {Glaser}\ \emph {et~al.}(2015)\citenamefont {Glaser},
  \citenamefont {Nguyen}, \citenamefont {Anderson}, \citenamefont {Lui},
  \citenamefont {Spiga}, \citenamefont {Millan}, \citenamefont {Morse},\ and\
  \citenamefont {Glotzer}}]{Glaser2015}%
  \BibitemOpen
  \bibfield  {author} {\bibinfo {author} {\bibfnamefont {J.}~\bibnamefont
  {Glaser}}, \bibinfo {author} {\bibfnamefont {T.~D.}\ \bibnamefont {Nguyen}},
  \bibinfo {author} {\bibfnamefont {J.~A.}\ \bibnamefont {Anderson}}, \bibinfo
  {author} {\bibfnamefont {P.}~\bibnamefont {Lui}}, \bibinfo {author}
  {\bibfnamefont {F.}~\bibnamefont {Spiga}}, \bibinfo {author} {\bibfnamefont
  {J.~A.}\ \bibnamefont {Millan}}, \bibinfo {author} {\bibfnamefont {D.~C.}\
  \bibnamefont {Morse}},\ and\ \bibinfo {author} {\bibfnamefont {S.~C.}\
  \bibnamefont {Glotzer}},\ }\bibfield  {title} {\enquote {\bibinfo {title}
  {Strong scaling of general-purpose molecular dynamics simulations on gpus},}\
  }\href {https://doi.org/https://doi.org/10.1016/j.cpc.2015.02.028} {\bibfield
   {journal} {\bibinfo  {journal} {Computer Physics Communications}\ }\textbf
  {\bibinfo {volume} {192}},\ \bibinfo {pages} {97 -- 107} (\bibinfo {year}
  {2015})}\BibitemShut {NoStop}%
\bibitem [{\citenamefont {Anderson}, \citenamefont {Lorenz},\ and\
  \citenamefont {Travesset}(2008)}]{Anderson2008}%
  \BibitemOpen
  \bibfield  {author} {\bibinfo {author} {\bibfnamefont {J.~A.}\ \bibnamefont
  {Anderson}}, \bibinfo {author} {\bibfnamefont {C.~D.}\ \bibnamefont
  {Lorenz}},\ and\ \bibinfo {author} {\bibfnamefont {A.}~\bibnamefont
  {Travesset}},\ }\bibfield  {title} {\enquote {\bibinfo {title} {General
  purpose molecular dynamics simulations fully implemented on graphics
  processing units},}\ }\href
  {https://doi.org/https://doi.org/10.1016/j.jcp.2008.01.047} {\bibfield
  {journal} {\bibinfo  {journal} {Journal of Computational Physics}\ }\textbf
  {\bibinfo {volume} {227}},\ \bibinfo {pages} {5342 -- 5359} (\bibinfo {year}
  {2008})}\BibitemShut {NoStop}%
\bibitem [{\citenamefont {Ziolek}\ \emph {et~al.}(2021)\citenamefont {Ziolek},
  \citenamefont {Smith}, \citenamefont {Pink}, \citenamefont {Dreiss},\ and\
  \citenamefont {Lorenz}}]{Ziolek2021}%
  \BibitemOpen
  \bibfield  {author} {\bibinfo {author} {\bibfnamefont {R.~M.}\ \bibnamefont
  {Ziolek}}, \bibinfo {author} {\bibfnamefont {P.}~\bibnamefont {Smith}},
  \bibinfo {author} {\bibfnamefont {D.~L.}\ \bibnamefont {Pink}}, \bibinfo
  {author} {\bibfnamefont {C.~A.}\ \bibnamefont {Dreiss}},\ and\ \bibinfo
  {author} {\bibfnamefont {C.~D.}\ \bibnamefont {Lorenz}},\ }\bibfield  {title}
  {\enquote {\bibinfo {title} {Unsupervised learning unravels the structure of
  four-arm and linear block copolymer micelles},}\ }\href@noop {} {\bibfield
  {journal} {\bibinfo  {journal} {Macromolecules}\ }\textbf {\bibinfo {volume}
  {54}},\ \bibinfo {pages} {3755--3768} (\bibinfo {year} {2021})}\BibitemShut
  {NoStop}%
\bibitem [{\citenamefont {Pedregosa}\ \emph {et~al.}(2011)\citenamefont
  {Pedregosa}, \citenamefont {Varoquaux}, \citenamefont {Gramfort},
  \citenamefont {Michel}, \citenamefont {Thirion}, \citenamefont {Grisel},
  \citenamefont {Blondel}, \citenamefont {Prettenhofer}, \citenamefont {Weiss},
  \citenamefont {Dubourg}, \citenamefont {Vanderplas}, \citenamefont {Passos},
  \citenamefont {Cournapeau}, \citenamefont {Brucher}, \citenamefont {Perrot},\
  and\ \citenamefont {Duchesnay}}]{scikit-learn}%
  \BibitemOpen
  \bibfield  {author} {\bibinfo {author} {\bibfnamefont {F.}~\bibnamefont
  {Pedregosa}}, \bibinfo {author} {\bibfnamefont {G.}~\bibnamefont
  {Varoquaux}}, \bibinfo {author} {\bibfnamefont {A.}~\bibnamefont {Gramfort}},
  \bibinfo {author} {\bibfnamefont {V.}~\bibnamefont {Michel}}, \bibinfo
  {author} {\bibfnamefont {B.}~\bibnamefont {Thirion}}, \bibinfo {author}
  {\bibfnamefont {O.}~\bibnamefont {Grisel}}, \bibinfo {author} {\bibfnamefont
  {M.}~\bibnamefont {Blondel}}, \bibinfo {author} {\bibfnamefont
  {P.}~\bibnamefont {Prettenhofer}}, \bibinfo {author} {\bibfnamefont
  {R.}~\bibnamefont {Weiss}}, \bibinfo {author} {\bibfnamefont
  {V.}~\bibnamefont {Dubourg}}, \bibinfo {author} {\bibfnamefont
  {J.}~\bibnamefont {Vanderplas}}, \bibinfo {author} {\bibfnamefont
  {A.}~\bibnamefont {Passos}}, \bibinfo {author} {\bibfnamefont
  {D.}~\bibnamefont {Cournapeau}}, \bibinfo {author} {\bibfnamefont
  {M.}~\bibnamefont {Brucher}}, \bibinfo {author} {\bibfnamefont
  {M.}~\bibnamefont {Perrot}},\ and\ \bibinfo {author} {\bibfnamefont
  {E.}~\bibnamefont {Duchesnay}},\ }\bibfield  {title} {\enquote {\bibinfo
  {title} {Scikit-learn: Machine learning in {P}ython},}\ }\href@noop {}
  {\bibfield  {journal} {\bibinfo  {journal} {Journal of Machine Learning
  Research}\ }\textbf {\bibinfo {volume} {12}},\ \bibinfo {pages} {2825--2830}
  (\bibinfo {year} {2011})}\BibitemShut {NoStop}%
\bibitem [{\citenamefont {Paszke}\ \emph {et~al.}(2019)\citenamefont {Paszke},
  \citenamefont {Gross}, \citenamefont {Massa}, \citenamefont {Lerer},
  \citenamefont {Bradbury}, \citenamefont {Chanan}, \citenamefont {Killeen},
  \citenamefont {Lin}, \citenamefont {Gimelshein}, \citenamefont {Antiga} \emph
  {et~al.}}]{pytorch}%
  \BibitemOpen
  \bibfield  {author} {\bibinfo {author} {\bibfnamefont {A.}~\bibnamefont
  {Paszke}}, \bibinfo {author} {\bibfnamefont {S.}~\bibnamefont {Gross}},
  \bibinfo {author} {\bibfnamefont {F.}~\bibnamefont {Massa}}, \bibinfo
  {author} {\bibfnamefont {A.}~\bibnamefont {Lerer}}, \bibinfo {author}
  {\bibfnamefont {J.}~\bibnamefont {Bradbury}}, \bibinfo {author}
  {\bibfnamefont {G.}~\bibnamefont {Chanan}}, \bibinfo {author} {\bibfnamefont
  {T.}~\bibnamefont {Killeen}}, \bibinfo {author} {\bibfnamefont
  {Z.}~\bibnamefont {Lin}}, \bibinfo {author} {\bibfnamefont {N.}~\bibnamefont
  {Gimelshein}}, \bibinfo {author} {\bibfnamefont {L.}~\bibnamefont {Antiga}},
  \emph {et~al.},\ }\href@noop {} {\enquote {\bibinfo {title} {Pytorch: An
  imperative style, high-performance deep learning library},}\ } (\bibinfo
  {year} {2019})\BibitemShut {NoStop}%
\bibitem [{\citenamefont {Statt}, \citenamefont {Kleeblatt},\ and\
  \citenamefont {Reinhart}(2021{\natexlab{b}})}]{stattData2021}%
  \BibitemOpen
  \bibfield  {author} {\bibinfo {author} {\bibfnamefont {A.}~\bibnamefont
  {Statt}}, \bibinfo {author} {\bibfnamefont {D.}~\bibnamefont {Kleeblatt}},\
  and\ \bibinfo {author} {\bibfnamefont {W.}~\bibnamefont {Reinhart}},\
  }\bibfield  {title} {\enquote {\bibinfo {title} {{Data for "Unsupervised
  learning of sequence- specific aggregation behavior for a model
  copolymer"}},}\ }\href {https://doi.org/10.5281/zenodo.5303221}
  {10.5281/zenodo.5303221} (\bibinfo {year} {2021}{\natexlab{b}})\BibitemShut
  {NoStop}%
\bibitem [{\citenamefont {Nogueira}(14  )}]{bayesopt}%
  \BibitemOpen
  \bibfield  {author} {\bibinfo {author} {\bibfnamefont {F.}~\bibnamefont
  {Nogueira}},\ }\href {https://github.com/fmfn/BayesianOptimization} {\enquote
  {\bibinfo {title} {{Bayesian Optimization}: Open source constrained global
  optimization tool for {Python}},}\ } (\bibinfo {year} {2014--})\BibitemShut
  {NoStop}%
\bibitem [{\citenamefont {Wen}\ \emph {et~al.}(2019)\citenamefont {Wen},
  \citenamefont {Liu}, \citenamefont {Shi}, \citenamefont {Huang},
  \citenamefont {Deng},\ and\ \citenamefont {Xiao}}]{wen_classification_2019}%
  \BibitemOpen
  \bibfield  {author} {\bibinfo {author} {\bibfnamefont {J.}~\bibnamefont
  {Wen}}, \bibinfo {author} {\bibfnamefont {Y.}~\bibnamefont {Liu}}, \bibinfo
  {author} {\bibfnamefont {Y.}~\bibnamefont {Shi}}, \bibinfo {author}
  {\bibfnamefont {H.}~\bibnamefont {Huang}}, \bibinfo {author} {\bibfnamefont
  {B.}~\bibnamefont {Deng}},\ and\ \bibinfo {author} {\bibfnamefont
  {X.}~\bibnamefont {Xiao}},\ }\bibfield  {title} {\enquote {\bibinfo {title}
  {A classification model for lncrna and mrna based on k-mers and a
  convolutional neural network},}\ }\href@noop {} {\bibfield  {journal}
  {\bibinfo  {journal} {BMC bioinformatics}\ }\textbf {\bibinfo {volume}
  {20}},\ \bibinfo {pages} {1--14} (\bibinfo {year} {2019})}\BibitemShut
  {NoStop}%
\bibitem [{\citenamefont {Solis-Reyes}\ \emph {et~al.}(2018)\citenamefont
  {Solis-Reyes}, \citenamefont {Avino}, \citenamefont {Poon},\ and\
  \citenamefont {Kari}}]{solis-reyes_open-source_2018}%
  \BibitemOpen
  \bibfield  {author} {\bibinfo {author} {\bibfnamefont {S.}~\bibnamefont
  {Solis-Reyes}}, \bibinfo {author} {\bibfnamefont {M.}~\bibnamefont {Avino}},
  \bibinfo {author} {\bibfnamefont {A.}~\bibnamefont {Poon}},\ and\ \bibinfo
  {author} {\bibfnamefont {L.}~\bibnamefont {Kari}},\ }\bibfield  {title}
  {\enquote {\bibinfo {title} {An open-source k-mer based machine learning tool
  for fast and accurate subtyping of {HIV}-1 genomes},}\ }\href
  {https://doi.org/10.1371/journal.pone.0206409} {\bibfield  {journal}
  {\bibinfo  {journal} {PLOS ONE}\ }\textbf {\bibinfo {volume} {13}},\ \bibinfo
  {pages} {e0206409} (\bibinfo {year} {2018})}\BibitemShut {NoStop}%
\bibitem [{\citenamefont {Wang}, \citenamefont {Wang},\ and\ \citenamefont
  {Xu}(2020)}]{wang_evolutionary_2020}%
  \BibitemOpen
  \bibfield  {author} {\bibinfo {author} {\bibfnamefont {H.}~\bibnamefont
  {Wang}}, \bibinfo {author} {\bibfnamefont {H.}~\bibnamefont {Wang}},\ and\
  \bibinfo {author} {\bibfnamefont {K.}~\bibnamefont {Xu}},\ }\bibfield
  {title} {\enquote {\bibinfo {title} {Evolutionary recurrent neural network
  for image captioning},}\ }\href
  {https://doi.org/https://doi.org/10.1016/j.neucom.2020.03.087} {\bibfield
  {journal} {\bibinfo  {journal} {Neurocomputing}\ }\textbf {\bibinfo {volume}
  {401}},\ \bibinfo {pages} {249--256} (\bibinfo {year} {2020})}\BibitemShut
  {NoStop}%
\bibitem [{\citenamefont {Auli}\ \emph {et~al.}(2013)\citenamefont {Auli},
  \citenamefont {Galley}, \citenamefont {Quirk},\ and\ \citenamefont
  {Zweig}}]{auli_joint_2013}%
  \BibitemOpen
  \bibfield  {author} {\bibinfo {author} {\bibfnamefont {M.}~\bibnamefont
  {Auli}}, \bibinfo {author} {\bibfnamefont {M.}~\bibnamefont {Galley}},
  \bibinfo {author} {\bibfnamefont {C.}~\bibnamefont {Quirk}},\ and\ \bibinfo
  {author} {\bibfnamefont {G.}~\bibnamefont {Zweig}},\ }\bibfield  {title}
  {\enquote {\bibinfo {title} {Joint {Language} and {Translation} {Modeling}
  with {Recurrent} {Neural} {Networks}},}\ }in\ \href
  {https://www.microsoft.com/en-us/research/publication/joint-language-and-translation-modeling-with-recurrent-neural-networks/}
  {\emph {\bibinfo {booktitle} {Proc. of {EMNLP}}}}\ (\bibinfo {year} {2013})\
  \bibinfo {note} {edition: Proc. of EMNLP}\BibitemShut {NoStop}%
\bibitem [{\citenamefont {Zhang}, \citenamefont {Chen},\ and\ \citenamefont
  {Qin}(2018)}]{8588934}%
  \BibitemOpen
  \bibfield  {author} {\bibinfo {author} {\bibfnamefont {X.}~\bibnamefont
  {Zhang}}, \bibinfo {author} {\bibfnamefont {M.~H.}\ \bibnamefont {Chen}},\
  and\ \bibinfo {author} {\bibfnamefont {Y.}~\bibnamefont {Qin}},\ }\bibfield
  {title} {\enquote {\bibinfo {title} {Nlp-qa framework based on lstm-rnn},}\
  }in\ \href {https://doi.org/10.1109/ICDSBA.2018.00065} {\emph {\bibinfo
  {booktitle} {2018 2nd International Conference on Data Science and Business
  Analytics (ICDSBA)}}}\ (\bibinfo {year} {2018})\ pp.\ \bibinfo {pages}
  {307--311}\BibitemShut {NoStop}%
\bibitem [{\citenamefont {Cho}\ \emph {et~al.}(2014)\citenamefont {Cho},
  \citenamefont {Van~Merri{\"e}nboer}, \citenamefont {Gulcehre}, \citenamefont
  {Bahdanau}, \citenamefont {Bougares}, \citenamefont {Schwenk},\ and\
  \citenamefont {Bengio}}]{cho2014learning}%
  \BibitemOpen
  \bibfield  {author} {\bibinfo {author} {\bibfnamefont {K.}~\bibnamefont
  {Cho}}, \bibinfo {author} {\bibfnamefont {B.}~\bibnamefont
  {Van~Merri{\"e}nboer}}, \bibinfo {author} {\bibfnamefont {C.}~\bibnamefont
  {Gulcehre}}, \bibinfo {author} {\bibfnamefont {D.}~\bibnamefont {Bahdanau}},
  \bibinfo {author} {\bibfnamefont {F.}~\bibnamefont {Bougares}}, \bibinfo
  {author} {\bibfnamefont {H.}~\bibnamefont {Schwenk}},\ and\ \bibinfo {author}
  {\bibfnamefont {Y.}~\bibnamefont {Bengio}},\ }\bibfield  {title} {\enquote
  {\bibinfo {title} {Learning phrase representations using rnn encoder-decoder
  for statistical machine translation},}\ }\href@noop {} {\bibfield  {journal}
  {\bibinfo  {journal} {arXiv preprint arXiv:1406.1078}\ } (\bibinfo {year}
  {2014})}\BibitemShut {NoStop}%
\bibitem [{\citenamefont {Hochreiter}\ and\ \citenamefont
  {Schmidhuber}(1997)}]{hochreiter_long_1997}%
  \BibitemOpen
  \bibfield  {author} {\bibinfo {author} {\bibfnamefont {S.}~\bibnamefont
  {Hochreiter}}\ and\ \bibinfo {author} {\bibfnamefont {J.}~\bibnamefont
  {Schmidhuber}},\ }\bibfield  {title} {\enquote {\bibinfo {title} {Long
  short-term memory},}\ }\href {https://doi.org/10.1162/neco.1997.9.8.1735}
  {\bibfield  {journal} {\bibinfo  {journal} {Neural Computation}\ }\textbf
  {\bibinfo {volume} {9}},\ \bibinfo {pages} {1735--1780} (\bibinfo {year}
  {1997})}\BibitemShut {NoStop}%
\bibitem [{\citenamefont {Ju}, \citenamefont {Zhang},\ and\ \citenamefont
  {Zhu}(2019)}]{ju2019study}%
  \BibitemOpen
  \bibfield  {author} {\bibinfo {author} {\bibfnamefont {Y.}~\bibnamefont
  {Ju}}, \bibinfo {author} {\bibfnamefont {M.}~\bibnamefont {Zhang}},\ and\
  \bibinfo {author} {\bibfnamefont {H.}~\bibnamefont {Zhu}},\ }\bibfield
  {title} {\enquote {\bibinfo {title} {Study on a new deep bidirectional gru
  network for electrocardiogram signals classification},}\ }in\ \href@noop {}
  {\emph {\bibinfo {booktitle} {3rd International Conference on Computer
  Engineering, Information Science \& Application Technology (ICCIA 2019)}}}\
  (\bibinfo {organization} {Atlantis Press},\ \bibinfo {year} {2019})\ pp.\
  \bibinfo {pages} {355--359}\BibitemShut {NoStop}%
\bibitem [{\citenamefont {Du}\ \emph {et~al.}(2018)\citenamefont {Du},
  \citenamefont {Pan}, \citenamefont {Wang},\ and\ \citenamefont
  {Ji}}]{du2018biomedical}%
  \BibitemOpen
  \bibfield  {author} {\bibinfo {author} {\bibfnamefont {Y.}~\bibnamefont
  {Du}}, \bibinfo {author} {\bibfnamefont {Y.}~\bibnamefont {Pan}}, \bibinfo
  {author} {\bibfnamefont {C.}~\bibnamefont {Wang}},\ and\ \bibinfo {author}
  {\bibfnamefont {J.}~\bibnamefont {Ji}},\ }\bibfield  {title} {\enquote
  {\bibinfo {title} {Biomedical semantic indexing by deep neural network with
  multi-task learning},}\ }\href@noop {} {\bibfield  {journal} {\bibinfo
  {journal} {BMC bioinformatics}\ }\textbf {\bibinfo {volume} {19}},\ \bibinfo
  {pages} {1--11} (\bibinfo {year} {2018})}\BibitemShut {NoStop}%
\bibitem [{\citenamefont {Maulud}\ and\ \citenamefont
  {Abdulazeez}(2020)}]{maulud_review_2020}%
  \BibitemOpen
  \bibfield  {author} {\bibinfo {author} {\bibfnamefont {D.}~\bibnamefont
  {Maulud}}\ and\ \bibinfo {author} {\bibfnamefont {A.~M.}\ \bibnamefont
  {Abdulazeez}},\ }\bibfield  {title} {\enquote {\bibinfo {title} {A {Review}
  on {Linear} {Regression} {Comprehensive} in {Machine} {Learning}},}\ }\href
  {https://doi.org/10.38094/jastt1457} {\bibfield  {journal} {\bibinfo
  {journal} {Journal of Applied Science and Technology Trends}\ }\textbf
  {\bibinfo {volume} {1}},\ \bibinfo {pages} {140--147} (\bibinfo {year}
  {2020})}\BibitemShut {NoStop}%
\bibitem [{\citenamefont {Graves}\ and\ \citenamefont
  {Schmidhuber}(2005)}]{1556215}%
  \BibitemOpen
  \bibfield  {author} {\bibinfo {author} {\bibfnamefont {A.}~\bibnamefont
  {Graves}}\ and\ \bibinfo {author} {\bibfnamefont {J.}~\bibnamefont
  {Schmidhuber}},\ }\bibfield  {title} {\enquote {\bibinfo {title} {Framewise
  phoneme classification with bidirectional lstm networks},}\ }in\ \href
  {https://doi.org/10.1109/IJCNN.2005.1556215} {\emph {\bibinfo {booktitle}
  {Proceedings. 2005 IEEE International Joint Conference on Neural Networks,
  2005.}}},\ Vol.~\bibinfo {volume} {4}\ (\bibinfo {year} {2005})\ pp.\
  \bibinfo {pages} {2047--2052 vol. 4}\BibitemShut {NoStop}%
\bibitem [{\citenamefont {Rajan}(2022)}]{rajan_efficient_2022}%
  \BibitemOpen
  \bibfield  {author} {\bibinfo {author} {\bibfnamefont {M.~P.}\ \bibnamefont
  {Rajan}},\ }\bibfield  {title} {\enquote {\bibinfo {title} {An {Efficient}
  {Ridge} {Regression} {Algorithm} with {Parameter} {Estimation} for {Data}
  {Analysis} in {Machine} {Learning}},}\ }\href
  {https://doi.org/10.1007/s42979-022-01051-x} {\bibfield  {journal} {\bibinfo
  {journal} {SN Computer Science}\ }\textbf {\bibinfo {volume} {3}},\ \bibinfo
  {pages} {171} (\bibinfo {year} {2022})}\BibitemShut {NoStop}%
\bibitem [{\citenamefont {Zhang}(2016)}]{ATM10170}%
  \BibitemOpen
  \bibfield  {author} {\bibinfo {author} {\bibfnamefont {Z.}~\bibnamefont
  {Zhang}},\ }\bibfield  {title} {\enquote {\bibinfo {title} {Variable
  selection with stepwise and best subset approaches},}\ }\href@noop {}
  {\bibfield  {journal} {\bibinfo  {journal} {Annals of translational
  medicine}\ }\textbf {\bibinfo {volume} {4}} (\bibinfo {year}
  {2016})}\BibitemShut {NoStop}%
\bibitem [{\citenamefont {Patel}\ and\ \citenamefont
  {Jokhakar}(2016)}]{7919549}%
  \BibitemOpen
  \bibfield  {author} {\bibinfo {author} {\bibfnamefont {S.~V.}\ \bibnamefont
  {Patel}}\ and\ \bibinfo {author} {\bibfnamefont {V.~N.}\ \bibnamefont
  {Jokhakar}},\ }\bibfield  {title} {\enquote {\bibinfo {title} {A random
  forest based machine learning approach for mild steel defect diagnosis},}\
  }in\ \href {https://doi.org/10.1109/ICCIC.2016.7919549} {\emph {\bibinfo
  {booktitle} {2016 IEEE International Conference on Computational Intelligence
  and Computing Research (ICCIC)}}}\ (\bibinfo {year} {2016})\ pp.\ \bibinfo
  {pages} {1--8}\BibitemShut {NoStop}%
\bibitem [{\citenamefont {Levenshtein}(1966)}]{levenshtein1966binary}%
  \BibitemOpen
  \bibfield  {author} {\bibinfo {author} {\bibfnamefont {V.~I.}\ \bibnamefont
  {Levenshtein}},\ }\bibfield  {title} {\enquote {\bibinfo {title} {Binary
  codes capable of correcting deletions, insertions, and reversals},}\ }in\
  \href@noop {} {\emph {\bibinfo {booktitle} {Soviet Physics Doklady}}},\
  Vol.~\bibinfo {volume} {10}\ (\bibinfo {organization} {Soviet Union},\
  \bibinfo {year} {1966})\ pp.\ \bibinfo {pages} {707--710}\BibitemShut
  {NoStop}%
\bibitem [{\citenamefont {Bachmann}(10  )}]{levenshtein}%
  \BibitemOpen
  \bibfield  {author} {\bibinfo {author} {\bibfnamefont {M.}~\bibnamefont
  {Bachmann}},\ }\href {https://github.com/maxbachmann/Levenshtein} {\enquote
  {\bibinfo {title} {Levenshtein},}\ } (\bibinfo {year} {2010--})\BibitemShut
  {NoStop}%
\bibitem [{\citenamefont {Bhattacharya}\ \emph {et~al.}(2022)\citenamefont
  {Bhattacharya}, \citenamefont {Kleeblatt}, \citenamefont {Statt},\ and\
  \citenamefont {Reinhart}}]{this_data}%
  \BibitemOpen
  \bibfield  {author} {\bibinfo {author} {\bibfnamefont {D.}~\bibnamefont
  {Bhattacharya}}, \bibinfo {author} {\bibfnamefont {D.}~\bibnamefont
  {Kleeblatt}}, \bibinfo {author} {\bibfnamefont {A.}~\bibnamefont {Statt}},\
  and\ \bibinfo {author} {\bibfnamefont {W.}~\bibnamefont {Reinhart}},\
  }\bibfield  {title} {\enquote {\bibinfo {title} {{Data for "Predicting
  aggregate morphology of sequence-defined macromolecules with Recurrent Neural
  Networks"}},}\ }\href {https://doi.org/10.5281/zenodo.6585654}
  {10.5281/zenodo.6585654} (\bibinfo {year} {2022})\BibitemShut {NoStop}%
\bibitem [{\citenamefont {Reinhart}\ and\ \citenamefont
  {Bhattacharya}(2022)}]{this_github}%
  \BibitemOpen
  \bibfield  {author} {\bibinfo {author} {\bibfnamefont {W.}~\bibnamefont
  {Reinhart}}\ and\ \bibinfo {author} {\bibfnamefont {D.}~\bibnamefont
  {Bhattacharya}},\ }\href {https://github.com/wfreinhart/sdmm-regression}
  {\enquote {\bibinfo {title} {sdmm-regression},}\ } (\bibinfo {year}
  {2022})\BibitemShut {NoStop}%
\end{thebibliography}%
\end{document}


\preprint{--}

\title{Electronic Supplementary Information for Predicting aggregate morphology of sequence-defined macromolecules with Recurrent Neural Networks}

\author{Debjyoti Bhattacharya}
\affiliation{Materials Science and Engineering, Pennsylvania State University, PA 16802 }

\author{Devon C. Kleeblatt}
\affiliation{Materials Science and Engineering, Pennsylvania State University, PA 16802 }

\author{Antonia Statt}
\affiliation{Materials Science and Engineering, Grainger College of Engineering, University of Illinois, Urbana-Champaign, IL 61801 }

\author{Wesley F. Reinhart}
\email[email:]{reinhart@psu.edu}
\affiliation{Materials Science and Engineering, Pennsylvania State University, PA 16802 }
\affiliation{Institute for Computational and Data Sciences, Pennsylvania State University, PA 16802 }

\date{\today}

\maketitle

The origin of the $63 \, 090$ unique sequences used in High Throughput Screening can be explained by permutations and combinations of the fixed number of A type beads within the monomer sequence.
The total number of possible sequences is $\frac{20!}{12!(8)!} = 125 \, 970$.
However, this treats forward and reverse sequences as distinct, thereby double-counting, hence it needs to be halved to $62 \, 985$ sequences.
Then, we need to add back the perfectly symmetrical sequences.
In order to count the perfectly symmetrical sequences, there would be 2 possible cases wherein the 10th and 11th position would be occupied by either AA or BB and the other 2 possible cases AB or BA are not possible because then the sequence would be asymmetrical (as it would then leave behind 7 A and 11 B to arrange in either of the sides of the 10th and 11th positions).
Hence, the possible symmetrical sequences for the case where AA occupies 10th and 11th positions would be $\frac{9!}{2\cdot6!(3)!}=42$.
Similarly, for the case where BB occupies the 10th and 11th positions, the number of possible sequences would be  $\frac{9!}{2\cdot5!(4)!}=63$.
Adding these symmetrical sequences back, gives us a total of $62 \, 985 + 42 + 63 = 63 \, 090$ possible sequences.

Note that in practice, we arrived at this number by exhaustive enumeration and explicitly checking for symmetry.
This explanation is only provided post hoc for the satisfaction of the interested reader.

\begin{figure}
    \centering
    \includegraphics[width=0.8\columnwidth]{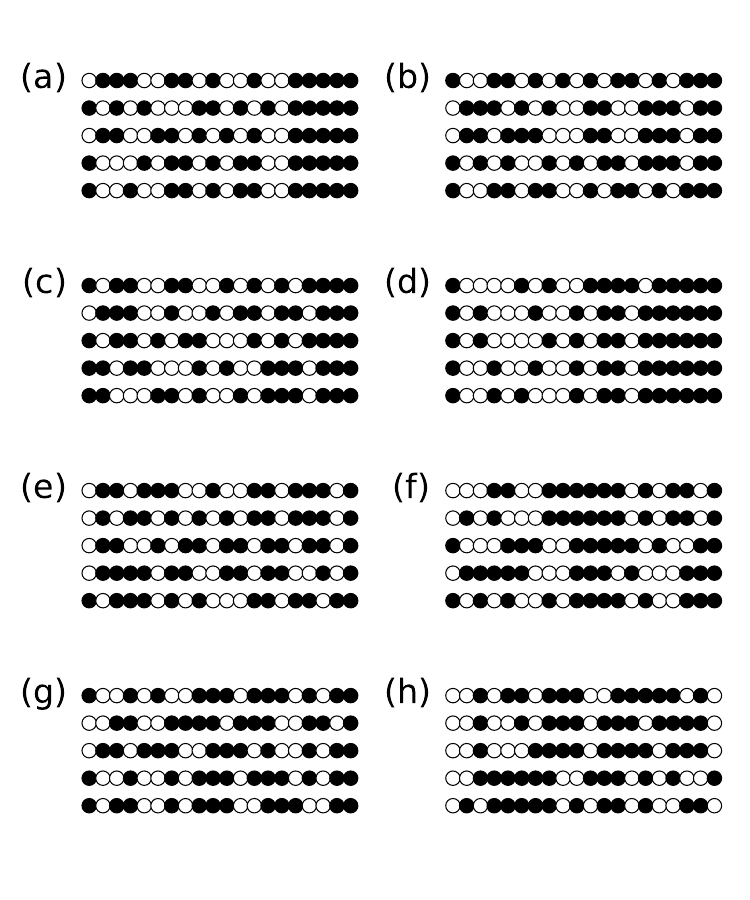}
    \caption{Ranked list of sequences used to generate Fig.~8(b) in the main text.
    }
    \label{fig:sequences-kmeans}
\end{figure}

\begin{figure}[h]
    \centering
    \includegraphics[width=\textwidth]{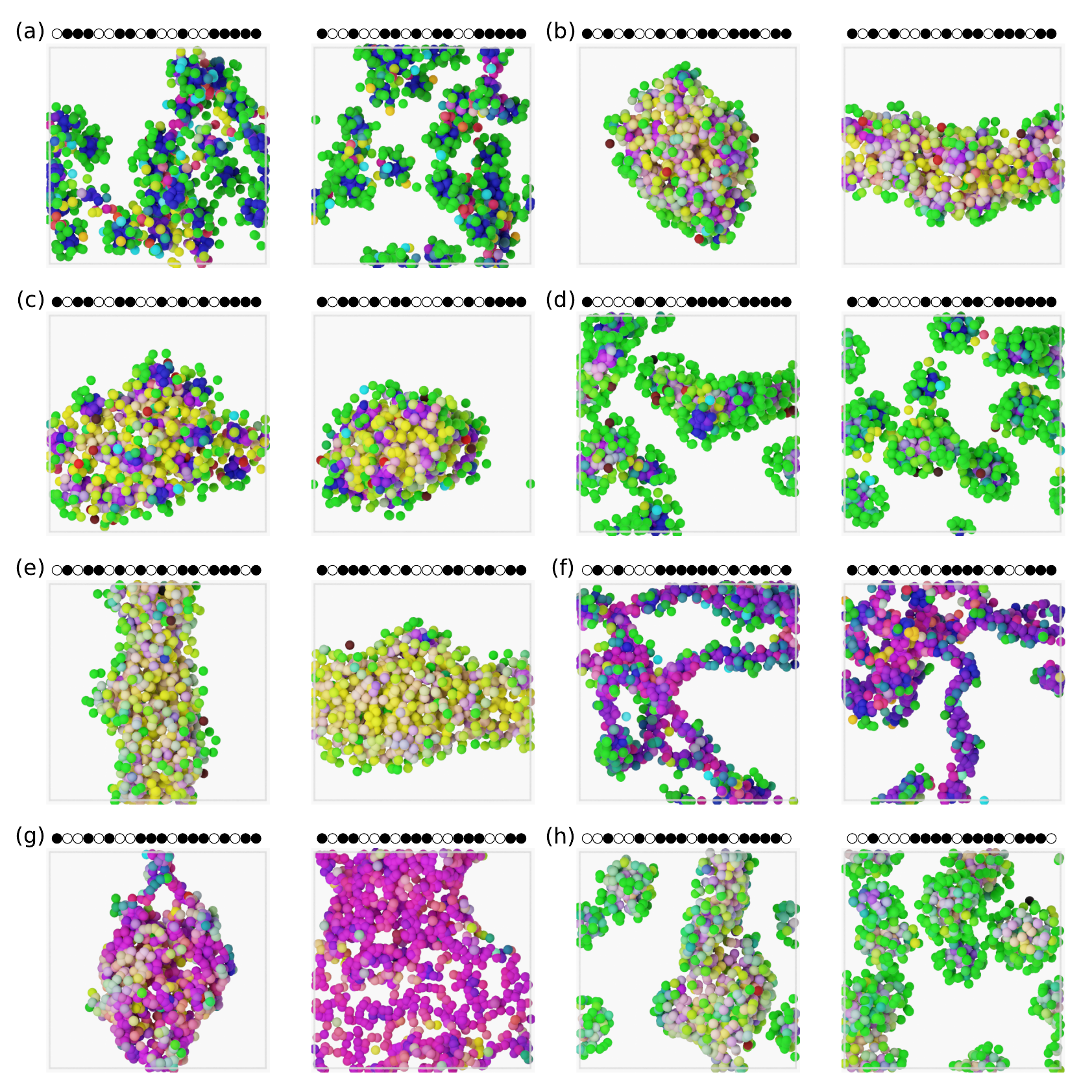}
    \caption{Snapshots from the simulations in Fig.~8(b) in the main text, corresponding to the targets selected by K-Means clustering.
    Coloring is determined by the local environment around each particle, as described in Ref.~74.
    The left panels are closest to the target in the batch of candidates (the best result of 25 samples), right panels are farthest away (the worst result of 25 samples).
    Labels correspond to those in Fig.~8(b).
    \label{fig:snapshots-archetypes}}
\end{figure}

\begin{figure}[h]
    \centering
    \includegraphics[width=0.48\textwidth]{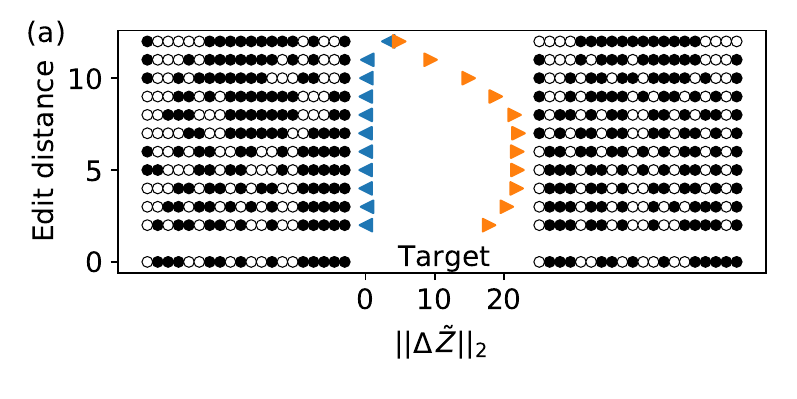}
    \includegraphics[width=0.48\textwidth]{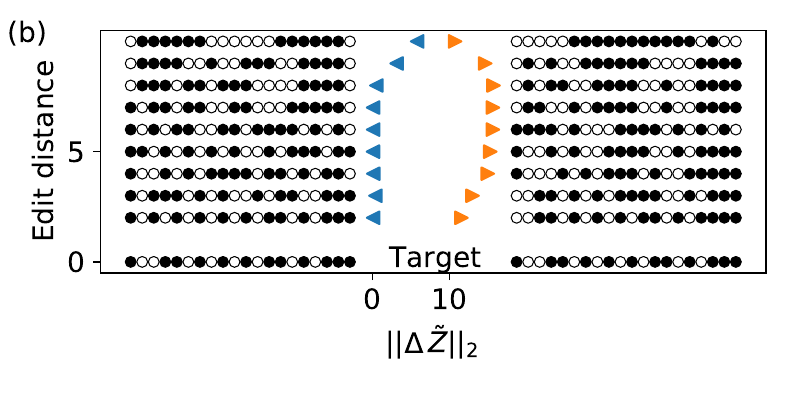}
    \includegraphics[width=0.48\textwidth]{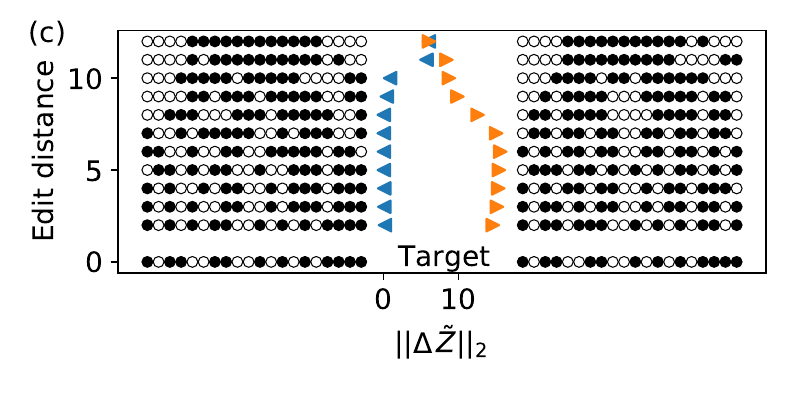}
    \includegraphics[width=0.48\textwidth]{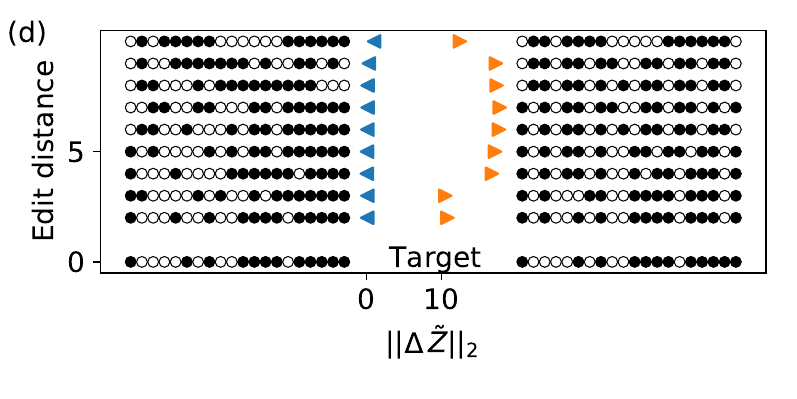}
    \includegraphics[width=0.48\textwidth]{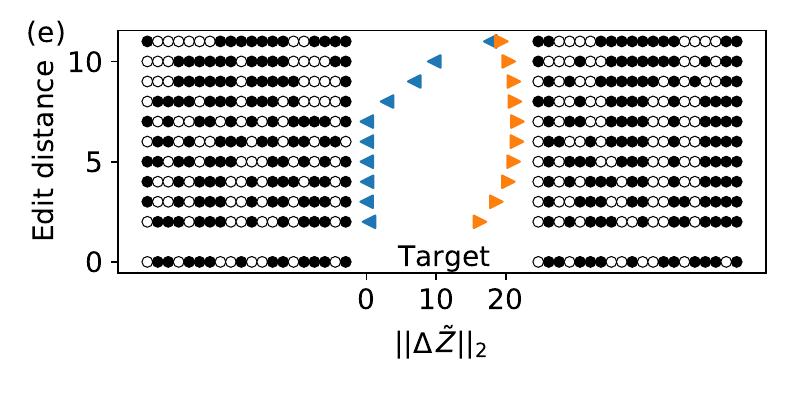}
    \includegraphics[width=0.48\textwidth]{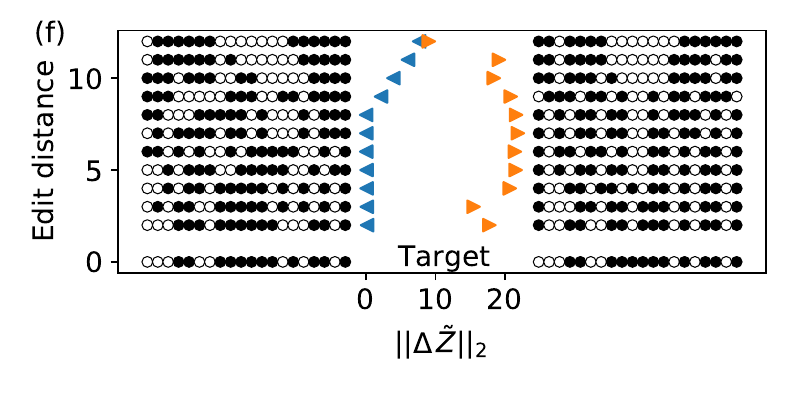}
    \includegraphics[width=0.48\textwidth]{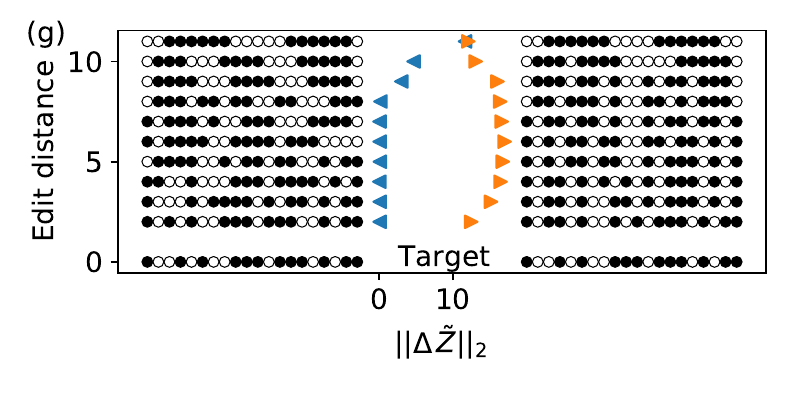}
    \includegraphics[width=0.48\textwidth]{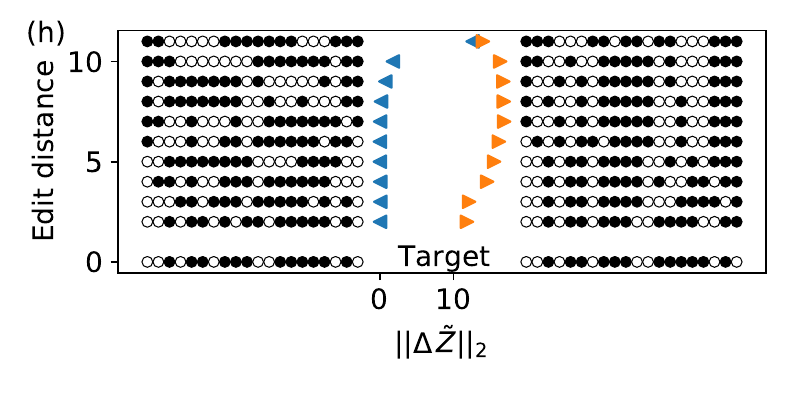}
    \caption{Contrastive analysis of RNN inference on selected sequences from Fig.~8(b).
    Symbols show Levenshtein edit distance versus predicted distance from target sequence in $Z$ space.
    Blue left triangles show minimum $||\Delta \tilde{Z}||_2$ at fixed edit distance from among all possible sequences (at fixed composition), while orange right triangles show maximum.}
\end{figure}
